\documentclass[twocolumn, floatfix, tighten, fleqn, useAMS, usenatbib]{aastex63}
\acceptjournal{ApJ}

\RequirePackage{silence}
\usepackage{float}
\usepackage{graphicx}	
\usepackage{amsmath}	
\usepackage{amssymb}	
\usepackage{amsfonts}
\usepackage{bm}		
\usepackage{mathtools}
\usepackage{booktabs}
\usepackage{listings}
\usepackage{color}
\usepackage{threeparttable}
\usepackage{multirow}
\usepackage[perpage,symbol*,flushmargin]{footmisc}
\usepackage{ae,aecompl}
\usepackage{array}
\usepackage{soul}

\usepackage[T1]{fontenc}
\usepackage[perpage,symbol*]{footmisc}
\usepackage{ae,aecompl}

\DeclareMathOperator{\sech}{sech}

\DeclareRobustCommand{\appropto}{\mathrel{\vcenter{
			\offinterlineskip\halign{\hfil$##$\cr 
				\propto\cr\noalign{\kern2pt}\sim\cr\noalign{\kern-2pt}}}}}

\defcitealias{Sellwood_2019}{S19} 

\begin{document}

\title{The global stability of M33 in MOND}
\shorttitle{The global stability of M33 in MOND} 

\shortauthors{I. Banik, I. Thies, G. Candlish, B. Famaey, R. Ibata \& P. Kroupa} 

\author[0000-0002-4123-7325]{Indranil Banik}
\affiliation{Helmholtz-Institut f\"ur Strahlen und Kernphysik (HISKP), University of Bonn, Nussallee 14$-$16, D-53115 Bonn, Germany}
\altaffiliation{Humboldt fellow}
\email{ibanik@astro.uni-bonn.de}

\author{Ingo Thies}
\affiliation{Helmholtz-Institut f\"ur Strahlen und Kernphysik (HISKP), University of Bonn, Nussallee 14$-$16, D-53115 Bonn, Germany}

\author[0000-0003-3180-9825]{Benoit Famaey}
\affiliation{Universit\'{e} de Strasbourg, CNRS UMR 7550, Observatoire astronomique de Strasbourg, 11 rue de l'Universit\'{e},\\ 67000 Strasbourg, France}

\author[0000-0001-9511-2782]{Graeme Candlish}
\affiliation{Instituto de F\'{i}sica y Astronom\'{i}a, Universidad de Valpara\'{i}so, Gran Breta\~{n}a 1111, Valpara\'{i}so, Chile}

\author[0000-0002-7301-3377]{Pavel Kroupa}
\affiliation{Helmholtz-Institut f\"ur Strahlen und Kernphysik (HISKP), University of Bonn, Nussallee 14$-$16, D-53115 Bonn, Germany}
\affiliation{Astronomical Institute, Faculty of Mathematics and Physics, Charles University in Prague, V Hole\v{s}ovi\v{c}k\'ach 2,\\ CZ-180 00 Praha 8, Czech Republic}

\author[0000-0002-3292-9709]{Rodrigo Ibata}
\affiliation{Universit\'{e} de Strasbourg, CNRS UMR 7550, Observatoire astronomique de Strasbourg, 11 rue de l'Universit\'{e},\\ 67000 Strasbourg, France}

\begin{abstract} 

The dynamical stability of disk galaxies is sensitive to whether their anomalous rotation curves are caused by dark matter halos or Milgromian Dynamics (MOND). We investigate this by setting up a MOND model of M33. We first simulate it in isolation for 6 Gyr, starting from an initial good match to the rotation curve (RC). Too large a bar and bulge form when the gas is too hot, but this is avoided by reducing the gas temperature. A strong bar still forms in 1 Gyr, but rapidly weakens and becomes consistent with the observed weak bar. Previous work showed this to be challenging in Newtonian models with a live dark matter halo, which developed strong bars. The bar pattern speed implies a realistic corotation radius of 3 kpc. However, the RC still rises too steeply, and the central line of sight velocity dispersion (LOSVD) is too high. We then add a constant external acceleration field of $8.4 \times 10^{-12}$ m/s$^2$ at $30^\circ$ to the disk as a first order estimate for the gravity exerted by M31. This suppresses buildup of material at the centre, causing the RC to rise more slowly and reducing the central LOSVD. Overall, this simulation bears good resemblance to several global properties of M33, and highlights the importance of including even a weak external field on the stability and evolution of disk galaxies. Further simulations with a time-varying external field, modeling the full orbit of M33, will be needed to confirm its resemblance to observations.

\end{abstract}

\keywords{Modified Newtonian Dynamics (1069) -- Triangulum Galaxy (1712) -- Barred spiral galaxies (136) -- Hydrodynamical simulations (767) -- Stellar dynamics (1596) -- Galaxy evolution (594)}

\section{Introduction}
\label{Introduction}

The ubiquity of thin disk galaxies belies their central role in a major astronomical mystery $-$ why do their rotation curves (RCs) go flat in their outskirts instead of following the expected Keplerian decline beyond the extent of their luminous matter \citep[e.g.][]{Babcock_1939, Rubin_Ford_1970, Rogstad_1972, Roberts_1975, Bosma_1981}? These acceleration discrepancies are historically related to the issue of disk galaxy stability. \citet{Hohl_1971} showed that a self-gravitating Newtonian disk is very unstable. Given the age of galaxy disks like that of our own Milky Way \citep{Knox_1999}, it is clear that something fundamental is missing from these simulations.

\citet{Ostriker_Peebles_1973} suggested the addition of a dominant dark matter (DM) halo around the disk, which can stabilize it. Such DM halos around galaxies are now considered an essential component of the Lambda-Cold Dark Matter ($\Lambda$CDM) cosmological paradigm \citep{Efstathiou_1990, Ostriker_Steinhardt_1995}. In this model, disks are not self-gravitating, but are mostly held together by a DM halo. The extra gravity from the halo elevates the disk RC, and can make it asymptotically flat in the outskirts. If the total gravity $g$ consists of a halo contribution $g_{_h}$ and a disk contribution $g_{_d}$, then the fractional change in $g$ from some disk surface density perturbation is reduced by a factor of $\left( g_{_d}/g \right)$ due to the halo \citep{Banik_2018_Toomre}. In this way, the halo mass interior to the radius of the disk can make the disk more stable and elevate its RC. 


\subsection{MOND}
\label{External_field_effect}

However, several problems remain to this day in matching detailed properties of RCs \citep[e.g.][]{Oman_2015, Desmond_2016, Desmond_2017} and the long-term stability of some disk galaxies \citep[e.g.][hereafter \citetalias{Sellwood_2019}]{Sellwood_2019}. Even if they were solved, agreement between theory and observations does not prove a theory correct $-$ there may be alternative explanations for the same data. In particular, the RC anomalies could be a sign that Newtonian dynamics breaks down on galaxy scales. It would after all not be too surprising if a theory developed exclusively using Solar System data cannot be reliably extrapolated by many orders of magnitude in distance and acceleration. The best developed proposal for how a breakdown might occur is Milgromian Dynamics \citep[MOND,][]{Milgrom_1983, Bekenstein_Milgrom_1984, Milgrom_2014}. In MOND, the gravitational field strength $g$ at distance $R$ from an isolated point mass $M$ transitions from the Newtonian ${GM/R^2}$ law at short range to:
\begin{eqnarray}
	g ~=~ \frac{\sqrt{GMa_{_0}}}{R} \, , \quad R \gg \sqrt{\frac{GM}{a_{_0}}} \, .
	\label{Deep_MOND_limit}
\end{eqnarray}
MOND introduces $a_{_0}$ as a fundamental acceleration scale of nature below which the deviation from Newtonian dynamics becomes significant. When $g \ll a_{_0}$, a system is said to be in the deep-MOND limit (DML) in which the point mass gravity declines only inversely with $R$, making the dynamics scale invariant \citep{Milgrom_2009_DML}. Empirically, $a_{_0} \approx 1.2 \times {10}^{-10}$ m/s$^2$ to match galaxy RCs \citep{Begeman_1991, Gentile_2011}. With this value of $a_{_0}$, MOND continues to this day to fit galaxy RCs very well using only their directly observed baryonic matter \citep[e.g.][]{Kroupa_2018, Li_2018, Sanders_2019}. The properties of polar ring galaxies \citep{Lughausen_2013} and shell galaxies \citep{Bilek_2015} are also consistent with MOND. It has been suggested that MOND is a manifestation of the quantum vacuum, and thus holds important clues on how to unify quantum mechanics with gravity \citep{Milgrom_1999, Verlinde_2016, Smolin_2017}. 


Equation \ref{Deep_MOND_limit} predicts that galaxy RCs should become asymptotically flat at a plateau with level $v_{_f}$, which depends only on the baryonic mass $M$ of the galaxy according to:
\begin{eqnarray}
	v_{_f} ~=~ \sqrt[4]{GMa_{_0}} \, .
	\label{BTFR}
\end{eqnarray}
The observational counterpart to this prediction is commonly known as the baryonic Tully-Fisher relation \citep[BTFR,][]{McGaugh_2000}, a generalization of the original luminous Tully-Fisher relation \citep{Tully_Fisher_1977}. Although this played an important role in the development of MOND, it was certainly not clear in the 1980s that the BTFR would continue to remain very tight once the gas mass was included in low mass galaxies, and more generally once galaxies with rather different properties were observed \citep{McGaugh_2012, Lelli_2016}.

MOND can also be applied to systems which do not have spherical symmetry. This requires the use of a generalized Poisson equation, which can be derived from a Lagrangian \citep{Bekenstein_Milgrom_1984, QUMOND}. We discuss this further in Section \ref{Simulation_setup} since we will need this for our numerical simulations. Unlike the standard Poisson equation, the MOND version is non-linear, as is readily apparent from Equation \ref{BTFR}. The fact that MOND gravity from different sources cannot be superposed creates an external field effect \citep[EFE,][]{Milgrom_1986}. As a result, the gravity from surrounding structures induces a weakening of the galaxy's gravitational field, but without affecting its inertial mass. This can cause the RC to decline at large distance \citep{Haghi_2016, Hees_2016}. In principle, it is possible to estimate this effect based on the environment of the galaxy in question \citep{Wu_2015}. Since large isolated galaxies are typically used for RC analyses, deviations from the BTFR are generally expected to be small. Nonetheless, \citet{Chae_2020} recently reported a highly significant detection of the predicted correlation between external gravitational environment and RC deviations from the isolated MOND prediction.

The EFE is more important for dwarfs near a massive host. It is particularly important to explain the diverse velocity dispersions of galaxies such as Dragonfly 2 \citep[DF2,][]{Famaey_2018, Kroupa_2018_Nature}, DF4 \citep{Haghi_2019}, or DF44 \citep{Bilek_2019, Haghi_2019_DF44}. The predicted dispersions for DF2 and DF4 are significantly affected by the EFE. However, the more isolated DF44 does indeed have a higher internal velocity dispersion despite a similar baryonic content. Some evidence for the EFE was also found in the dwarf satellite galaxies of M31 \citep{McGaugh_2013} and in the Milky Way satellite Crater 2 \citep{McGaugh_2016, Caldwell_2017}. Without the EFE, it is not possible to escape from a mass as its potential is logarithmically divergent (Equation \ref{Deep_MOND_limit}). Thus, the fact that MOND with the EFE can accommodate the Galactic escape velocity curve also argues in favour of the EFE \citep{Famaey_2007, Banik_2018_escape}. 

Recently, the second data release of the Gaia mission \citep{Gaia_2018} has been used to convincingly demonstrate that MOND cannot function without some form of EFE. The evidence comes from the velocity distribution of wide binary stars in the Solar neighbourhood $-$ at separations $\ga 5$~kAU, MOND should affect their orbits \citep{Hernandez_2012}. The predicted boost to their orbital velocity is rather large without the EFE, but a more modest 20\% if the Galactic EFE is included \citep{Banik_2018_Centauri}. Observations of wide binaries completely rule out MOND without the EFE \citep{Pittordis_2019}. The more subtle effect with the EFE is still allowed observationally, and should be testable with future data releases and/or further analysis. The main source of contamination is probably hierarchical triples, which should be accounted for statistically \citep{Belokurov_2020, Clarke_2020}.

In addition to these small scale tests, MOND might also be strongly constrained by the success of $\Lambda$CDM in fitting large scale observations like the cosmic microwave background \citep[CMB,][]{Planck_2020}. However, the excellent fit they obtain is based on an expansion rate history that requires a local Hubble constant below the observed value \citep[e.g.][]{Riess_2020}. This could be due to our position within a large local supervoid underdense by $\approx 30\%$ out to a radius of $\approx 300$~Mpc \citep{Keenan_2013}. Though observed in multiple surveys, such a large and deep underdensity is incompatible with $\Lambda$CDM at ${6.04 \sigma}$ \citep*{Haslbauer_2020}. Those authors showed that such a void could arise naturally in a MOND cosmology supplemented by light sterile neutrinos, as proposed by \citet{Angus_2009}. This model would have a nearly standard expansion history, primordial light element abundances, and CMB anisotropies. However, it would produce more structure than $\Lambda$CDM, in line with earlier analytic expectations \citep{Sanders_1998_cosmology}. This allows for the existence of large `Hubble bubbles' with enhanced apparent expansion rate, sufficient to explain several key local Universe observables at only $2.53\sigma$ tension using a background expansion history fixed to the baseline assumption of \citet{Planck_2020} and without placing us at a special location in the void. MOND with sterile neutrinos therefore appears to account for astronomical observations ranging from the kpc scales of galaxies all the way to the Gpc scale of the local supervoid, without causing any obvious problems in the early Universe. The large scale implications of MOND should therefore be considered in future work, especially in light of upcoming surveys and ongoing tensions explaining current observations within $\Lambda$CDM.

\subsection{Disk galaxy stability}
\label{Disk_galaxy_stability}

In this contribution, we deal with numerical simulations of disk galaxies, focusing in particular on the Local Group galaxy M33. In principle, this regime allows us to avoid the use of a relativistic MOND theory, whose weak-field limit should give a classical modification of gravity.\footnote{A relativistic MOND theory was recently developed in which gravitational waves travel at the speed of light \citep{Skordis_2019}.} Previous analytic and numerical studies indicate that MOND can stabilize a self-gravitating disk without DM \citep{Milgrom_1989, Brada_1999}. Essentially, this is because the deep-MOND force $\propto \sqrt{M}$ rather than $M$ (Equation \ref{Deep_MOND_limit}), limiting the extra gravity created by an overdensity. This effect saturates in the deep-MOND limit (DML), so MOND can provide only a limited amount of additional stability even for a galaxy with arbitrarily low surface density. This might explain why LSBs often have spiral features \citep{McGaugh_1995}, usually taken as a sign that disk self-gravity is important \citep{Lin_1964, Mousumi_2020}. 



The global stability of disks thus offers a promising way to break the CDM-modified gravity degeneracy in galaxies. M33 is one of the nearest galaxies with a rather low surface brightness. \citetalias{Sellwood_2019} recently conducted isolated $N$-body and hydrodynamical (hydro) simulations of M33 in the CDM context. As with many other galaxies, its observed RC required them to use a cored DM profile. They found it rather difficult to explain the weak observed bar together with the two-armed spiral \citep{Corbelli_2007_image} in the presence of a live CDM halo. Weak bars are also observed in many other galaxies \citep{Cuomo_2019}. They are problematic in the CDM picture because a bar in the baryons creates a response in the live DM halo, which would cause a resonant effect \citep{Athanassoula_2002}. This was demonstrated explicitly by \citetalias{Sellwood_2019}, who compared their live halo simulations to a `frozen halo' simulation in which the DM particles provide a fixed gravitational potential.\footnote{This can be thought of as the DM particles providing gravity but not moving.} \citetalias{Sellwood_2019} showed that M33 would be more stable in a frozen halo, leading to a weak bar similar to that in the observed M33. Such a model is of course unphysical in the $\Lambda$CDM context, but useful to better understand the dynamics.

Interestingly, a frozen halo is somewhat similar to gravity theories like MOND where galaxies lack hypothetical DM. This is because features in the disk very rapidly become irrelevant to the potential as one moves away from its plane. Consequently, perturbations in the disk have very little effect on the Milgromian `phantom DM' (PDM) halo as its density depends only on the local potential.\footnote{PDM is the DM density distribution that would be needed for the Newtonian gravitational field to equal the MOND one of the baryons alone. The PDM distribution of thin exponential disks was visualized in e.g. \citet[][]{Lughausen_2013, Lughausen_2015}.} Indeed, this is why a bare disk can be stable in MOND. The lack of an actual DM halo automatically removes the possibility of bar-halo angular momentum exchange. However, disk self-gravity is enhanced in MOND compared to the case of a frozen DM halo. The presence of an EFE would slightly decrease this enhancement, but would also decrease the spheroidal potential support from the PDM. In general, it would also break the axisymmetry and up-down symmetry of the problem. Clearly, numerical simulations are needed to properly understand how disk stability is influenced by a fundamental modification to the gravity law.

Several MOND $N$-body simulations have considered isolated disks, starting with the seminal work of \citet{Brada_1999} which used a custom potential solver but restricted particle motion to a plane. \citet{Tiret_2007} conducted a series of 3D simulations, finding rapid growth of the bar strength due to the enhanced role of disk self-gravity. However, the bar then weakened with time as the vertical velocity dispersion increased (see their section 5.1). The basic picture remained the same when those authors included gas as sticky particles in their simulations, though including dissipation in this way weakened the bar somewhat \citep{Tiret_2008_gas}. It was also shown that it is difficult to form a significant bulge through secular processes in clumpy disks \citep{Combes_2014}, potentially explaining the abundance of disks with a low bulge fraction \citep{Kormendy_2010}. MOND simulations of disk galaxies have also been used to understand how they interact with each other \citep{Tiret_2008, Renaud_2016, Bilek_2018}, and with the gravitational field of a galaxy cluster \citep{Candlish_2018}. These works leave no doubt that MOND can stabilize a bare self-gravitating disk provided its surface density is low enough that its dynamics are significantly non-Newtonian, i.e. the central surface density
\begin{eqnarray}
	\Sigma_0 ~\la~ \Sigma_M \equiv \frac{a_{_0}}{2 \mathrm{\pi} G} \, ,
	\label{Sigma_MOND}
\end{eqnarray}
corresponding to a vertical Newtonian gravity $\la a_{_0}$.

Moreover, the formation of disk galaxies from collapsing gas clouds was recently investigated for the first time in the MOND context \citep{Wittenburg_2020}. This naturally led to the formation of exponential disks after 10~Gyr of evolution. Somewhat too compact stellar bulges were also formed, but this problem could perhaps be avoided with a more realistic formation scenario including gas accretion. In particular, it has previously been shown that gas-rich clumpy galaxies in the early universe do not secularly form bulges from the coalescence of clumps in the MOND context \citep{Combes_2014}. Gas accretion would be naturally included in cosmological MOND simulations, which are currently under way (N. Wittenburg et al., in preparation).

In this contribution, we follow up the work of \citetalias{Sellwood_2019} in conventional gravity by conducting MOND hydro simulations of a galaxy with initial parameters chosen to match the photometry and RC of M33, with and without the EFE from M31. These global properties can be matched in MOND \citep{Sanders_1996} despite claims by \citet{Corbelli_2007}, in particular in the presence of a small gradient in the stellar mass-to-light ratio \citep[see figure 22 of][]{Famaey_McGaugh_2012}. To investigate the dynamical stability of our models, we simulate M33 for at least 6~Gyr using the publicly available code Phantom of RAMSES \citep[\textsc{por},][]{Lughausen_2015}. \textsc{por} adapts the potential solver of the \textsc{ramses} code \citep{Teyssier_2002} to solve the rather computer-friendly quasilinear formulation of MOND \citep[QUMOND,][]{QUMOND}. We analyze our M33 simulations similarly to \citetalias{Sellwood_2019}. 


In the following, we describe the initial conditions and setup of our isolated simulations (Section \ref{Setup}). We then present our results and analyses in Section \ref{Results}, where we also compare with observations. We extend our model by including the EFE from M31 in an approximate manner (Section \ref{Including_EFE}). Our conclusions are presented in Section \ref{Conclusions}. Videos of our isolated simulations with frames every 10~Myr are publicly available, along with the algorithm we use to extract data on gas cells.\footnote{\href{https://seafile.unistra.fr/d/843b0b8ba5a648c2bd05/}{https://seafile.unistra.fr/d/843b0b8ba5a648c2bd05/}}

\section{Initial conditions \& simulation setup}
\label{Setup}

We first conduct two isolated hydro simulations of M33 in MOND, varying the initial gas temperature $T$ to assess the impact of how we model the gas component (Section \ref{Including_gas}). We also run a stellar-only comparison simulation with the same total surface density profile. Our initial setup has the M33 disk spin vector pointing along the positive $z$-axis of our Cartesian $\left( x, y, z \right)$ co-ordinate system. We frequently make use of cylindrical polar co-ordinates $\left(r, z \right)$, where $r \equiv \sqrt{x^2 + y^2}$.

In MOND, encounters between M33 and other galaxies are less likely to end in a damaging merger due to the absence of dynamical friction from DM halos. A merger is possible for an almost head-on encounter, but geometrically a flyby is more likely. In this case, the other galaxy should still be visible. The nearest large galaxy to M33 is M31, but combining the orbit modeling of \citet{Patel_2017} with the latest proper motion measurements of both galaxies indicates that a recent close interaction is very unlikely \citep[figure 4 of][]{Van_der_Marel_2019}. There may have been an interaction ${\approx 6.5}$~Gyr ago \citep{Thor_2020}, but M33 should still be stable in its current configuration as its dynamical time is much shorter (Section \ref{Simulation_setup}). We also ignore the EFE at first, but we consider its possible influence in subsequent simulations (Section \ref{Including_EFE}).



Our simulations use QUMOND \citep{QUMOND}, which is based on a Lagrangian that obeys the usual symmetries and thus conservation laws. Its numerical implementation requires only the standard linear Poisson solver (Equation \ref{QUMOND_governing_equation}). The non-linearity of MOND is handled using a non-linear algebraic step, but crucially there is no need for a non-linear grid relaxation stage, a key part of the older aquadratic Lagrangian formulation \citep[AQUAL,][]{Bekenstein_Milgrom_1984}. Both AQUAL and QUMOND can be implemented using a publicly available numerical $N$-body solver called \textsc{raymond} \citep{Candlish_2015}, which has clarified that the two formulations give rather similar results \citep{Candlish_2016}. Not only do the two approaches agree exactly in spherical symmetry, the differences would be very small even in quite non-spherical situations like a point mass in a dominant external field, as can be shown analytically \citep[section 7.2 of][]{Banik_2018_Centauri}. Therefore, our results can probably be generalized to AQUAL as well.

\subsection{M33 mass distribution}
\label{M33_mass_distribution}

\begin{table}
	\centering
	\begin{tabular}{cccc}
		\hline
		Component & Mass & Scale length & Aspect ratio \\
		\hline
		Stellar disk & $4.7 \times 10^9 \, M_\odot$ & 1.78~kpc & 0.1 \\
		Gas disk & $1.8 \times 10^9 \, M_\odot$ & 4~kpc & 0.1 \\
		\hline
	\end{tabular}
	\caption{Parameters of our M33 model, adopted from the MOND RC fit in \citet{Sanders_1996} which was based on near-infrared photometry obtained by \citet{Regan_Vogel_1994}. The stellar disk scale length agrees with the determination of \citet{Kam_2015}. Part of this disk is modeled as gas, giving a total gas fraction of 0.3. This leads to the gas having a component distributed similarly to the stars, which we expect because of stellar evolution. The gas is assumed to be isothermal at temperature $T = 10^5 K$ (100~kK) or 25~kK, though $T$ includes the turbulent motions. Radially, each component is modeled as an exponential disk with $\sech^2$ vertical density profile. The aspect ratio is the ratio of their characteristic scales, e.g. the stellar disk density $\propto \sech^2 \left(z/ 0.178 \, \text{kpc} \right)$. In our hydro simulations, we then adjust the gas disk scale height at each radius using a Newton-Raphson algorithm in order to best achieve equilibrium (Section \ref{Including_gas}). Our stellar-only model includes the gas disk as collisionless particles, so all our simulations have the same total surface density profile.}
	\label{Parameters}
\end{table}

\begin{figure}
	\centering
	\includegraphics[width = 8.5cm] {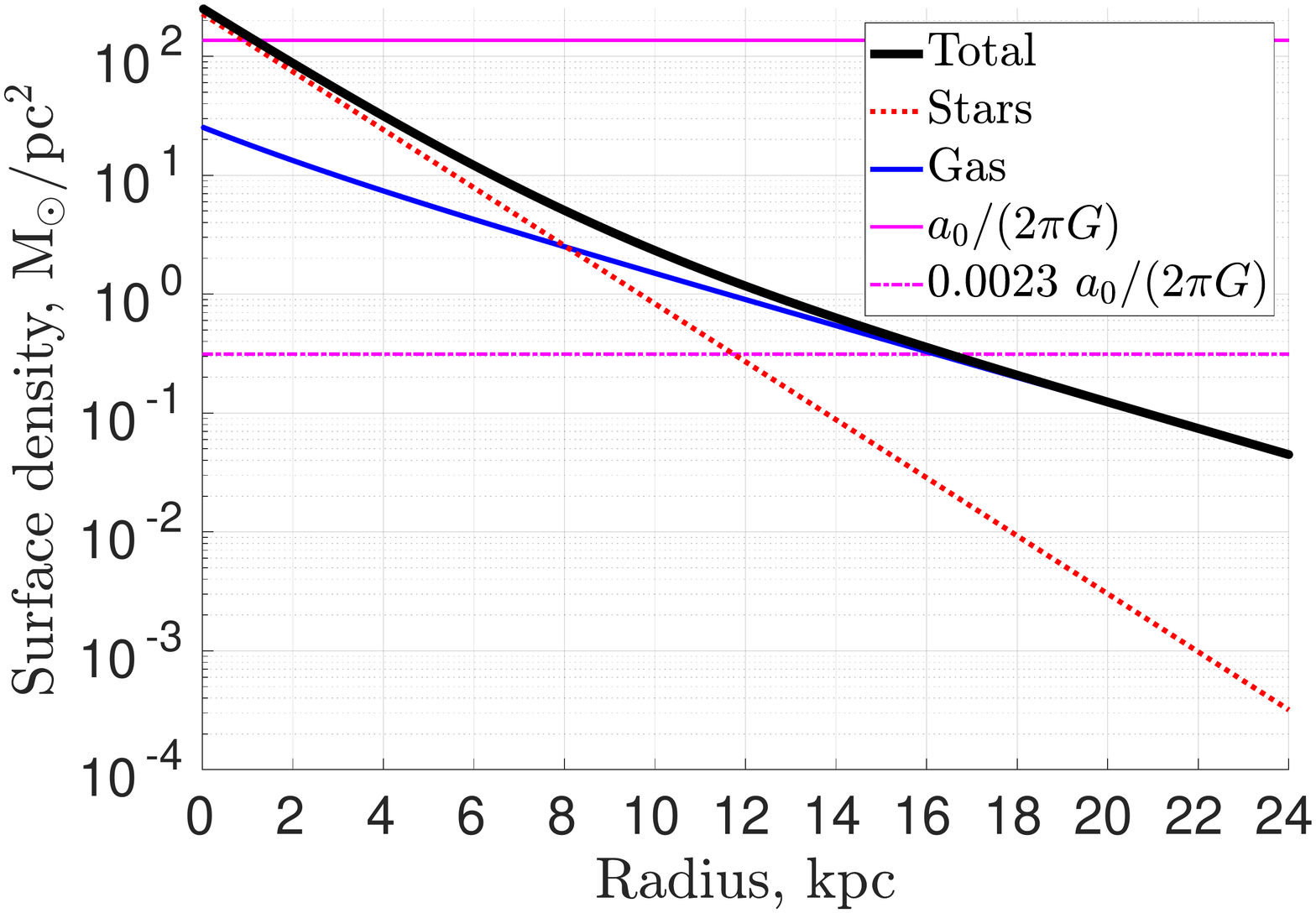}
	\caption{The total surface density profile of our M33 model (black) and its component parts $-$ stars (red) and gas (blue). The horizontal magenta solid line near the top represents the MOND critical surface density (Equation \ref{Sigma_MOND}). M33 has a lower surface density almost everywhere, so MOND should greatly impact its dynamical properties. The lower dot-dashed magenta line shows the vertical component of the Newtonian-equivalent external field from M31 if we put this at $30^\circ$ to the disk plane (Equation \ref{g_ext_value}). Notice that the EFE could have a percent-level effect on the gravitational field in the central regions of M33 (Section \ref{Including_EFE}).}
	\label{M33_surface_density}
\end{figure}

We treat M33 as a combination of two radially exponential disks with different scale lengths. No bulge is present initially, though one could form if the initial setup is unstable. The more extended disk is purely gas while the less extended disk is mostly stars, with a small proportion of gas to account for gas lost from stars as they evolve. Both disks have a $\sech^2$ vertical profile. Table \ref{Parameters} summarizes the parameters of our M33 model, whose initial surface density profile is shown in Figure \ref{M33_surface_density}. All our simulations start with the same total surface density profile and RC, with our hydro models all having the same initial mixture of stars and gas.

\subsection{Initializing a Milgromian disk}
\label{DICE}

We set up a MOND disk by adapting the Newtonian code Disk Initial Conditions Environment \citep[\textsc{dice},][]{Perret_2014} $-$ our modified version is publicly available.\footnote{\url{https://bitbucket.org/SrikanthTN/bonnpor/src/master/} \\ Separate folders are used for the stellar-only and hydro versions.} \textsc{dice} offers the advantage that the Jeans equations are not solved using the potential, which is difficult to define for an isolated system in MOND. Rather, \textsc{dice} uses only the Newtonian gravity $\bm{g}_{_N}$, which it calculates using the principle of superposition accelerated by a fast Fourier transform. To get the true gravity $\bm{g}$, we use the algebraic MOND approximation:
\begin{eqnarray}
    \bm{g} ~&\approx&~ \nu \bm{g}_{_N} \, \text{, where} \\
    \nu ~&=&~ \frac{1}{2} + \sqrt{\frac{1}{4} + \frac{a_{_0}}{g_{_N}}} \, .
    \label{ALM}
\end{eqnarray}
We employ the notation $v \equiv \left| \bm{v} \right|$ for any vector $\bm{v}$. The `simple' form of the $\nu$ function is used to interpolate between the Newtonian and MOND regimes \citep{Famaey_Binney_2005} since it provides a good fit to a variety of data on galactic and extragalactic dynamics \citep{Gentile_2011, Banik_2018_Centauri}. It is rather similar to the function used by \citet{McGaugh_Lelli_2016} to fit the Spitzer Photometry and Accurate Rotation Curve dataset \citep[SPARC,][]{SPARC}. In the QUMOND approach, $\nu$ depends only on $g_{_N}$ and is thus readily computable once standard techniques are used to obtain $\bm{g}_{_N}$.

The algebraic MOND approximation is exactly correct in spherical symmetry and works rather well in axisymmetric problems \citep{Angus_2012, Jones_2018}. It is expected to work particularly well just outside the disk \citep{Banik_2018_Toomre}. However, it becomes inaccurate within the disk due to the steep vertical gradient in $\nu$ caused by that in $\bm{g}_{_{N,z}}$. Naively applying Equation \ref{ALM} would imply a rapid change in $\bm{g}_{_r}$ with $z$, something that is physically unrealistic as it would cause $\nabla \times \bm{g} \neq 0$, allowing energy to be gained by a particle going around a closed loop.\footnote{This is not true of QUMOND if applied rigorously since it is derivable from a Lagrangian, therewith obeying the usual conservation laws \citep{QUMOND}.} To avoid this, we fix the value of $g_{_{N,z}}$ entering the calculation of $\nu$ (Equation \ref{ALM}) to $2 \tanh \left( 2 \right) \mathrm{\pi} G \Sigma$ if $\left| z \right|$ is small enough that the fraction of the local column density $\Sigma \left( r \right)$ at even smaller $\left| z \right|$ falls below $\tanh \left( 2 \right)$. This is based on the assumption that $g_{_{N,z}} = 2 \mathrm{\pi} G \Sigma$ at the disk `surface', which is valid for a thin disk. Once $\nu$ is calculated in this revised way, we set $\bm{g} = \nu \bm{g}_{_N}$.

To adjust the gravity law, we also need to change how the Toomre $Q$ parameter is calculated $-$ this is used to ensure the velocity dispersion is consistent with local stability \citep{Toomre_1964}. Classically, the Toomre stability condition (his equation 65) is that:
\begin{eqnarray}
	Q \equiv \frac{\sigma_r \Omega_r}{3.36 \, G \Sigma} \geq 1 \, ,
	\label{Q_Newton}
\end{eqnarray}
where $\sigma_r$ is the radial velocity dispersion, $\Omega_r$ is the radial epicyclic frequency, and $\Sigma$ is the disk surface density.

The QUMOND generalization of the Toomre condition was derived analytically in \citet{Banik_2018_Toomre}. Briefly, the effective value of the gravitational constant $G$ should be enhanced according to:
\begin{eqnarray}
    G ~\to~ G \nu \left( 1 + \frac{K_0}{2} \right) \, , \quad K_0 ~\equiv~ \frac{\partial \ln \nu}{\partial \ln g_{_N}} \, .
    \label{K_0_definition}
\end{eqnarray}
Their derivation requires $\bm{g}_{_N}$ to include a vertical component of $2 \mathrm{\pi} G \Sigma$ along with whatever radial component exists at the desired $r$. Since the Toomre condition is only concerned with local stability, we use the $Q_{lim}$ option in \textsc{dice} to require that $Q \geq 1.25$ everywhere, thus leaving a modest safety margin. Results are similar if a slightly lower floor is imposed on $Q$ (Section \ref{Different_Q_lim}).

\begin{figure}
	\centering
	\includegraphics[width = 8.5cm] {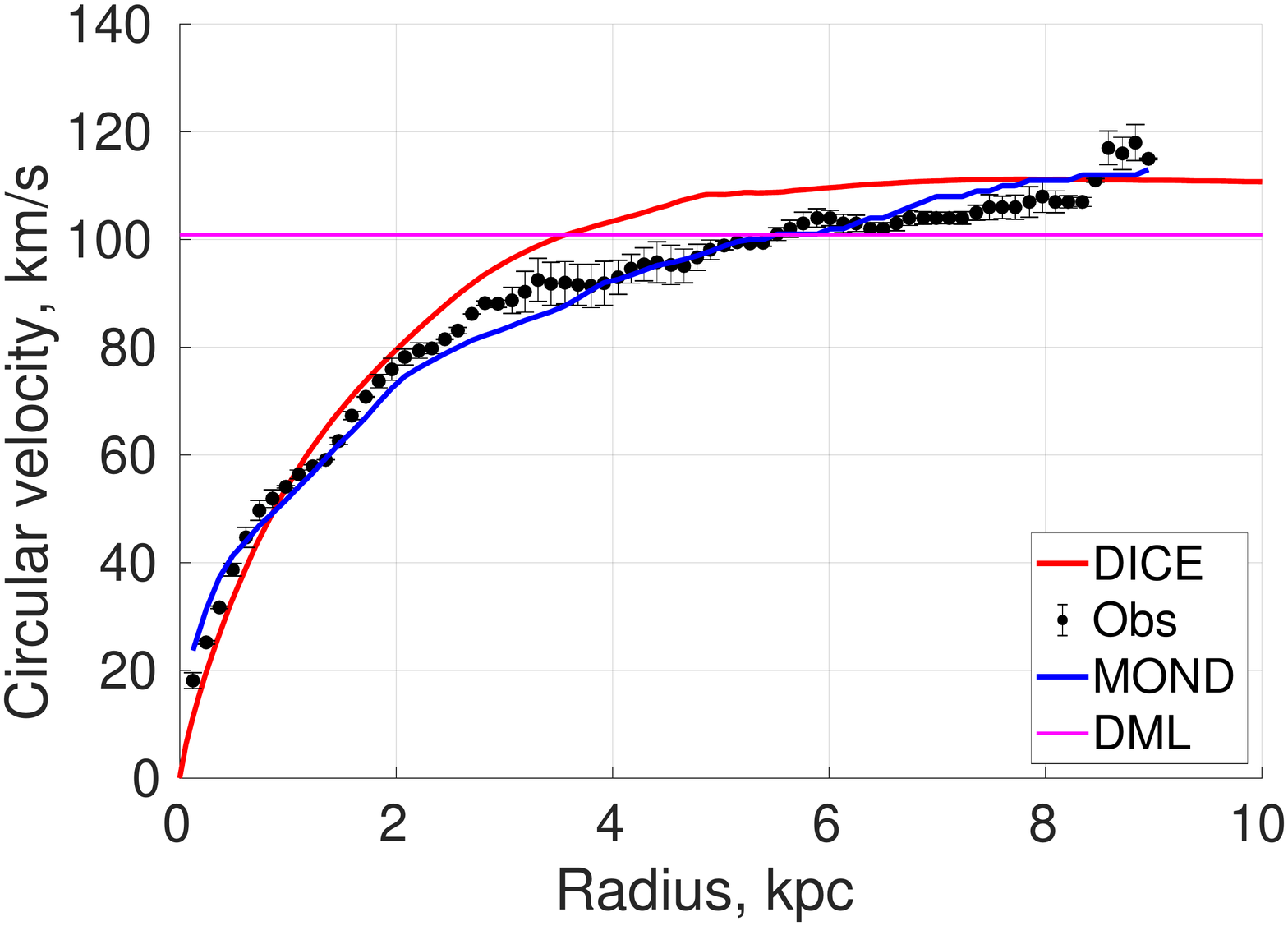}
	\caption{The RC of M33 according to the observations of \citealt{Sanders_1996} (points with error bars), their detailed MOND fit based on photometry (blue curve), and our \textsc{dice} template (red), which applies Equation \ref{ALM} to the double exponential profile described in Table \ref{Parameters}. The prediction of Equation \ref{Deep_MOND_limit} is shown as a magenta line. Beyond 25~kpc, the \textsc{dice} RC gently decreases to this level (not shown). Although this RC is used to set up the disk in all our simulations, the disks in our hydro models do not begin exactly in radial equilibrium (Section \ref{Including_gas}).}
	\label{M33_rotation_curve}
\end{figure}

Figure \ref{M33_rotation_curve} shows the M33 RC resulting from our modified \textsc{dice} algorithm. We also show the observed RC and the MOND fit based on a much more detailed surface density profile than we assumed (Figure \ref{M33_surface_density}). As expected, the MOND fit works rather well despite lacking the flexibility afforded by a DM halo. Since M33 is in the DML almost everywhere, its RC rises along with the enclosed mass (Equation \ref{Deep_MOND_limit}) and lacks a region with an approximately Keplerian decline. Due to the disk geometry, our calculated RC nonetheless rises slightly above the asymptotic value of 101~km/s (Equation \ref{BTFR}) before gently declining toward it. Despite our rather simplified mass model for M33, its observed RC is rather similar to that of our \textsc{dice} template.


\subsection{Including gas}
\label{Including_gas}

We model the gas component of M33 as initially isothermal at $T = 10^5$~K ($\equiv 100$~kK) or 25~kK. We will see that the cooler model is much more realistic, so we use it for further analyses and simulations. Since we are only interested in the large scale behaviour of M33, we suppress star formation and metallicity-dependent cooling. We do allow gas in our simulations to cool, but only down to its initial temperature. Allowing further cooling would require us to allow star formation, which is beyond the scope of this project. Our rather high $T$ partly compensates for the fact that feedback is not included in our simulations.

The use of a constant $T$ simplifies our initial setup because it is already in thermal equilibrium. Even so, ensuring dynamical equilibrium is non-trivial and we only approximately achieve this, as discussed next.

\subsubsection{Thickness profile}
\label{Gas_thickness_profile}

The inclusion of gas is complicated because there are also stars. We fix the thickness of the stellar component and iteratively adjust that of the gas, following a similar approach to equation 11 of \citet{Corbelli_2003}.

We start with the Newtonian result of \citet{Spitzer_1942} that an isothermal gas slab with sound speed $c_s$ has a $\sech^2$ vertical profile with characteristic thickness $h_g$ given by:
\begin{eqnarray}
    {c_s}^2 ~=~ \mathrm{\pi}G h_g \Sigma_g ~=~ \frac{g_{_{N,z}} h_g}{2} \, ,
	\label{Spitzer_result}
\end{eqnarray}
where the gas surface density $\Sigma_g$ gives rise to a vertical Newtonian gravity of $g_{_{N,z}} = 2 \mathrm{\pi} G \Sigma_g$ at the disk `surface', i.e. as $z \to \infty$. We generalize this to MOND by supposing that a $\sech^2$ vertical profile remains a good description if we modify Equation \ref{Spitzer_result} to:
\begin{eqnarray}
    {c_s}^2 ~=~ \frac{g_{_{N,z}} h_g}{2} \nu \left( \sqrt{{g_{_{N,r}}}^2 + {g_{_{N,z}}}^2}\right) \, .
    \label{cs_MOND}
\end{eqnarray}

The issue now arises of what value to use for $g_{_{N,z}}$ in the presence of a stellar disk with surface density $\Sigma_*$ and scale height $h_*$. Since this will be found in a different way to evaluating $g_{_{N,z}}$ at a particular position, we use $\widetilde{g}_{_{N,z}}$ to denote the appropriate value of $g_{_{N,z}}$ in Equation \ref{cs_MOND}. Thus, the generalization of Equation \ref{Spitzer_result} to a stellar $+$ gas disk is:
\begin{eqnarray}
	{c_s}^2 ~\equiv~ \frac{\widetilde{g}_{_{N,z}} h_g}{2} \nu \left( \sqrt{{g_{_{N,r}}}^2 + {\widetilde{g}_{_{N,z}}}\!^2}\right) \, .
	\label{Gas_NR}
\end{eqnarray}
If $h_g = h_*$ and the stellar disk is also isothermal, we can treat Equation \ref{Spitzer_result} as applying to the combined stellar $+$ gas disk since there is no practical distinction between the individual components. Thus, it is clear that
\begin{eqnarray}
	\widetilde{g}_{_{N,z}} ~=~ 2 \mathrm{\pi} G \left( \Sigma_g + \Sigma_* \right) \, , \quad h_g = h_* \, .
	\label{gnz_equal_height}
\end{eqnarray}
In general, $h_g \neq h_*$. We make the assumption that each component still follows a $\sech^2$ vertical profile. We then find what fraction of the stellar disk is enclosed at $\left| z \right| < h_g$. For the gas disk, this fraction is $\tanh \left( 1 \right)$. These fractions are unequal if $h_g \neq h_*$. In this case, we assume that the appropriate generalization of Equation \ref{gnz_equal_height} is:
\begin{eqnarray}
	\widetilde{g}_{_{N,z}} ~=~ 2 \mathrm{\pi} G \left( \Sigma_g + \Sigma_* \frac{\tanh \left( h_g/h_* \right)}{\tanh \left( 1 \right)} \right) \, .
	\label{gnz_unequal_height}
\end{eqnarray}
This approach can easily be generalized to stellar disks with multiple components that have different scale heights and/or vertical profiles different from $\sech^2$.

At large radii, the surface density of an exponential disk becomes very small. Since the gas disk is isothermal, this weakening of the vertical gravity causes the disk to flare \citep[as observed for the Milky Way,][]{Sanchez_2008}. As a result, it is no longer reasonable to assume that the vertical gravity comes mostly from disk material at similar $r$. Instead, it mostly comes from material at much smaller $r$. For an observer at large $r$, this material can be approximated as a point mass at the centre of the galaxy. Following our previous approach, the important quantity is the resulting vertical Newtonian gravity at a height of $h_g$. This is approximately given by $g_{_{N,r}} \left( h_g/r \right)$ for $r \gg h_g$. To prevent our correction term diverging at small $r$, we put in a softening length of $h_g$. We are then left with the vertical gravity receiving an extra contribution of $g_{_{N,r}}$ at small $r$, contrary to the idea of applying a `geometric correction' that only becomes significant many disk scale lengths out. We fix this with an extra factor of $\tanh \left( r/h_* \right)$, leading to our final approximation that:
\begin{eqnarray}
	\widetilde{g}_{_{N,z}} ~&=&~ 2 \mathrm{\pi} G \left( \Sigma_g + \Sigma_* \frac{\tanh \left( h_g/h_* \right)}{\tanh \left( 1 \right)} \right) \nonumber \\
	&+&~ g_{_{N,r}} \frac{h_g}{\sqrt{r^2 + {h_g}^2}} \tanh \left( \frac{r}{h_*}\right) \, .
	\label{gnz_final}
\end{eqnarray}

\begin{figure}
	\centering
	\includegraphics[width = 8.5cm] {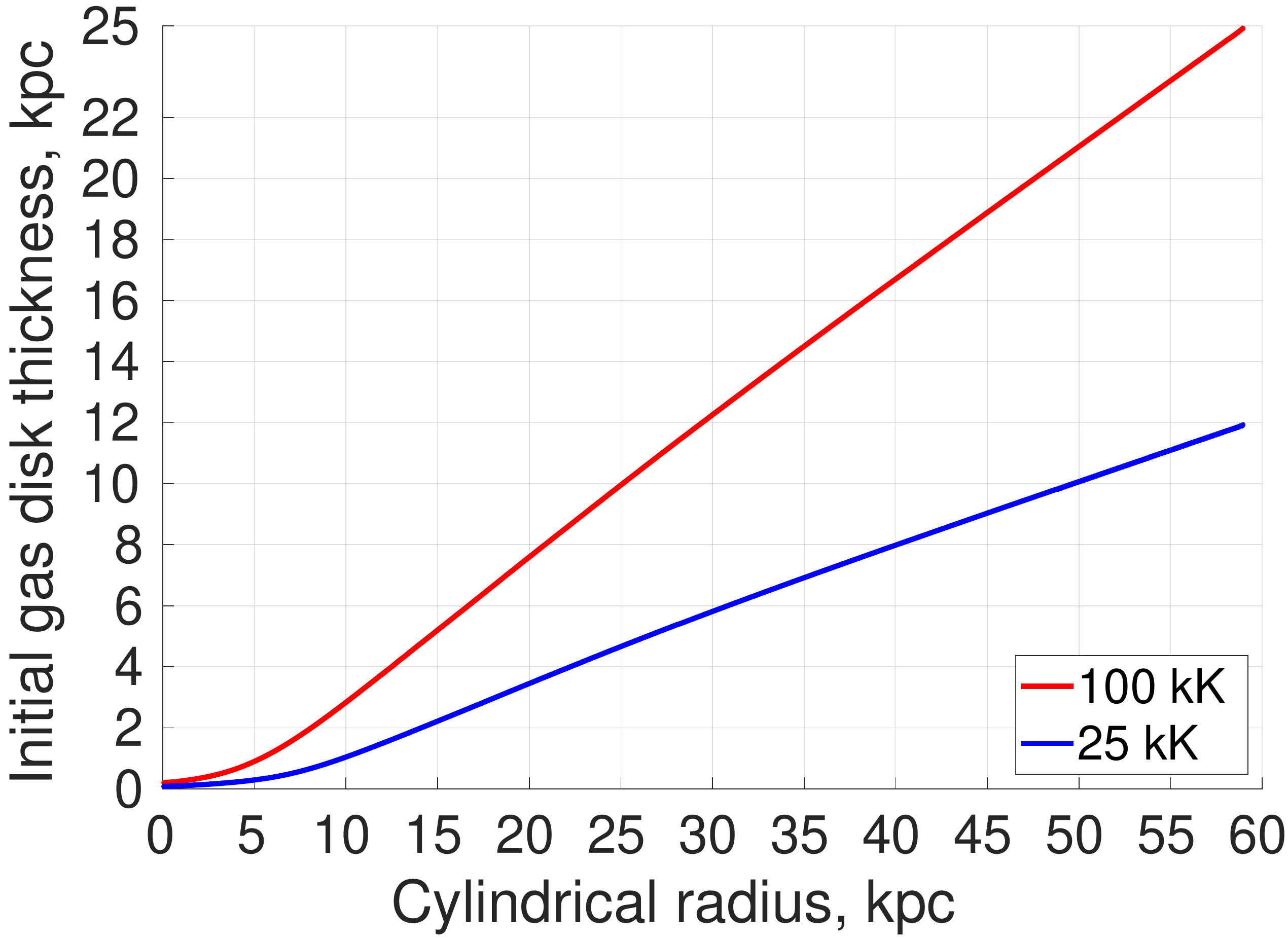}
	\caption{Initial $\sech^2$ scale height of our simulated M33 gas disk as a function of $r$, with the adopted temperature shown in the legend. This thickness is much larger than assumed when calculating the RC in \textsc{dice} (Section \ref{Parameters}). This weakens the radial gravity at intermediate radii, implying that our hydro simulations are not set up exactly in radial equilibrium (see text).}
	\label{Gas_initial_thickness}
\end{figure}

We can now solve Equation \ref{Gas_NR} by varying $h_g$ using a Newton-Raphson algorithm, each time recalculating $\widetilde{g}_{_{N,z}}$ as above. Once we have obtained $h_g$ at some $r$, we use it as an initial guess for $h_g$ at the next $r$ in our grid, minimizing the number of steps required to reach convergence. In this way, we build up the gas thickness profiles $h_g \left( r \right)$ shown in Figure \ref{Gas_initial_thickness}, which we incorporate into our hydro simulations as described next.

\subsubsection{Density and velocity field}

We set up the gas density and velocity field in \textsc{por} using our estimated gas disk thickness profile $h_g \left(r \right)$ from our modified \textsc{dice} algorithm (Section \ref{Gas_thickness_profile}). Practically, this involves adjustments to \textsc{por}, which we now describe. In principle, these adjustments are unrelated to the gravity law $-$ this only enters the algorithm by way of the RC and thickness profile, which are provided externally by \textsc{dice}. Other codes could in principle be used to prepare these input files.

At each $r$, we assume the gas follows a $\sech^2$ vertical profile with characteristic height $h_g \left(r \right)$. As a result, the gas density is:
\begin{eqnarray}
    \rho_g \left( r \right) ~=~ \frac{\Sigma_g}{2 \, h_g} \sech^2 \left( \frac{z}{h_g} \right) \, .
    \label{rho_gas}
\end{eqnarray}

We next assign the gas velocity field. Consider a gas parcel with density $\rho$ and isothermal sound speed $c_s$ rotating around a galaxy at cylindrical radius $r$ with speed $v_a$, which in general can differ from the circular velocity $v_c \equiv \sqrt{-r g_r}$ due to pressure gradients. The equation of radial hydrostatic equilibrium is:
\begin{eqnarray}
    {v_c}^2 ~=~ {v_a}^2 - {c_s}^2 \left( \frac{\partial \ln \rho_g}{\partial \ln r} \right) \, .
    \label{Pressure_correction}
\end{eqnarray}
Because $h_g$ depends on $r$, it is non-trivial to ensure even radial hydrostatic equilibrium of the gas. We approximately achieve this in the disk mid-plane by realizing that $\rho_g \propto \Sigma_g/h_g$ (Equation \ref{rho_gas}), implying that:
\begin{eqnarray}
    \frac{\partial \ln  \rho_g}{\partial \ln r} ~=~ \frac{\partial \ln \Sigma_g}{\partial \ln r} - \frac{\partial \ln h_g}{\partial \ln r} \, , \quad z = 0 \, .
\end{eqnarray}
The radial gradient of $h_g$ is found by centred differencing. We avoid numerical difficulties at the origin by exploiting the fact that:
\begin{eqnarray}
    \frac{\partial \ln  h_g}{\partial \ln r} ~=~ \frac{r}{h_g} \frac{\partial h_g}{\partial r} \, .
\end{eqnarray}

The gas velocity field is set up by applying the above pressure correction to the RC calculated by \textsc{dice} based on a constant gas disk thickness of 0.4~kpc (Table \ref{Parameters}). Since the actual thickness is generally larger (Figure \ref{Gas_initial_thickness}), our algorithm is slightly out of equilibrium even in the radial direction. However, we expect the disk to settle down within a few dynamical times.

\subsection{Simulation setup}
\label{Simulation_setup}

We simulate M33 using \textsc{por} \citep{Lughausen_2015} in non-cosmological particle-in-cell mode. \textsc{por} solves the field equation of QUMOND, which involves obtaining the Newtonian gravity $\bm{g}_{_N}$ and then solving:
\begin{eqnarray}
	\nabla \cdot \bm{g} ~=~ \nabla \cdot \left( \nu \bm{g}_{_N} \right) \, .
	\label{QUMOND_governing_equation}
\end{eqnarray}
The generalization to non-zero EFE is given in Equation \ref{QUMOND_governing_equation_EFE}, with the simple form of the interpolating $\nu$ function (Equation \ref{ALM}) used in all cases.

Our adopted cubic box size has sides of 512~kpc, much larger than M33. This is necessary because \textsc{por} assumes that the gravitational potential $\Phi$ satisfies the boundary condition appropriate for a point mass $M$, namely:
\begin{eqnarray}
	\Phi ~=~ \sqrt{GMa_{_0}} \ln R \, ,
	\label{Boundary_Phi_iso}
\end{eqnarray}
where $\bm{R}$ is the position relative to the centre of mass in any unit of length. The simulations are advanced for at least 6~Gyr, which represents ${\approx 50}$ revolutions for a typical extent of 2~kpc (Table \ref{Parameters}) and rotation velocity of 100~km/s (Equation \ref{BTFR}). This should provide ample time for our M33 models to settle into dynamical equilibrium, though results for the first ${\approx 1}$~Gyr might depend on details of our initialization procedure, especially for our hydro simulations (Section \ref{Including_gas}).

To provide adequate spatial resolution, we refine the simulation volume into $levelmin = 7$ up to $levelmax = 13$ refinement levels, i.e. there are at least $2^7$ pixels across each of the three spatial dimensions. \textsc{ramses} uses adaptive mesh refinement (AMR) to improve the resolution in regions with a high density. We allow up to $13 - 7 = 6$ levels of refinement, giving a maximum spatial resolution of $512/2^{13}$~kpc $=$ 62.5~pc, sufficient to resolve each disk scale length into many pixels (Table \ref{Parameters}). Our chosen refinement conditions are $mass\_sph = 10^3 \, M_\odot$ and $m\_refine = 20$. Refined regions operate at a reduced timestep as defined by the runtime parameter $nsubcycle$, which we set to $\left( 1, 1, 2, 2 \right)$. We also set the Poisson parameter $epsilon = 10^{-4}$, defining the convergence condition for the iterative Poisson solver. Note that a standard Poisson solver is sufficient for QUMOND simulations as the non-linearity of MOND is handled in an algebraic step (Equation \ref{QUMOND_governing_equation}). Further details of \textsc{ramses} can be found in \citet{Teyssier_2002}, which also gives the default values for parameters that we do not adjust.

Our simulations use $10^6$ particles for M33, which we divide amongst its inner and outer exponential disks in proportion to their mass. Since all particles have the same mass within each component, this ensures that masses are also equal between the components. In our stellar-only simulation, we do not implement the procedures described in Section \ref{Including_gas} since these relate to the gas component. In this case only, we advance our $\textsc{por}$ simulation in $N$-body mode (the runtime parameter $hydro$ is set to false).

To include gas with a gas fraction of 0.3, we remove all particles in the more extended component and apply a uniform fractional reduction to the masses of particles in the less extended component. We then put this removed mass back in as gas using a modified \texttt{condinit} routine, which sets the properties of each gas cell using the procedures described in Section \ref{Including_gas}. The RC it uses is obtained from \textsc{dice}, which is slightly different to that generated by the mass distribution. However, the difference should be small because it arises only from the initial gas disk being thicker than assumed for the \textsc{dice} RC calculation.

As we are only interested in the large scale behaviour of M33, we suppress star formation and metallicity-dependent cooling. We allow the gas to cool, but impose a temperature floor of $T2\_star = T2\_ISM$, where $T2\_ISM$ is the initial temperature (100~kK or 25~kK). Feedback processes are not included in our simulations, differences between which are summarized in Table \ref{Simulations_list}.

\begin{table}
	\centering
	\begin{tabular}{cccccc}
		\hline
		 & External & & Box & & \\
		& field & $level$- & size & & \\
		Gas $T$ & direction & $max$ & (kpc) & $Q\_lim$ & Section \\
		\hline
		Stars only & Isolated & 13 & 512 & 1.25 & \ref{Results} \\
		100 kK & Isolated & 13 & 512 & 1.25 & \ref{Results} \\
		25 kK & Isolated & 13 & 512 & 1.25 & \ref{Results} \\ [5pt]
		25 kK & $30^\circ$ to disk & 12 & 1024 & 1.25 & \ref{Including_EFE} \\
		25 kK & $30^\circ$ to disk & 13 & 512 & 1.25 & \ref{Numerical_convergence} \\
		25 kK & $30^\circ$ to disk & 12 & 1024 & 1.1 & \ref{Different_Q_lim} \\
		25 kK & $30^\circ$ to disk & 12 & 1024 & 1 & \ref{Different_Q_lim} \\
		25 kK & In disk & 12 & 1024 & 1.25 & \ref{Different_g_ext_directions} \\
		25 kK & Spin axis & 12 & 1024 & 1.25 & \ref{Different_g_ext_directions} \\
		\hline
	\end{tabular}
	\caption{Summary of the simulations presented in this contribution, and the section in which each is mostly discussed. If included, the EFE has a strength of $0.07 \, a_{_0}$. In our stellar-only simulation, the gas is treated as stellar particles, thus preserving the total surface density profile in all models. We also ran an isolated simulation at 10~kK, but abandoned it as it became very unstable (not shown).}
	\label{Simulations_list}
\end{table}

\section{Results of isolated simulations}
\label{Results}

We extract \textsc{por} data on the simulated particles into plain text format using an algorithm that we make publicly available.\footnote{\href{https://github.com/GFThomas/MOND/tree/master/extract\_por}{https://github.com/GFThomas/MOND/tree/master/extract\_por}} For extracting the gas distribution, we use a modified version of the \textsc{rdramses} algorithm\footnote{\href{http://www.astro.lu.se/~florent/rdramses.php}{http://www.astro.lu.se/$\sim$florent/rdramses.php} \\ Note that we obtain reliable results only if \textsc{rdramses} is operated on just one core, though the simulation it analyses need not be. Our modified version is available at \href{https://seafile.unistra.fr/d/843b0b8ba5a648c2bd05/}{https://seafile.unistra.fr/d/843b0b8ba5a648c2bd05/}} to create a list of positions, velocities, and masses for all gas cells. This lets us analyze particles and gas in the same way, with gas cells treated as particles at the cell centres. We then subtract the combined centre of mass position and velocity in all analyses, correcting for a very small numerical drift over the course of our simulations. This drift is more significant once we include the EFE (Section \ref{Including_EFE}). Apart from images of our simulations, all our quantitative analyses focus only on stars and gas at $\left| z \right| < 20$~kpc to ensure robustness against material ejected to large distance.

\subsection{Rotation curve}
\label{Rotation_curve}

The evolution of our M33 model can be seen in its RC, which we calculate based on the radial gravity $g_r$ acting on particles within 1~kpc of the disk plane. The circular velocity is calculated as:
\begin{eqnarray}
	v_c ~=~ \sqrt{-r g_r} \, .
	\label{vc_equation}
\end{eqnarray}

\begin{figure}
	\centering
	\includegraphics[width = 8.5cm] {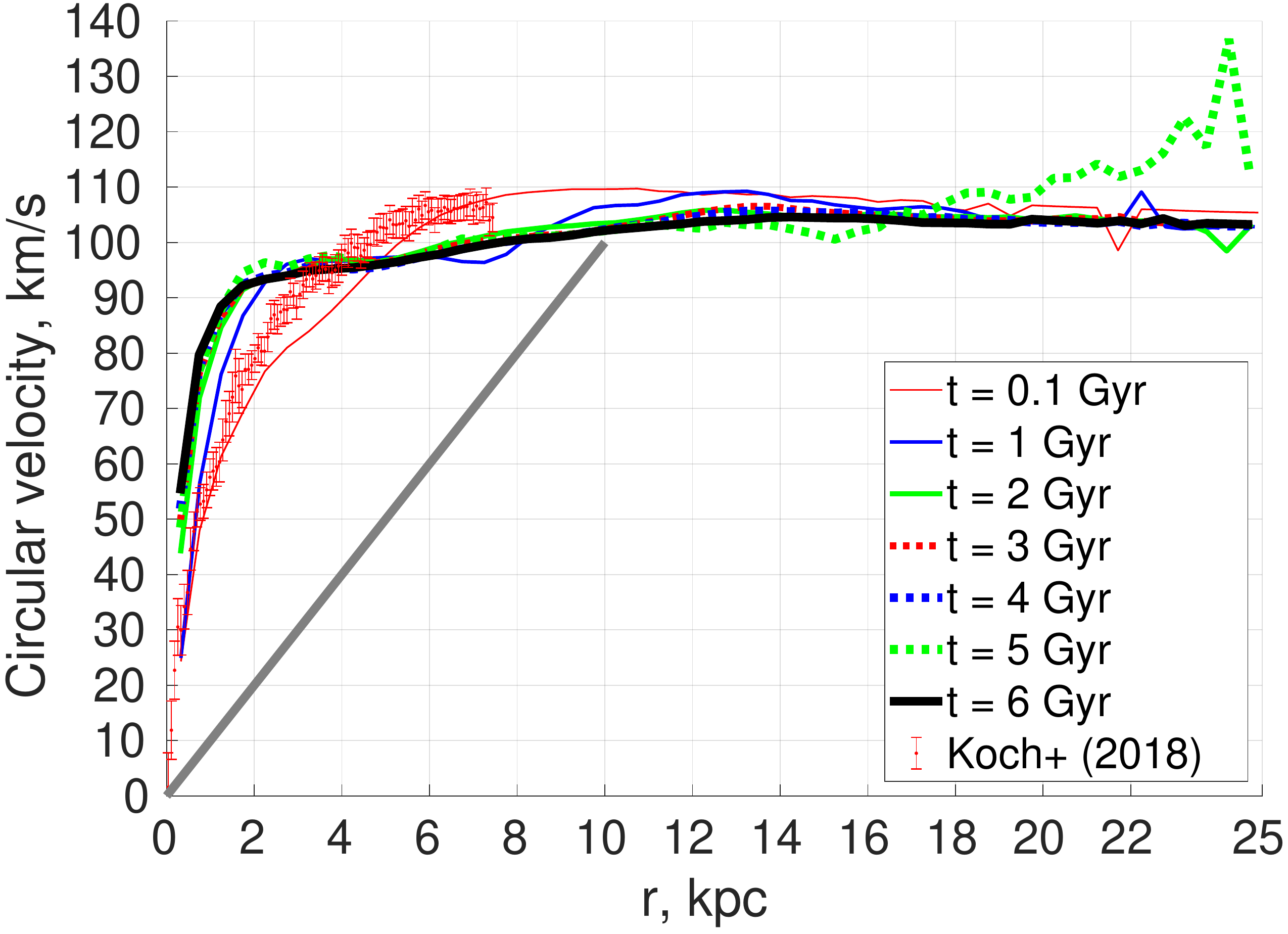}
	\includegraphics[width = 8.5cm] {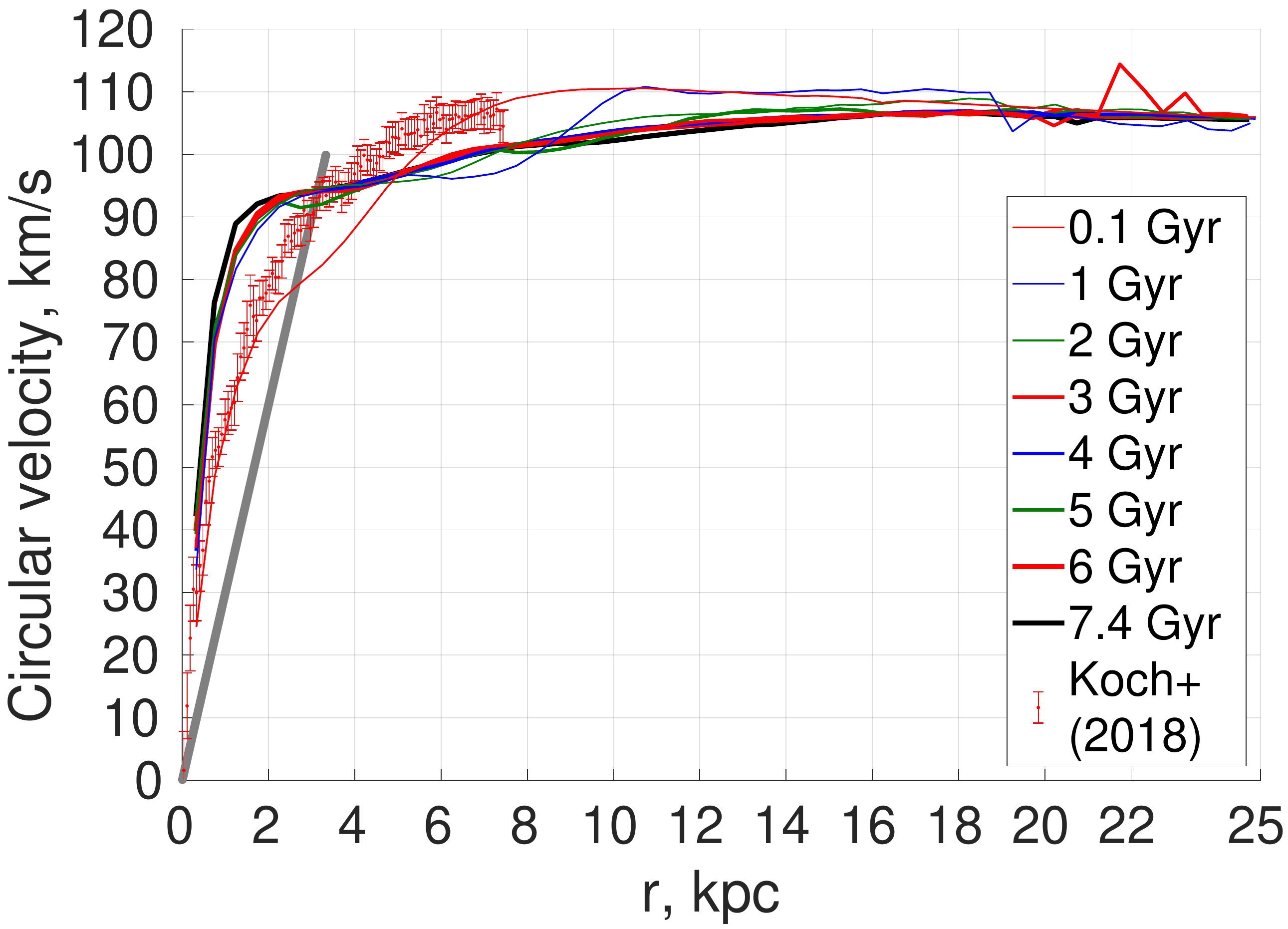}
	\caption{The simulated RC of M33 in our isolated hydro models based on the radial gravity felt by particles within 1~kpc of its disk plane (Equation \ref{vc_equation}). The top (bottom) panel shows results for $T = 100$~kK (25~kK). Different curves correspond to different times, which we indicate in the legend. The data points are from \citet{Koch_2018_error} based on an M33 distance of 840~kpc \citep[][and references therein]{Kam_2015}. The grey line represents an angular frequency of 10~km/s/kpc (30~km/s/kpc) for our 100~kK (25~kK) model, which is roughly the bar pattern speed in each case (Section \ref{Bar_pattern_speed}). Although the 100~kK and 25~kK RCs are quite similar, the disks have very different morphologies (Section \ref{Cylindrical_view}).}
	\label{M33_rotation_curves_iso}
\end{figure}

The simulated RCs of our isolated hydro models are shown in Figure \ref{M33_rotation_curves_iso}. The mass distribution becomes more centrally concentrated, boosting the RC at low $r$. The decrease at intermediate $r$ is caused by outward spreading of the disk, especially of its gas component. These changes mostly occur within the first Gyr, after which the RC changes very little.

For both gas temperatures, the final simulated RC rises much more steeply than observed in the central regions of M33 ($r \la 4$~kpc). This is not much dependent on which observations are used $-$ figure 5 of \citet{Koch_2018} shows that their study gives similar results to the previous studies of \citet{Corbelli_2014} and \citet{Kam_2017}, with the RC similar also to the more recent study of \citet{Utomo_2019}. This RC mismatch is undoubtedly a serious shortcoming of our isolated models, so we investigate it in more detail to try and understand the reason behind the discrepancy. We will see that it is related to the radial redistribution of material being too efficient, a process which is disrupted by including the EFE such that better agreement is obtained in this case (Section \ref{Including_EFE}).

\subsection{Cylindrically projected view and the bulge}
\label{Cylindrical_view}

\begin{figure*}
	\centering
	\includegraphics[width = 17.2cm] {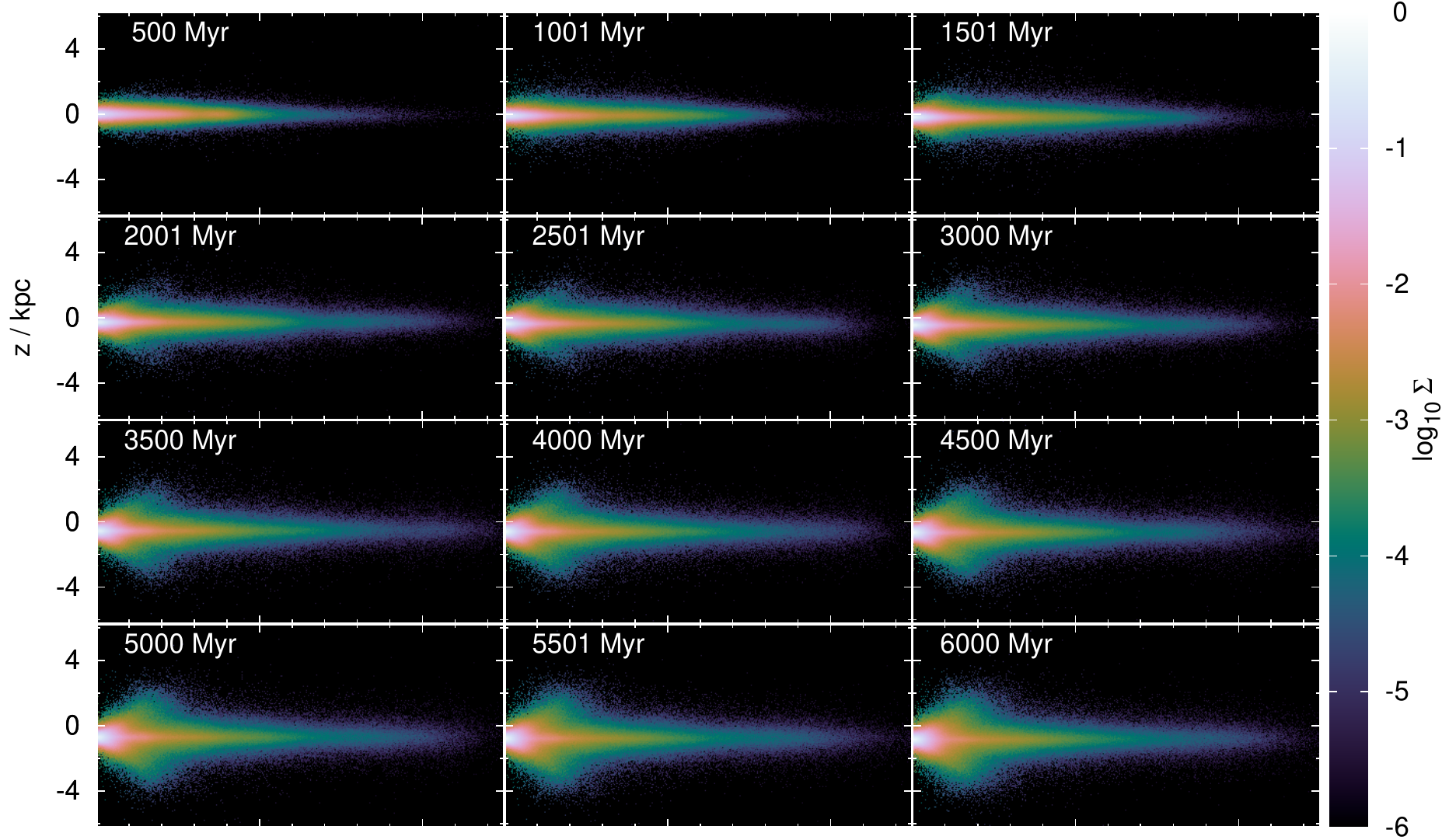}
	\includegraphics[width = 17.7cm] {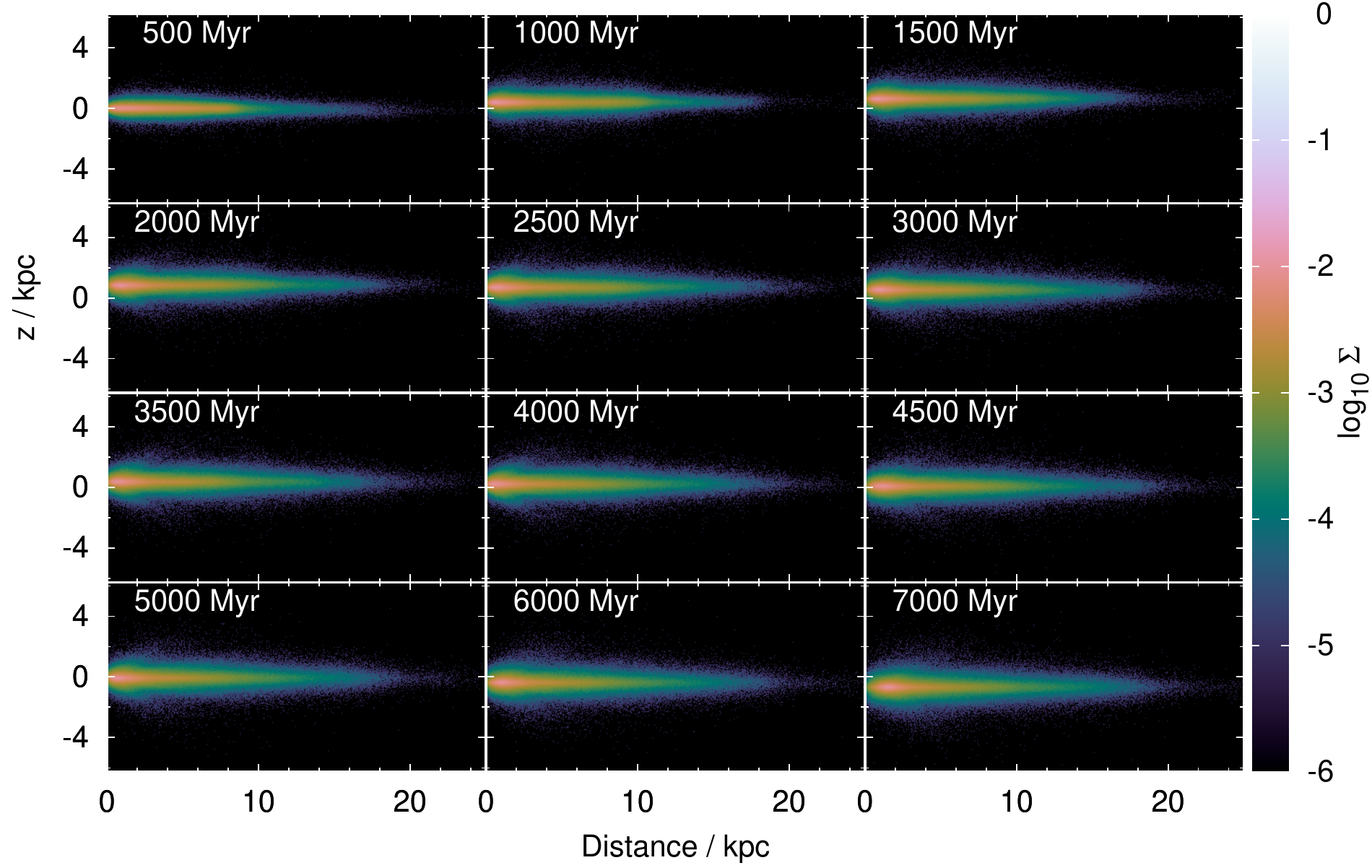}
	\caption{The stellar particles in our isolated hydro simulation of M33 with $T = 100$~kK (top) and 25~kK (bottom), shown using a cylindrical $rz$ projection (Equation \ref{Cylindrical_projection}). Values are shown in units of $1000 \, M_\odot$/pc$^2$. The time elapsed since the start of the simulation is shown in the top left corner of each frame. Analogous results for the gas are shown in Appendix \ref{Gas_edge_on}. Notice the significant central bulge in the hotter model, and its absence in the cooler model.}
	\label{Particles_radial}
\end{figure*}

To better understand the behaviour of our simulated M33, we show it in a cylindrical $rz$ projection where the intensity of each pixel is:
\begin{eqnarray}
	I\left( r, z \right) ~=~ \frac{1}{r} \int_0^{2\mathrm{\pi}} \rho \left( r, \phi, z \right) \, d\phi \, .
	\label{Cylindrical_projection}
\end{eqnarray}
The density $\rho$ actually consists of many discrete particles and gas cells. We find the  contribution to each pixel by summing up the total mass in a finite range of $r$ and $z$ regardless of the cylindrical polar angle $\phi$. This is much better than a traditional edge-on (e.g. $xz$) projection because flaring at large $r$ can make it very difficult to see the crucially important central regions of the disk. While this may indeed occur in an external galaxy viewed close to edge-on, we can get a better view of our simulations.

Figure \ref{Particles_radial} shows the stellar particles in our isolated hydro simulations using this cylindrical projection. The initially very thin stellar disk remains rather thin for $\approx 1$~Gyr before starting to thicken. By the end of our 100~kK simulation (6~Gyr), though the disk remains rather thin further out, an X-shaped bulge is apparent in the central 2~kpc (top panel). This is related to the rather thick gas disk (Figure \ref{Gas_initial_thickness}) being unable to stabilize the much thinner stellar disk. We obtain similar results in our stellar-only simulation, where the gas component is treated as collisionless particles.

The situation is different at a lower temperature of 25~kK, which is closer to the 12~kK adopted in section 3.2 of \citetalias{Sellwood_2019}. Our 25~kK model retains a thin stellar disk for at least 7~Gyr (bottom panel of Figure \ref{Particles_radial}), demonstrating the importance of dissipation in the gas component despite its sub-dominant contribution to the central surface density (Figure \ref{M33_surface_density}).

Although \citet{Corbelli_2007_image} could obtain acceptable RC fits with a very small bulge having a scale length of only 0.15~kpc, their section 4.1 indicates that they ``could not exclude a larger bulge using dynamical arguments''. This is also apparent from our results $-$ the RC appears similar in both our 25~kK and 100~kK models (Figure \ref{M33_rotation_curves_iso}), even though only the 100~kK model has a significant bulge (Figure \ref{Particles_radial}). The possibility of a central bulge can also be addressed using photometry. The surface brightness of M33 does in fact show a rise toward its central regions, more so than an inward extrapolation of an exponential law fitted to larger radii \citep{Bothun_1992}. This suggests the presence of a 2~kpc bulge \citep{Regan_Vogel_1994}. However, other studies found that M33 has only a very small bulge \citep{Kormendy_1993, Gebhardt_2001}. More recent Spitzer photometry at $3.6 \, \mu$m suggests a bulge scale length of only $0.4$~kpc and a bulge fraction of just 4\% \citep[section 5.1.2 of][]{Kam_2015}. Thus, M33 does not have the sort of bulge evident in Figure \ref{Particles_radial}, indicating that our 100~kK model is unable to correctly reproduce this important aspect of the observations. However, qualitative agreement is gained using a temperature of 25~kK.

The importance of hydro effects in the gas is in line with some previous CDM simulations \citep{Shlosman_1993, Berentzen_1998}. Including hydro effects apparently did not much influence the simulations conducted by \citetalias{Sellwood_2019}, at least not to the point of obtaining a good match to observations. Nonetheless, the gas stabilizes their $\Lambda$CDM simulations somewhat $-$ their section 3.2 indicates that after 2~Gyr, the bar strength $A_2/A_0$ (defined in Section \ref{Bar_strength}) is $\approx 0.2$ with hydro treatment of the gas, but their nominal collisionless model reaches $A_2/A_0 \approx 0.3$ by this time (see the `full mass' curve in their figure 5). Although dissipation in the gas was insufficient on its own to stabilize the M33 disk and prevent it forming a strong bar, hydro effects could well be sufficient when combined with a different gravity law and the lack of angular momentum exchange with a live DM halo. A reduction in $A_2/A_0$ upon hydro treatment of the gas is also evident in the MOND model shown in figure 4 of \citet{Tiret_2008_gas}, especially in their Sc galaxy. Thus, it is not surprising that the gas should help to stabilize the disk to some extent, especially in a model where it constitutes a larger fraction of the mass because no CDM component is assumed.

\subsubsection{Velocity dispersion}
\label{Vertical_behaviour}

\begin{figure}
	\centering
	\includegraphics[width = 8.5cm] {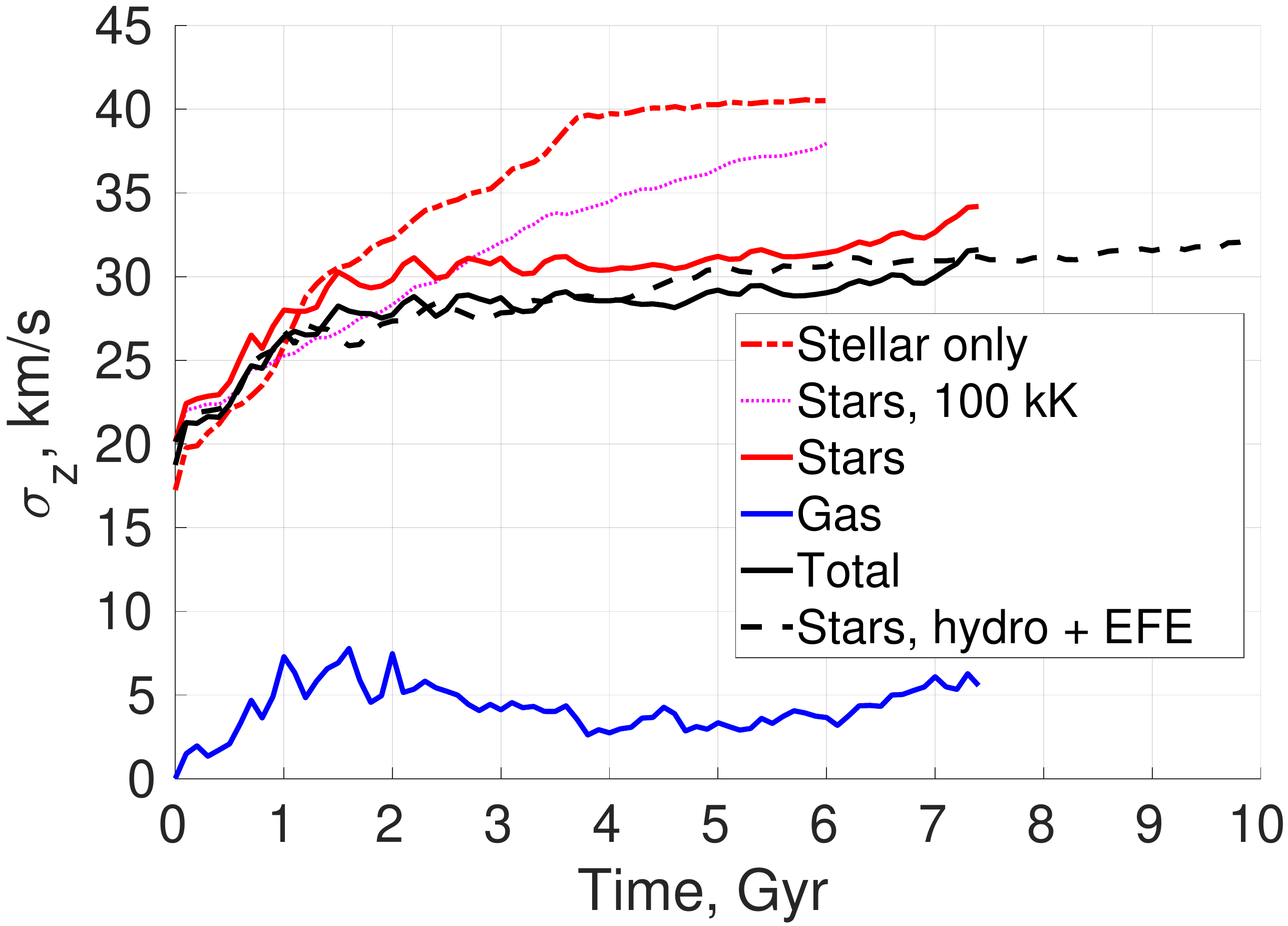}
	\caption{Time evolution of the simulated vertical velocity dispersion $\sigma_z$ of material with $r = \left( 0.5 - 3 \right)$~kpc. `Stellar-only' refers to our purely stellar $N$-body simulation. Other results are for hydro simulations at 25~kK, except for the pink dotted line (100~kK). The dotted black curve shows results when including the EFE (Section \ref{Including_EFE}), but other models are isolated. The 1D velocity dispersion of the gas is $\sigma_g \approx 22$~km/s at $T = 100$~kK but only $\sigma_g \approx 11$~km/s at $T = 25$~kK, possibly explaining the difference in stability properties of the stellar disk (see text).}
	\label{sigma_z}
\end{figure}

In both models, there is at least a gradual increase in thickness. This is mirrored in the mass-weighted vertical velocity dispersion $\sigma_z$ in the central region of M33 (Figure \ref{sigma_z}). Since this region is dominated by stars (Figure \ref{M33_surface_density}), $\sigma_z$ is not much affected by whether we consider the gas. In our purely stellar simulation, $\sigma_z$ almost doubles (from 20~km/s to 40~km/s) over a 6~Gyr period, though it is rising much more slowly towards the end. The same trend is apparent if gas is included at 100~kK, though the final $\sigma_z$ is reduced slightly to $\approx 35$~km/s. A more significant change arises in our model at 25~kK $-$ the stellar $\sigma_z$ now stabilizes at $\approx 30$~km/s after only $\approx 2$~Gyr, suggesting the model is now stable. This is also approximately when the stellar disk stops thickening (Figure \ref{Particles_radial}). A mild amount of disk heating at this early stage in the simulation could well be due to difficulties setting up the gas and stellar disks in equilibrium with each other (Section \ref{Including_gas}).

The difference in behaviour due to the gas temperature can be understood by relating the above values to the 1D velocity dispersion of the gas. This is $\sigma_g = \sqrt{kT/\left( \mu m_p \right)} = 21.7$~km/s for $T = 100$~kK, where $k$ is the Boltzmann constant, $m_p$ is the mass of a hydrogen atom, and the mean molecular weight for neutral gas with a 25\% primordial helium mass fraction is $\mu = 7/4$. We see that $\sigma_g$ is similar to the initial stellar $\sigma_z$, so the gas is unable to help stabilize the stellar component. However, reducing $T$ to 25~kK reduces $\sigma_g$ to 10.9~km/s, which is much less than the stellar $\sigma_z$. Therefore, gas in our cooler model can in principle help to stabilize the stellar disk.

To facilitate a more direct comparison with observations, we find the line of sight (LOS) velocity dispersion $\sigma_{_{LOS}}$ of the stars. We assume that M33 is inclined by $i = 50^\circ$ with respect to the plane of our sky \citep[figure 1 of][]{Corbelli_2007}. Although mild warping is evident, this inclination is particularly accurate for the central few kpc of M33. It is also consistent with the $i = 52^\circ \pm 2^\circ$ obtained in section 4.2.1 of \citet{Kam_2015}. We project the position and velocity of each stellar particle onto our LOS, which we take to be along the direction $\left( \sin i, 0, \cos i \right)$. This lets us construct an `image' of M33 based on the sky-projected position of each particle. We divide this image into squares of side 1~kpc, with the centre of M33 coincident with that of the central pixel.

Since each pixel contains a finite number of particles, there is a non-zero correlation between their mean LOS velocity $\overline{v}_{_{LOS}}$ and that of each particle. The real M33 has many more particles, so this is a numerical artefact which should be corrected. The standard approach is to enhance the measured variance by a factor of $N/\left( N - 1 \right)$ for a pixel with $N$ particles. For particles with arbitrary masses $m_i$ and LOS velocity $v_{_{LOS, i}}$, the appropriate generalization of this result is:
\begin{equation}
	\sigma_{_{LOS}} = \sqrt{\frac{\sum\limits_{i=1}^N m_i \sum\limits_{i=1}^N m_i v_{_{LOS,i}}^2 - \left( \sum\limits_{i=1}^N m_i v_{_{LOS, i}} \right)^2}{\left( \sum\limits_{i=1}^N m_i \right)^2 - \sum\limits_{i=1}^N {m_i}^2}} \, .
	\label{sigma_LOS_mass_weighted}
\end{equation}

Once we have obtained $\overline{v}_{_{LOS}}$ and $\sigma_{_{LOS}}$, we perform iterative ${5\sigma}$ outlier rejection with stringent convergence conditions, one of which is that the number of `accepted' particles must be the same as on the previous iteration and must exceed 150. We do not report $\sigma_{_{LOS}}$ for pixels with fewer accepted particles as these results are more affected by Poisson noise. We found that the $\sigma$-clipping gives very similar results for any threshold number of standard deviations $\geq 2.5$, including in the case of not applying outlier rejection at all. We nonetheless apply a conservative $5\sigma$ outlier rejection to ensure the robustness of our procedure. We also check in all cases that the $v_{_{LOS}}$ values in the central pixel closely follow a Gaussian of width $\sigma_{_{LOS}}$. An example is given in Section \ref{Bulge_EF}.

\begin{figure}
	\centering
	\includegraphics[width = 8.5cm] {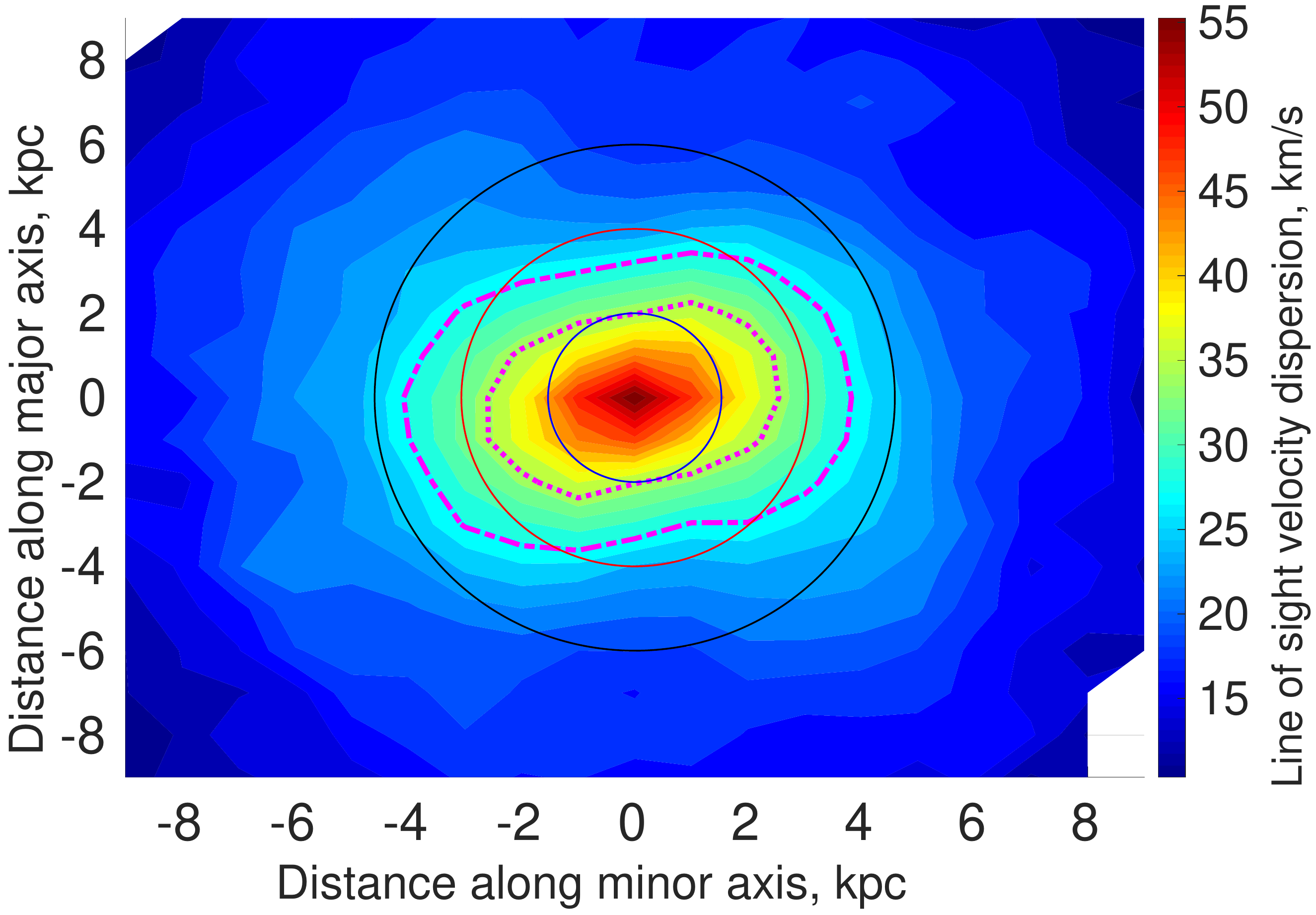}
	\caption{The line of sight velocity dispersion of stars in different parts of the sky-projected image of M33, shown for the final snapshot at 7.4~Gyr for our isolated 25~kK simulation. Thin ellipses show the projected appearance of circles in the disk plane with radii of 2, 4, and 6~kpc, approximately integer multiples of the stellar disk scale length (Table \ref{Parameters}). The magenta dotted (dot-dashed) contours show $\sigma_{_{LOS}} = 35$ (28)~km/s, bracketing the observational range \citep[section 3.2 of][]{Corbelli_2007_image}. The central $\sigma_{_{LOS}}$ is 57~km/s, well above this range. Similar results are obtained if viewing M33 from within the $\bm{y}\bm{z}$ plane, with a central $\sigma_{_{LOS}}$ differing by $< 0.1$~km/s (not shown). Values are $\approx 5$~km/s higher for the final snapshot of our isolated 100~kK simulation (not shown). In both cases, the $v_{_{LOS}}$ values in the central pixel closely follow a Gaussian with the calculated $\sigma_{_{LOS}}$, as shown explicitly for a better fitting model in Section \ref{Bulge_EF}.}
	\label{sigma_LOS}
\end{figure}

Figure \ref{sigma_LOS} shows our simulated stellar $\sigma_{_{LOS}}$ map for our isolated 25~kK model. We include magenta contours bracketing the observational range of ${\left( 28 - 35 \right)}$~km/s, which does not consider the very central regions of M33 \citep[section 3.2 of][]{Corbelli_2007_image}. Assuming only a very small region near the centre is excluded observationally, we expect to get a central $\sigma_{_{LOS}}$ approximately within this range. However, the central pixel in Figure \ref{sigma_LOS} has a much higher $\sigma_{_{LOS}} = 57$~km/s. Thus, the simulated stellar $\sigma_{_{LOS}}$ is higher than observed. The situation is worse for the 100~kK model (not shown), where $\sigma_{_{LOS}} = 62$~km/s. This is related to the simulation developing a significant central bulge (Figure \ref{Particles_radial}) and higher $\sigma_z$ (Figure \ref{sigma_z}).

Since the stellar and gas RCs trace the same potential, any mismatch between them is partly caused by asymmetric drift, which in turn depends on the size and shape of the stellar velocity dispersion tensor. This technique could yield constraints on $\sigma_z$, as attempted by \citet{Corbelli_2007_image} in their section 4.1. They did not obtain conclusive results because they found that part of the mismatch must be due to non-circular motions caused by the bar, which we discuss next. Their only conclusion was that ${\sigma_z \ga 20}$~km/s, consistent with our simulations (Figure \ref{sigma_z}).

The stars in our 25~kK simulation form a rather thin disk. The gas has a thicker distribution (Appendix \ref{Gas_edge_on}). This is mainly due to the difficulties we encountered setting up the gas component, which forced us to make it rather hot (Section \ref{Including_gas}). Nonetheless, \citet{Koch_2018} mention in their section 5.1 that there is evidence for a rotationally lagging gas component, which could well be a thick gas disk. Directly detecting this would be difficult due to the inclination of M33, which makes it hard to know how far any detected gas lies from the disk mid-plane. A better comparison with observations would require a simulation at even lower $T$. In this respect, we attempted to reduce $T$ further, but found that the disk becomes very unstable at 10~kK. This might be addressed by raising $levelmax$, but ultimately one would need to include the associated star formation and stellar feedback. A higher initial gas fraction might then be required to match the observed value for M33, depending on the resulting star formation rate.

It is thus unclear what effect an even thinner gas disk would have. On the one hand, it would increase the surface density enclosed within the stellar disk, making it less stable. On the other hand, since reducing $T$ from 100~kK to 25~kK helps to stabilize our model, further reductions might reduce $\sigma_z$ further and make the disk even thinner. Any developing instabilities in the stellar disk could be more easily absorbed by the gas disk if it is allowed to cool further, which numerically would involve reducing $T2\_star$ (Section \ref{Simulation_setup}).


\begin{figure*}
	\centering
	\includegraphics[width = 17.7cm] {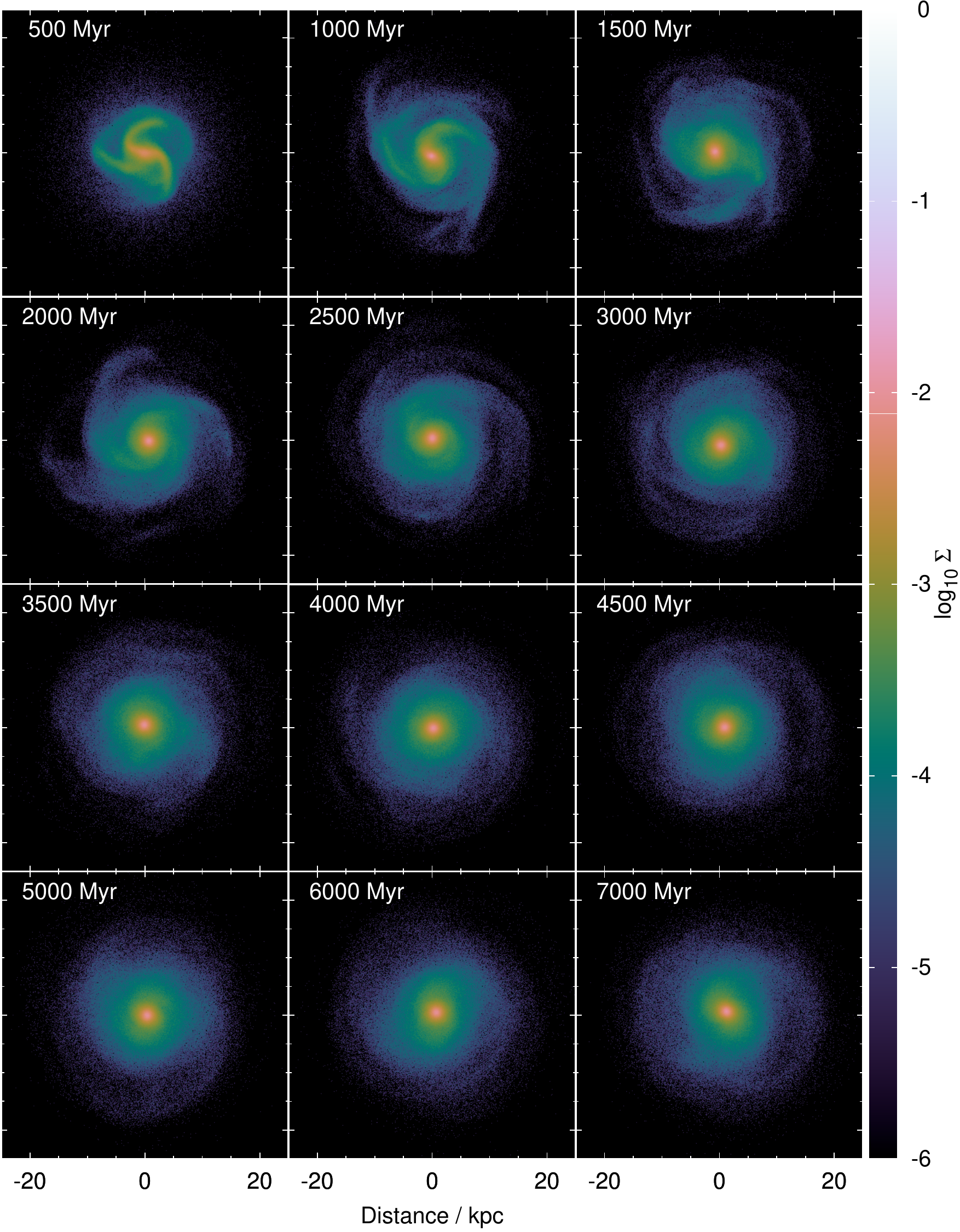}
	\caption{Face-on view of the stellar mass distribution in our isolated hydro model of M33 at 25~kK, with the surface density $\Sigma$ shown in $1000 \, M_\odot$/pc$^2$. Results for the analogous model at 100~kK are shown in Appendix \ref{100kK_iso_face_on}. Despite the significant difference in morphology, the RCs are nearly identical (Figure \ref{M33_rotation_curves_iso}).}
	\label{Particles_face_on}
\end{figure*}

\begin{figure*}
	\centering
	\includegraphics[width = 17.7cm] {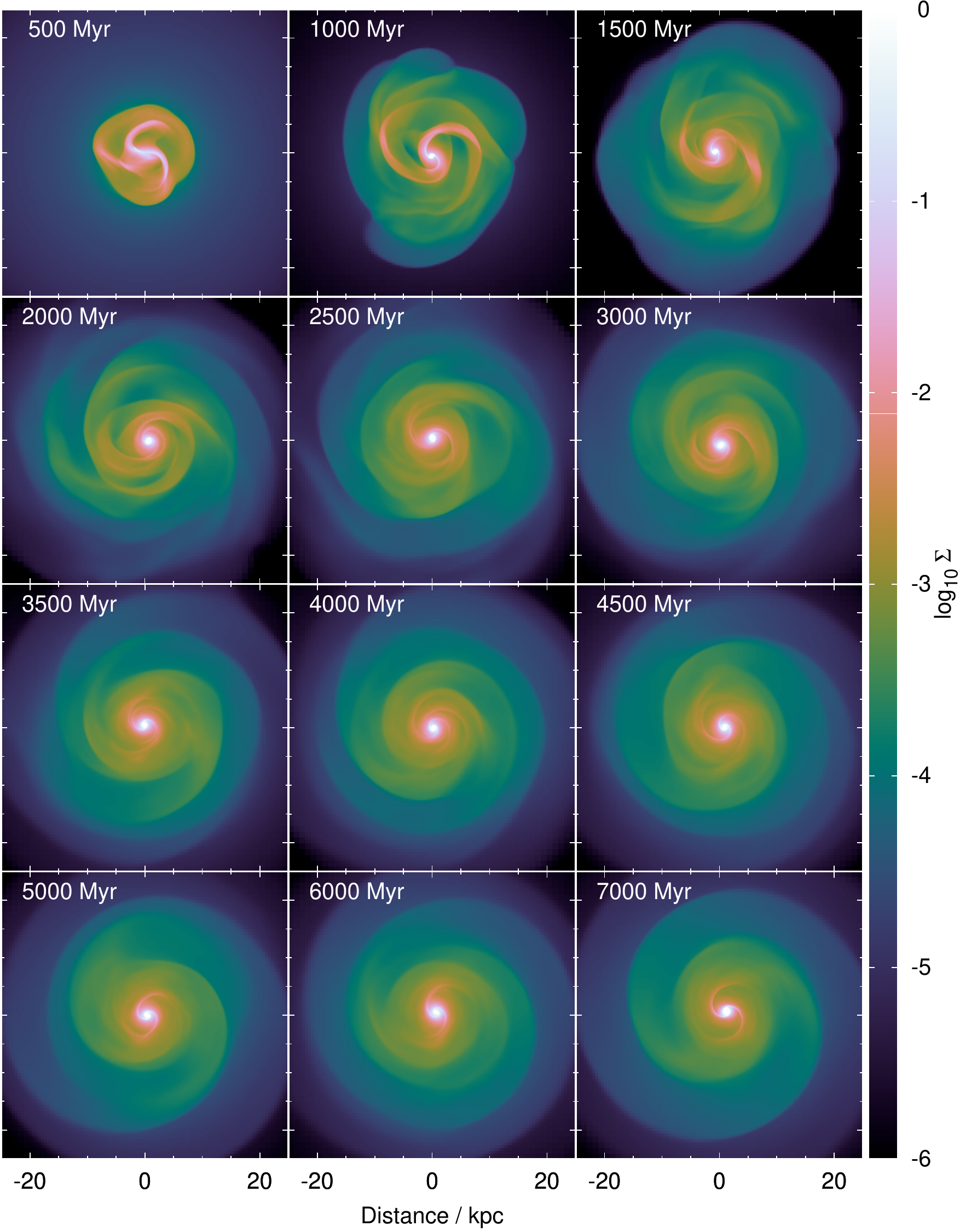}
	\caption{Similar to Figure \ref{Particles_face_on}, but for the gas. Notice the bisymmetric spiral towards the end. Results for the 100~kK model are shown in Appendix \ref{100kK_iso_face_on}.}
	\label{Gas_face_on}
\end{figure*}

\subsection{Face-on view and the M33 bar}
\label{Face_on_view}

Figure \ref{Particles_face_on} shows face-on ($xy$) views of the stellar particles in our 25~kK hydro simulation every 500~Myr. A weak two-armed spiral is apparent after $\approx 3$ Gyr, similar to the actual M33 \citep[figure 11 of][]{Corbelli_2007_image}. A fairly strong bar develops quite rapidly, but then weakens as the simulation progresses. This is also evident when viewing the gas face-on (Figure \ref{Gas_face_on}).

Results for the 100~kK simulation are shown in Appendix \ref{100kK_iso_face_on}. There is a very strong bar even after 6~Gyr, to the extent that the whole galaxy is essentially one giant bar. An accurate treatment of the gas component is therefore important to get a simulated M33 resembling the observed one in a MOND context. In what follows, we focus mostly on our 25~kK models.

\subsubsection{Bar strength}
\label{Bar_strength}

To quantify the strength of our simulated bar, we follow the approach of \citetalias{Sellwood_2019} and perform a Fourier decomposition of the mass distribution within cylindrical radii of $r = \left( 0.5 - 3 \right)$~kpc. We begin by calculating:
\begin{eqnarray}
	\label{Fourier_mode_equation}
	A_0 &\equiv& \frac{1}{2} \sum\limits_i m_i \, \left(i \text{ labels different particles} \right) ,\\
	A_{\mathrm{m}} &\equiv& \sqrt{{\left( \sum\limits_i m_i \cos \mathrm{m} \phi_i \right)}^2 + {\left( \sum\limits_i m_i \sin \mathrm{m} \phi_i \right)}^2} ,\,  \mathrm{m} \geq 1 \, .\nonumber
\end{eqnarray}
This quantifies the presence of non-axisymmetry, i.e. dependence on the azimuthal angle $\phi \equiv \sin^{-1} \left( y/r \right)$. We then find the strength of the $\mathrm{m}$\textsuperscript{th} Fourier mode using:
\begin{eqnarray}
	\text{Strength of Fourier mode } \mathrm{m} ~\equiv~ \frac{A_{\mathrm{m}}}{A_0} \, .
\end{eqnarray}

\begin{figure}
	\centering
	\includegraphics[width = 8.5cm] {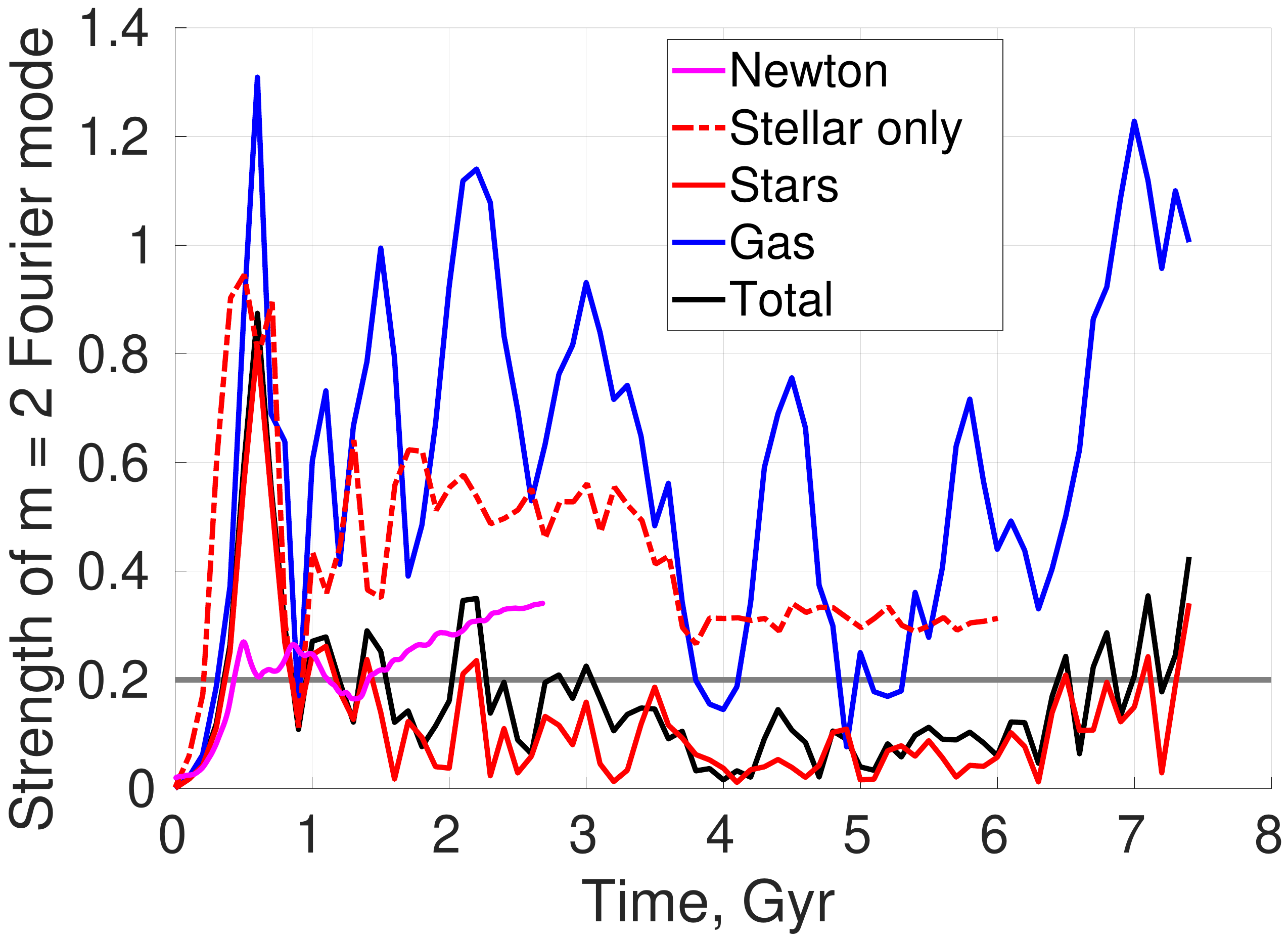}
	\caption{Strength of the $\mathrm{m} = 2$ azimuthal Fourier mode for material with $r = \left( 0.5 - 3 \right)$ kpc (Equation \ref{Fourier_mode_equation}). The observational estimate of 0.2 is shown as a horizontal grey line \citep[section 4.3 of][]{Corbelli_2007_image}. The magenta curve shows the most realistic (`full mass') Newtonian simulation published in \citetalias{Sellwood_2019}, which included a live DM halo but only ran for 2.7~Gyr. The hydro model shown here is our isolated 25~kK run. The 100~kK run (not shown) gives results only slightly below the stellar-only model, as expected for a gas disk too thick to much affect the stellar disk (Appendix \ref{Gas_edge_on}).}
	\label{Mode_2_strength}
\end{figure}

Of particular importance is the ${\mathrm{m} = 2}$ Fourier mode, whose strength we show for the stellar distribution (Figure \ref{Mode_2_strength}). We use a magenta line to overplot the nominal model of \citetalias{Sellwood_2019}, which they labeled `full mass' in their figure 5. This simulation contains a live DM halo, but only ran for 2.7~Gyr compared to our 7.4~Gyr. In a $\Lambda$CDM context, it is rather unlikely that the bar would get weaker if evolved for longer \citep[e.g.][]{Tiret_2007}. Their figure 7 indicates that a temporary dip in bar strength is possible after a few dynamical times, but the bar then rapidly recovers its strength. A temporary dip is indeed evident in the \citetalias{Sellwood_2019} results reproduced here, but the bar then strengthens after $\approx 1$~Gyr, suggesting that there is no scope for subsequent weakening. This led those authors to conclude that their model cannot match the observed rather weak bar of M33, whose strength has been estimated at 0.2 \citep[section 4.3 of][]{Corbelli_2007_image}.

Our simulation initially forms a much stronger bar than the simulations of \citetalias{Sellwood_2019}. However, it reaches peak strength after $\la 1$~Gyr and starts getting weaker. This is similar to the behaviour evident in figure 4 of \citet{Tiret_2008_gas}, who ran their simulations for $\left( 7 - 8 \right)$~Gyr. By the end of our 25~kK simulation, $A_2/A_0 \approx 0.2$, similar to that observed in M33. There are extended periods with a very weak bar ($A_2/A_0 \la 0.1$). While this is mitigated to some extent by the generally more important $\mathrm{m} = 1$ mode (Section \ref{Other_harmonics}), it is clear from the face-on view that there are indeed extended periods with very small departures from axisymmetry (Figure \ref{Particles_face_on}).

The bar is stronger in our purely stellar simulation ($A_2/A_0 \approx 0.3$, see Figure \ref{Mode_2_strength}). Including gas at 100~kK leads to similar overall behaviour to the stellar-only model, but with the final $A_2/A_0 \approx 0.2$ (not shown). In both cases, the $A_2/A_0$ ratio in our analysis does not indicate a genuinely weak bar comparable to observations. Rather, the face-on view shows that the entire disk is essentially one giant bar, with semi-minor axis comparable to the 3~kpc aperture used to calculate $A_2/A_0$ (Appendix \ref{100kK_iso_face_on}). The very long bar it reveals implies a large corotation radius, since bars should be shorter than corotation \citep{Contopoulos_1980}. We discuss this next by finding the bar pattern speed.

\subsubsection{Bar pattern speed}
\label{Bar_pattern_speed}

Fourier decomposing the stellar distribution allows us to extract the bar pattern speed $\Omega_p$. For this purpose, we first define an angle $\theta_{\mathrm{m}}$ for each Fourier mode, where
\begin{eqnarray}
	\tan \left( \mathrm{m} \theta_{\mathrm{m}} \right) ~\equiv~ \frac{\sum_i m_i \sin \left( \mathrm{m} \phi_i \right)}{\sum_i m_i \cos \left( \mathrm{m} \phi_i \right)} \, .
	\label{Mode_phase}
\end{eqnarray}
We then consider how $\theta_2$ changes with time, adding multiples of $\mathrm{\pi}$ using the method of exact fractions in order to minimize the change between successive snapshots. 

\begin{figure}
	\centering
	\includegraphics[width = 8.5cm] {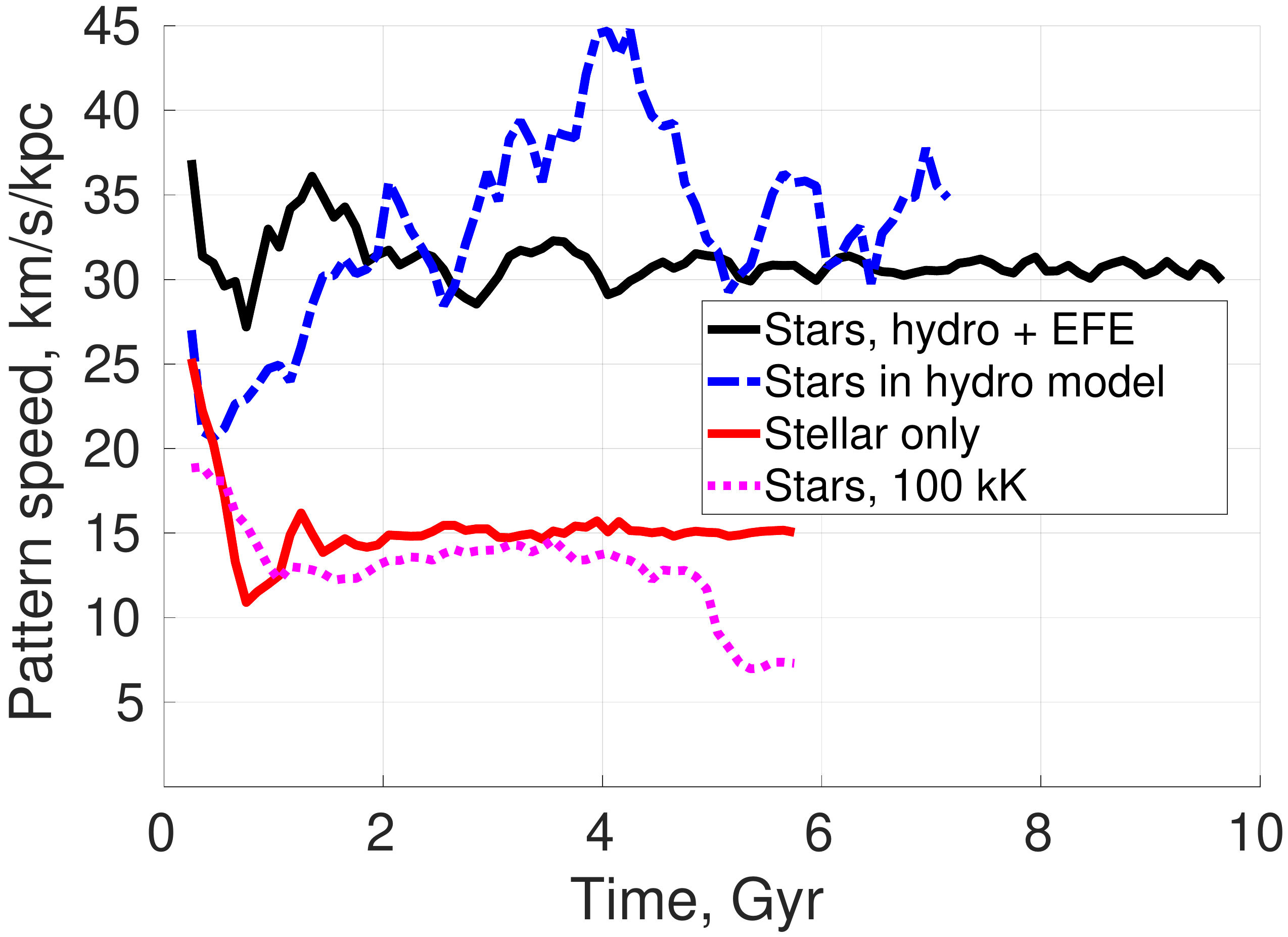}
	\caption{Bar pattern speed for stars with $r = \left( 0.5 - 3 \right)$ kpc in our isolated stellar-only and 25~kK hydro simulations (solid red and dotted blue curves, respectively). Our 100~kK simulation (dotted pink) behaves similarly to the stellar-only case. The black curve shows the pattern speed of the stars in our simulation with the EFE (Section \ref{Including_EFE}). In all hydro simulations, results are very similar when using the total mass distribution instead (not shown).}
	\label{M33_pattern_speed}
\end{figure}

The bar rotates in a prograde sense at the speed shown in Figure \ref{M33_pattern_speed}. Our hydro models at 25~kK yield $\Omega_p \approx 30$~km/s/kpc. Since galactic bars are typically fast and end close to their corotation radius \citep[e.g.][]{Guo_2019}, we use Figure \ref{M33_rotation_curves_iso} to show a line at this gradient. It intersects the RC at $r \approx 3$~kpc regardless of whether we use the simulated or observed RC, suggesting a bar intermediate in length between the stellar and gas disk scale lengths (Table \ref{Parameters}). Since our $\Omega_p$ calculation is based on material at $r = \left( 0.5 - 3 \right)$~kpc, the estimated corotation radius of 3~kpc for our 25~kK models should be rather reliable.

This estimate can be compared with the observational result given in section 3.2.7 of \citet{Elmegreen_1992} that the corotation radius is $0.4 \, r_{25}$, with their table 1 indicating $r_{25} = 30\arcmin$ for NGC 598 $\equiv$ M33. For our adopted distance of 840~kpc \citep[][and references therein]{Kam_2015}, this corresponds to an observed corotation radius of 2.9~kpc. This is based on the length of one star formation ridge, but the presence of other ridges that extend out to larger $r$ means the result of \citet{Elmegreen_1992} is not very secure. Section 5.2 of \citet{Corbelli_2019} gives a corotation radius of $\left( 4.7 \pm 0.3 \right)$~kpc for the spiral pattern in M33. Since bars generally have a larger $\Omega_p$, it could well be that their estimated corotation radius exceeds that of \citet{Elmegreen_1992} because the latter result corresponds to the bar. Although a comparison with observations would be quite valuable, corotation radii are rather difficult to measure for a bar as weak as that in M33.

Turning to our less realistic isolated models with only stars or with $T = 10^5$~K, both yield $\Omega_p \approx 10$~km/s/kpc (Figure \ref{M33_pattern_speed}). A line at this gradient would intersect the flat portion of the RC at $r \approx 10$ kpc, suggesting a very long bar (Figure \ref{M33_rotation_curves_iso}). This is actually quite consistent with Appendix \ref{100kK_iso_face_on}, where we see that the whole galaxy is essentially one giant bar when viewed face-on. However, this is very different to the observed M33 \citep[e.g. figure 11 of][]{Corbelli_2007_image}.

Despite these serious problems, the simulated $\Omega_p$ in these models is quite consistent with the observational estimate of $\left( 10 \pm 2 \right)$~km/s/kpc given in section 4.3 of \citet{Corbelli_2007_image}. Those authors assumed the bar ends at its own corotation radius and has a length of $\left( 0.6 - 1.5 \right)$~kpc. However, this agreement is not a success of the models $-$ the RC of \citet{Koch_2018_error} reaches an amplitude of 72~km/s at 1.5~kpc, so a 1.5~kpc long bar ending at corotation has $\Omega_p = v_c/r = 48$~km/s/kpc. Under this assumption, a shorter bar would have an even higher $\Omega_p$ (Figure \ref{M33_rotation_curves_iso}). Clearly, the $\Omega_p$ estimate of \citet{Corbelli_2007_image} is seriously discrepant with subsequent analyses. Its main problem is that it seems to equate $\Omega_p$ with the local slope of the RC instead of the angular frequency implied by the RC. In other words, in their pattern speed estimate, \citet{Corbelli_2007_image} probably assumed that a bar ending at corotation has $\Omega_p = dv_c/dr$, instead of the correct $\Omega_p = v_c/r$. Thus, the agreement between their estimated $\Omega_p$ and that of our stellar-only or 100~kK models appears to be coincidental.

\subsubsection{Other harmonics}
\label{Other_harmonics}

\begin{figure}
	\centering
	\includegraphics[width = 8.5cm] {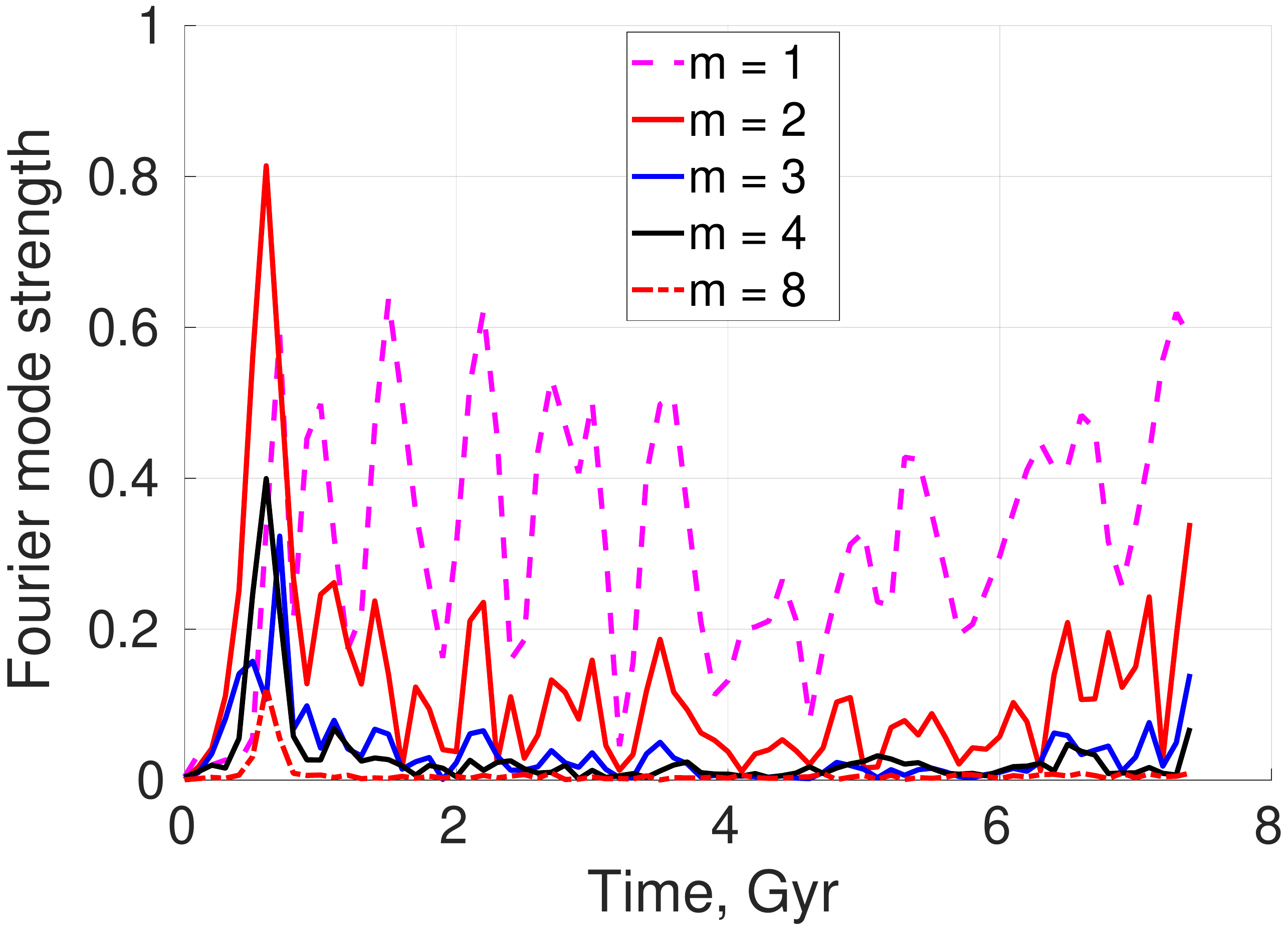}
	\caption{Strengths of different Fourier modes (Equation \ref{Fourier_mode_equation}) for stellar particles at ${r = \left( 0.5 - 3 \right)}$ kpc in our isolated 25~kK hydro model. Some modes can have strength ${>1}$ if multiple modes are present and their phases (Equation \ref{Mode_phase}) are fortuitously aligned. Results remain rather similar when including the gas (not shown).}
	\label{Mode_strengths_stellar_only}
\end{figure}

We can use Equation \ref{Fourier_mode_equation} to find the strengths of modes with any $\mathrm{m} \geq 1$. Following \citet{Tiret_2007}, we repeat our mode strength calculations for $\mathrm{m} = 1, 2, 3, 4$, and 8. Our results are shown in Figure \ref{Mode_strengths_stellar_only} based on the distribution of stellar particles with $r = \left( 0.5 - 3 \right)$~kpc in our 25~kK isolated model. Results are similar if the total mass distribution is used instead (not shown). It is clear that the $\mathrm{m} = 1$ mode is dominant except for brief periods when it is overtaken by the $\mathrm{m} = 2$ mode.

It is difficult to know whether the strong $\mathrm{m} = 1$ mode is a problem for our model, since \citet{Elmegreen_1992} stated in their section 4 that background intensity gradients make it difficult to reliably obtain the $\mathrm{m} = 1$ mode strength observationally, causing them to not plot this at all. Some asymmetry between different sides of M33 is required to explain the $\approx 10$~km/s differences in the RC inferred from its approaching and receding halves \citep{Kam_2015, Kam_2017}. In any case, the dominant modes of non-axisymmetry in the stellar distribution are clearly $\mathrm{m} \leq 2$, with very little power in higher harmonics. Moreover, Figure \ref{Gas_face_on} shows a very symmetric two-armed spiral in the final snapshot of the gas. In a more advanced simulation, this may well cause a prominent spiral arm traced by newly formed stars, perhaps resembling the observed two-armed spiral structure of M33 \citep[e.g. figure 11 of][]{Corbelli_2007_image}. Therefore, our isolated 25~kK model seems to resemble the observed M33 in many respects. It is also possible to avoid a dominant $\mathrm{m} = 1$ mode in one of our models with the EFE (Section \ref{Different_g_ext_directions}).


The number of spiral arms is related to the wavelength most unstable to amplification by disk self-gravity. In general, we expect this to be shorter in Newtonian models with an appropriate DM halo, leading to a larger expected number of arms \citep[section 3.3 of][]{Banik_2018_Toomre}. While this is clearly a positive point for our isolated MOND simulation of M33 at 25~kK, the presence of M31 and its possible EFE on M33 motivate us to consider relaxing the assumption of isolation in the next section. This is especially relevant in light of the theoretical requirement for the EFE in MOND \citep{Milgrom_1986}, the clear disagreement between data on local wide binaries and MOND expectations if the Galactic EFE were artificially removed \citep{Pittordis_2019}, and the recent detection of the EFE through accurate RC measurements of galaxies in different environments \citep{Chae_2020}. There is no classical analogue to the EFE, which violates the strong equivalence principle.


\section{Including the EFE from M31}
\label{Including_EFE}

In MOND, M31 can affect M33 through the EFE (Section \ref{External_field_effect}). Since the M31 RC has ${v_{_f} \approx 225}$~km/s \citep{Carignan_2006} and its distance to M33 is ${d \approx 200}$~kpc \citep[table 1 of][]{Patel_2017}, the external field on M33 is mainly the M31 gravity $g_{ext} = {v_{_f}}^2/d = 0.07 \, a_{_0}$, which lies well within the DML. This is much less than the central surface gravity of M33, which slightly exceeds $a_{_0}$ (Section \ref{M33_mass_distribution}). Therefore, the {\it a priori} expectation would be that M31 does not have a big effect on M33 via the EFE. This is nevertheless worth investigating for the above reasons.

To estimate the possible impact of the EFE on M33, we conduct another hydro simulation with the same disk template (and thus parameters) as our isolated model, but this time including a constant external field of $0.07 \, a_{_0}$ directed at $30^\circ$ to the disk plane:
\begin{eqnarray}
	\frac{\bm g_{ext}}{0.07 \, a_{_0}} ~=~ \begin{bmatrix}
	\cos 30^\circ \\
	0 \\
	\sin 30^\circ
	\end{bmatrix} \, .
	\label{g_ext_value}
\end{eqnarray}
This can be thought of as placing M31 at $\left(173.2, 0, 100 \right)$~kpc relative to M33. We use $30^\circ$ as a representative angle $-$ this is the median inclination of a randomly chosen vector to a fixed plane. Other $\bm{g}_{ext}$ orientations are considered in Section \ref{Different_g_ext_directions}.

Note however that the magnitude and direction of $\bm{g}_{ext}$ would have been different at earlier times due to orbital motion of M33 around M31. It would be interesting to consider time variation of $\bm{g}_{ext}$ by directly including M31 in the simulation volume as a collection of static particles, similar to the technique used in \citet{Thomas_2017} to minimize the computational cost. However, this goes beyond the scope of the present work, which intends to check if the EFE can influence disk stability. Given the relatively slow orbit of M33 around M31 compared to the rotation period of M33 (Section \ref{Simulation_setup}), time variation of $\bm{g}_{ext}$ should have only a small effect on our results. This is discussed further in Section \ref{External_field_warp}.

To simplify our analysis, we neglect the effect of tides from M31 even though we consider its EFE. This is because the tidal stress on M33 is of order $g_{ext} \left( r_d/d \right) $, where $r_d \approx 2$~kpc is the M33 disk scale length (Table \ref{Parameters}), and $d \approx 200$~kpc is the distance to M31 \citep{Patel_2017}. The tidal stress from M31 is thus two orders of magnitude less significant than the EFE. Since even the EFE is rather sub-dominant in the central regions of M33 (Figure \ref{M33_surface_density}), neglecting the tidal stress from M31 should be a very good approximation at the present epoch.

\subsection{Numerical implementation}
\label{Numerical_implementation_EFE}

Including the EFE in our \textsc{por} simulations requires adjusting the PDM calculation and the boundary condition for the potential, which we now describe. The simulation setup is otherwise the same as before (Section \ref{Setup}), apart from some changes to the resolution settings (Section \ref{Reduced_resolution_EFE}). We also use the same disk template as the RC should not be affected very much in the region of interest (see below). Given our results in Section \ref{Results}, we use $T = 25$~kK for all our models with the EFE.

\subsubsection{Changes to the gravity solver}
\label{Gravity_solver_EFE}

The gravity solver is adjusted to impose a constant external gravitational field on the simulated domain. Thus, $\bm{g}_{ext}$ adds an additional source of gravity to Equation \ref{QUMOND_governing_equation}, altering $\nu$. In the QUMOND approach, the precise way in which this occurs depends on the Newtonian-equivalent external field $\bm{g}_{_{N, ext}}$, which is what the external field would have been in Newtonian gravity. Since M31 can be thought of as a distant point mass, we assume $\bm{g}_{_{N, ext}}$ is parallel to $\bm{g}_{ext}$, with their magnitudes related by Equation \ref{ALM}:
\begin{eqnarray}
	g_{_{N, ext}} ~=~ \frac{{g_{ext}}^2}{g_{ext} + a_{_0}} ~=~ 4.6 \times 10^{-3} \, a_{_0} \, .
\end{eqnarray}
To retain our previous notation that $\bm{g}_{_N}$ refers to the Newtonian gravity of M33 alone, we generalize Equation \ref{QUMOND_governing_equation} to:
\begin{eqnarray}
	\label{QUMOND_governing_equation_EFE}
	\nabla \cdot \bm{g} ~&=&~ \nabla \cdot \left[ \bm{g}_{_{N, tot}} \nu \left( \bm{g}_{_{N, tot}} \right) \right] \, , \\
	\bm{g}_{_{N, tot}} ~&\equiv&~ \bm{g}_{_N} + \bm{g}_{_{N, ext}} \, .
\end{eqnarray}
In a more detailed model, $\bm{g}_{_N}$ and $\bm{g}_{_{N, ext}}$ would be directed towards M31, making their magnitude and direction time-dependent. The EFE generally has the effect of reducing $\nu$, making the system more Newtonian. However, partial cancellation between internal and external fields is also possible, which would raise $\nu$ for a weak EFE.

In addition to changing $\nu$ inside the simulation box, the EFE also changes the boundary condition $-$ Equation \ref{Boundary_Phi_iso} is only valid for a point mass if the EFE can be neglected. $\Phi$ has a rather complicated behaviour if internal and external fields are comparable \citep[e.g.][]{Thomas_2018, Banik_2019_spacecraft}. However, it is possible to obtain $\Phi$ analytically if the boundary is external field dominated, i.e. if $g_{_{N, ext}} \gg g_{_N}$ \citep[section 3 of][]{Banik_2015}. Treating the matter distribution as a point mass $M$, they obtained the solution:
\begin{eqnarray}
	\Phi ~=~ -\frac{GM \nu_{ext}}{R} \left( 1 + \frac{K_0}{2} \sin^2 \theta \right) \, ,
	\label{Boundary_Phi_EFE}
\end{eqnarray}
where $\theta$ is the angle between $\bm{g}_{_{N, ext}}$ and the position $\bm{R}$, while $K_0$ was defined earlier (Equation \ref{K_0_definition}). This solution was confirmed numerically by direct integration of Equation \ref{QUMOND_governing_equation} \citep[section 2.2 of][]{Banik_2018_escape}. The external field dominates for distances $R \gg R_{_{EFE}}$, where the transition radius in the DML is:
\begin{eqnarray}
	R_{_{EFE}} ~=~ \frac{\sqrt{GMa_{_0}}}{g_{ext}} \, .
	\label{r_EFE}
\end{eqnarray}
Since the M33 mass $M = 6.5 \times 10^9 \, M_\odot$ (Table \ref{Parameters}), the external field from M31 dominates beyond $R_{_{EFE}} = 41$~kpc, much larger than M33 (Figure \ref{M33_surface_density}). We therefore increased the box size to 1024~kpc, ensuring a minimum distance of 512~kpc $\gg R_{_{EFE}}$ from M33 to the edge. At this distance, it would be quite accurate to treat M33 as a point mass. Therefore, we replace Equation \ref{Boundary_Phi_iso} with Equation \ref{Boundary_Phi_EFE} for our simulations with the EFE.

\subsubsection{Changes to resolution settings}
\label{Reduced_resolution_EFE}

Our preceding discussion shows that including a weak external field is quite computationally expensive because the box size must be large enough for the boundary to be external field dominated. Due to the high computational cost and memory requirement of our simulation, we reduce the maximum refinement level to $levelmax = 12$. The most resolved regions thus have a resolution of $1024/2^{12}$~kpc $=$ 250~pc, which is still much smaller than the scale length of the M33 disk (Table \ref{Parameters}). We maintain the previous setting that the minimum refinement level is $levelmin = 7$, so even the most poorly resolved regions have a cell size of 8~kpc.

To check how these changes affect our results, we then run a higher resolution simulation for 6.1~Gyr with the EFE (Section \ref{Numerical_convergence}). The simulation was not advanced further due to significant numerical drift of M33 in roughly the direction opposite to $\bm{g}_{ext}$, with the barycentre position changing approximately quadratically over time at an implied acceleration of $3.4 \times 10^{-3} \, a_{_0}$. The results of both simulations are sufficiently similar that it is clear the lower resolution simulation has reached numerical convergence. Our higher resolution simulation with the EFE has exactly the same resolution settings and box size as our isolated 25~kK simulation and also uses the same disk template, so any differences can be attributed solely to the EFE. However, to discuss the evolution of M33 over a longer timespan and to minimize the role of edge effects, we focus mainly on our lower resolution EFE simulation, which we advance for 9.9~Gyr. The numerical drift of M33 is much smaller in this case, with an implied acceleration of just $3.9 \times 10^{-4} \, a_{_0}$. This is probably due to the larger box size making Equation \ref{Boundary_Phi_EFE} more accurate on the boundary.\footnote{The numerical issue of barycentre drift can be explored with fewer computations using the ring library procedure described in \citet{Banik_2018_escape} $-$ a point mass embedded in a uniform $\bm{g}_{ext}$ experiences a non-zero acceleration despite the use of a spherical domain. This can be reduced to a very small level by improving the resolution and enlarging the simulated region.}

\subsection{Results}
\label{Results_EFE}

\subsubsection{Rotation curve and surface density}

\begin{figure}
	\centering
	\includegraphics[width = 8.5cm] {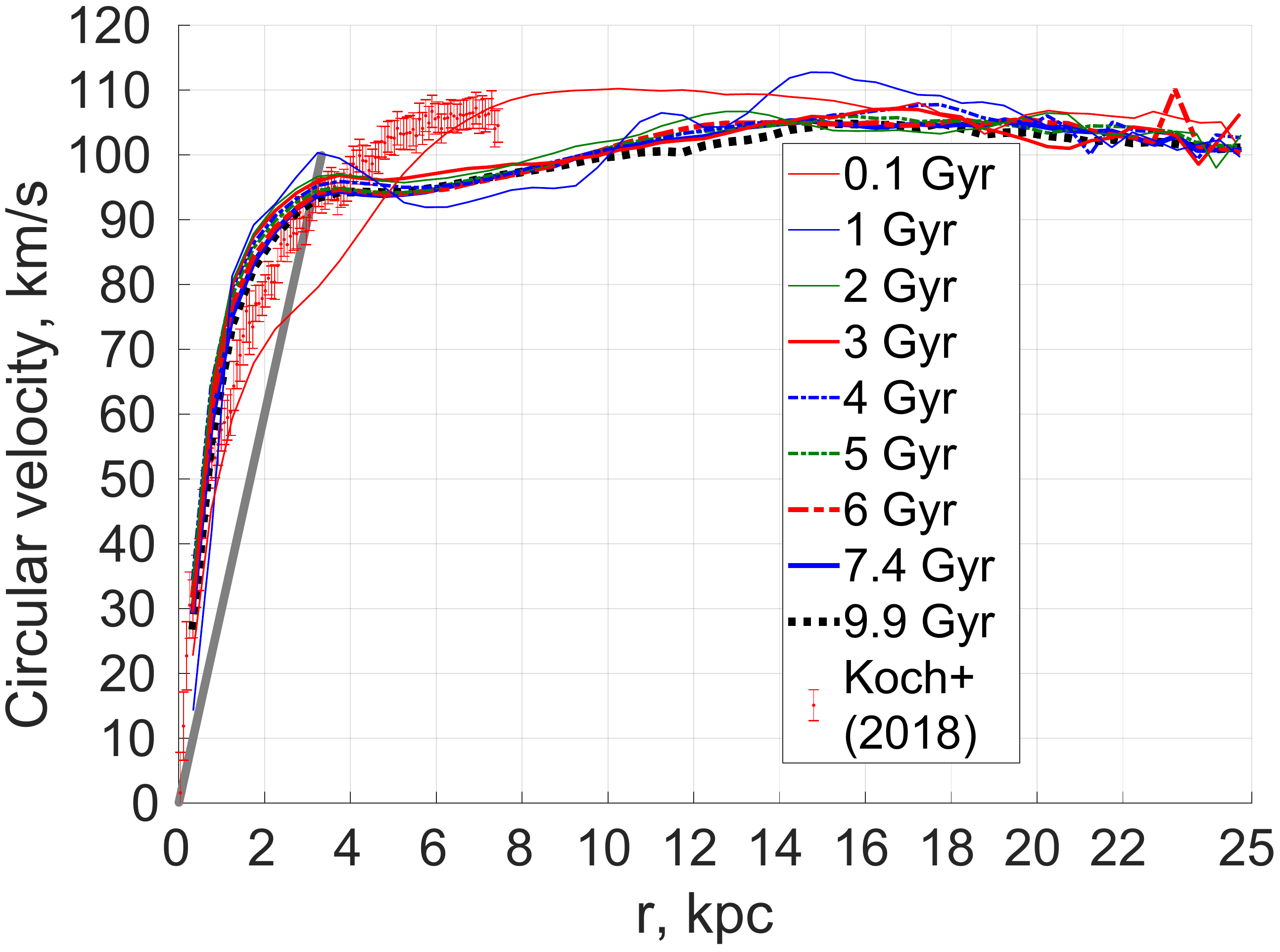}
	\caption{Similar to Figure \ref{M33_rotation_curves_iso}, but now showing the RC of M33 in our model with the intermediately aligned EFE. The final RC (dotted black) is now quite close to the observations.}
	\label{M33_rotation_curve_EF}
\end{figure}

One of the main problems with our isolated hydro simulations is that the RC rises too steeply at $r \la 4$~kpc (Figure \ref{M33_rotation_curves_iso}). This problem is largely resolved once we include the EFE (Figure \ref{M33_rotation_curve_EF}). In this model, the inner RC is still evolving somewhat by the end of the simulation, which further reduces the already small difference with the observed RC. Clearly, even a sub-dominant EFE can have an important effect on the stability and secular evolution of galactic disks in MOND. We discuss this further in Section~\ref{Broader_implications}.

\begin{figure}
	\centering
	\includegraphics[width = 8.5cm] {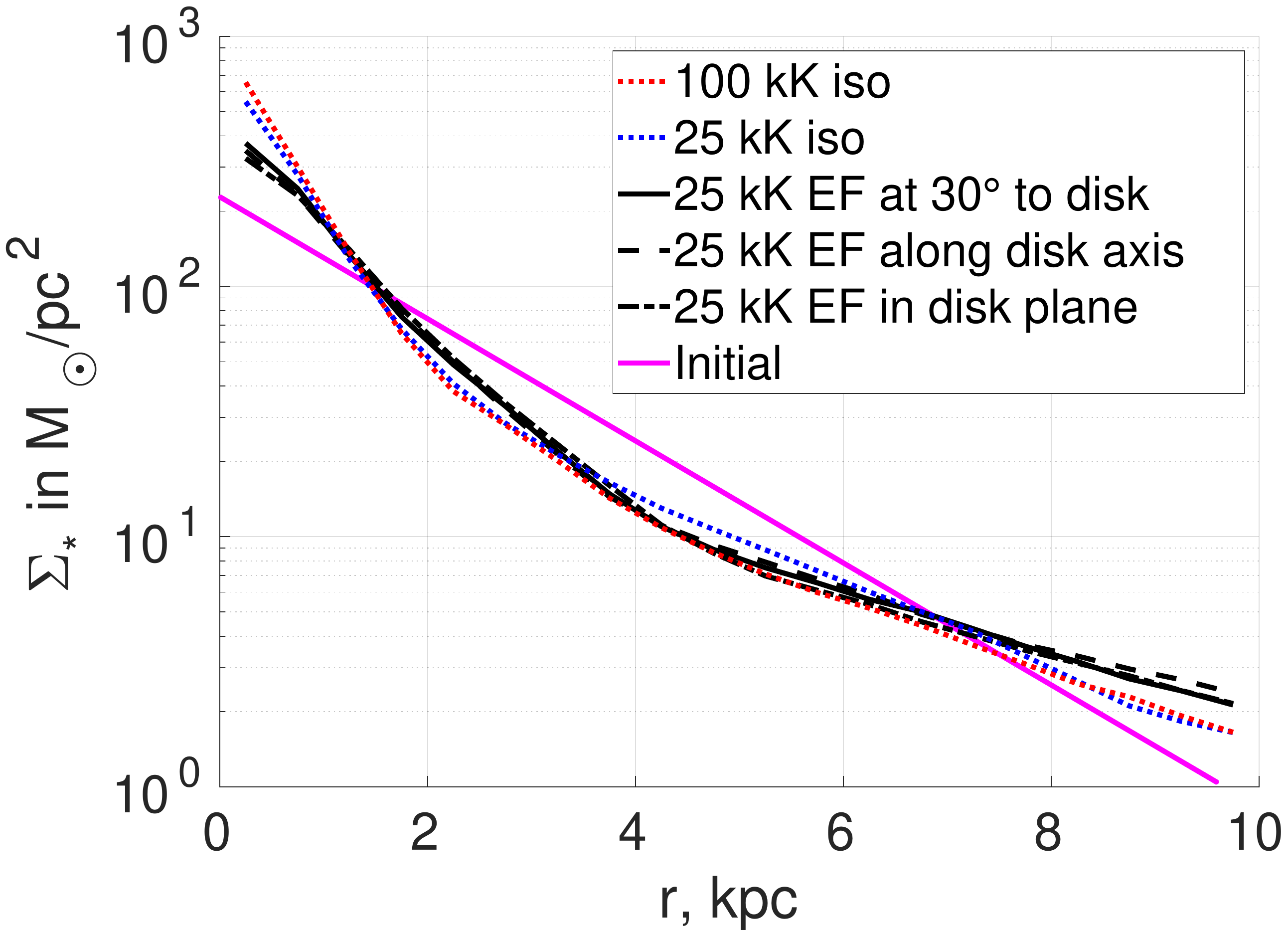}
	\caption{The stellar surface density distribution in our isolated models at 100~kK (red dotted line) and 25~kK (blue dotted line). Results with the EFE included according to Equation \ref{g_ext_value} are shown in solid black, with other black lines indicating special external field orientations (Section \ref{Different_g_ext_directions}). The initial distribution (Figure \ref{M33_surface_density}) is shown with a featureless pink line.}
	\label{Sigma_star_profile}
\end{figure}

Since the EFE is not very significant at $r \la 2$~kpc (Figure \ref{M33_surface_density}), changes in the RC arise mainly from differences in the matter distribution. We therefore investigate how the surface density profile of the stellar component differs between our hydro models with and without the EFE (Figure \ref{Sigma_star_profile}). In the isolated models, the central $\Sigma_*$ is lower when $T$ is reduced from 100~kK to 25~kK. This is expected given the cooler model prevents the formation of a substantial central bulge (Figure \ref{Particles_radial}). However, the difference is not enough to reconcile the RC with observations (Figure \ref{M33_rotation_curves_iso}). Including the EFE further suppresses the radial inflow of stars, causing $\Sigma_*$ to rise less steeply towards lower $r$. This is not much dependent on the choice of direction for the EFE, which we vary in Section \ref{Different_g_ext_directions}.

Our simulation with the `fiducial' $\bm{g}_{ext}$ (Equation \ref{g_ext_value}) still displays a small RC discrepancy at $r \la 4$~kpc. In addition to percent-level changes in the M33 distance, this could in principle be addressed by changing the initial configuration. Our adopted disk scale lengths (Table \ref{Parameters}) are based on M33 today, but its surface density profile was probably different several Gyr ago. To get a shallower RC, the disk should have been more extended. For the same disk mass, this would reduce $\Sigma$ and thus the disk self-gravity, putting it deeper into the MOND regime and making the disk more stable (Section \ref{Disk_galaxy_stability}). A time-varying external field also needs to be considered in further works.

Moreover, the RC itself could have small observational issues in the central region. The empirically determined RC of a galaxy is mainly sensitive to the radial velocities along its major axis. Even with rather weak spiral arms, it is quite possible for the inferred RC to differ from the true one at the 10~km/s level \citep[figures 3 and 4 of][]{Medina_2020}. If the galaxy has a $180^\circ$ rotational symmetry, this would not be recognisable simply by comparing its approaching and receding halves, though doing so for M33 does reveal differences of this order \citep[figure 10 of][]{Kam_2015}. Other issues include a slight $r$-dependent inclination, i.e. a warp. \citet{Koch_2018} obtained $i = 55.08^\circ \pm 1.56^\circ$ assuming fixed $i$ for the whole disk (see their table 2). If $i = 50^\circ$ at the centre \citep[as in the earlier work of][]{Corbelli_2007}, the simulated RC would appear to exceed the observed RC by 7\% even if the simulation matches the actual RC. This would still leave a mild disagreement at $r \approx 5$~kpc, which could be due to a warp. The tilted ring fit shown in figure 2 of \citet{Corbelli_2007} shows that at $r = 5$~kpc, the inclination could be $56^\circ$ even though it is only $50^\circ$ at the centre. We postulate that a warp might partially mask a rapid flattening of the RC at $r \approx 3$~kpc.

At larger radii, figure 16 of \citet{Kam_2017} shows reasonably good agreement with our simulated RC, which is essentially flat at $v_f = 101$~km/s (Equation \ref{BTFR}). This prediction also applies when including the EFE, since $\bm{g}_{ext}$ is sub-dominant at distances $\la 40$~kpc (Equation \ref{r_EFE}). Thus, the RC at $\left(10 - 25 \right)$~kpc is not such a strong test of our model $-$ most of the mass lies closer in. As a result, our isolated models also yield a similar RC in this distance range despite behaving rather differently at low $r$ (Figure \ref{M33_rotation_curves_iso}). In a different galaxy, the EFE could be more significant in the observationally accessible part of the RC \citep[Section \ref{External_field_effect}, see also][]{Chae_2020}.

\subsubsection{Bulge and velocity dispersion}
\label{Bulge_EF}

\begin{figure*}
	\centering
	\includegraphics[width = 17.7cm] {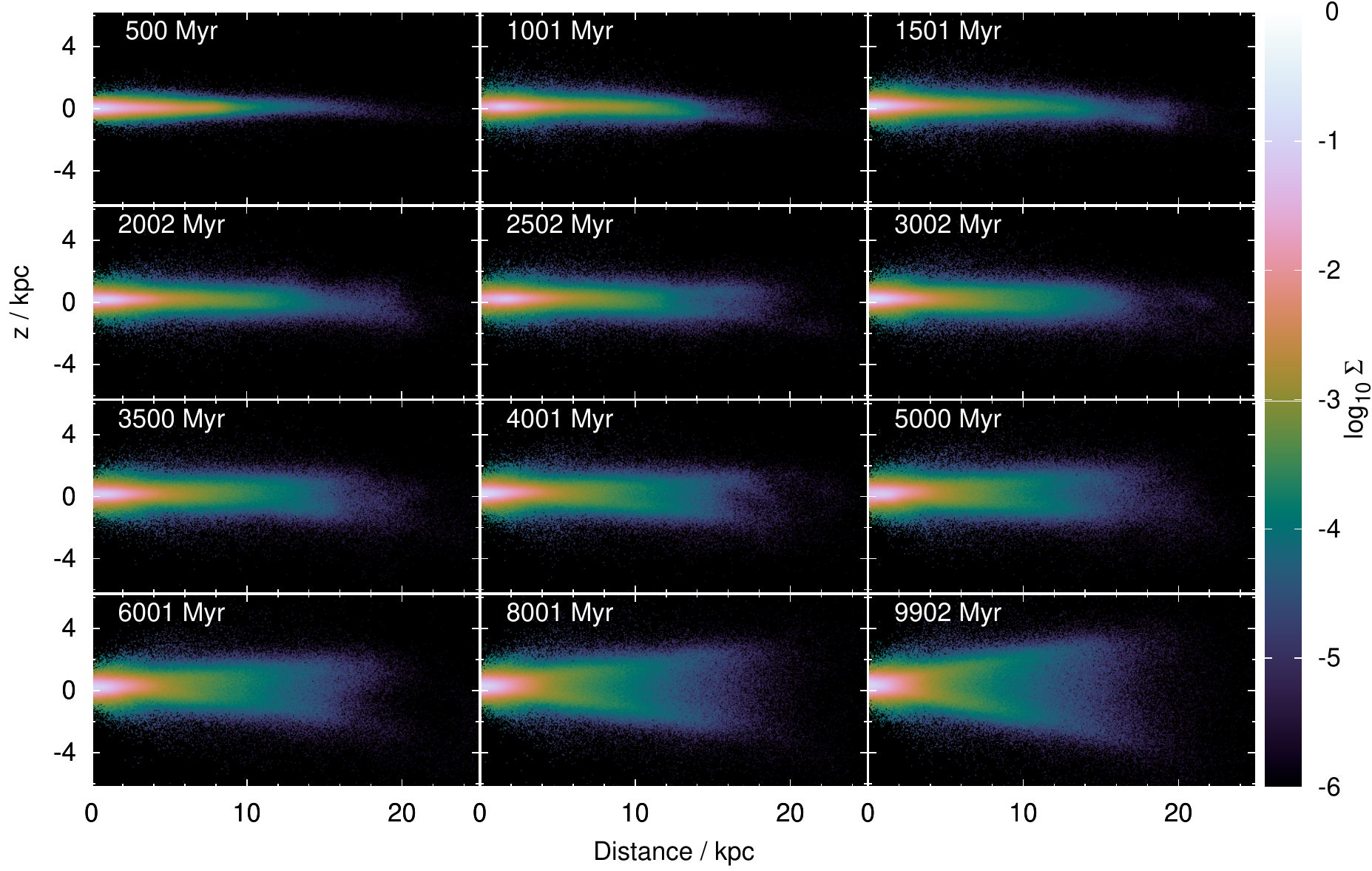}
	\caption{Similar stellar $rz$ view to Figure \ref{Particles_radial}, but for our hydro simulation with the EFE. Notice the absence of a central bulge even after 10~Gyr, similarly to our analogous isolated model (bottom panel of Figure \ref{Particles_radial}). The bifurcation at large $r$ is caused by precession of the whole disk (Section \ref{Warp}).}
	\label{Particles_radial_EF}
\end{figure*}

A bulge is readily apparent after just 3~Gyr in our isolated 100~kK simulation, but the formation of a bulge is suppressed at 25~kK (Figure \ref{Particles_radial}). As might be expected from Figure \ref{Sigma_star_profile}, a bulge also does not arise in our models with the EFE. This is evident in Figure \ref{Particles_radial_EF}, which shows the stellar component in a cylindrical $rz$ projection (Equation \ref{Cylindrical_projection}).

\begin{figure}
	\centering
	\includegraphics[width = 8.5cm] {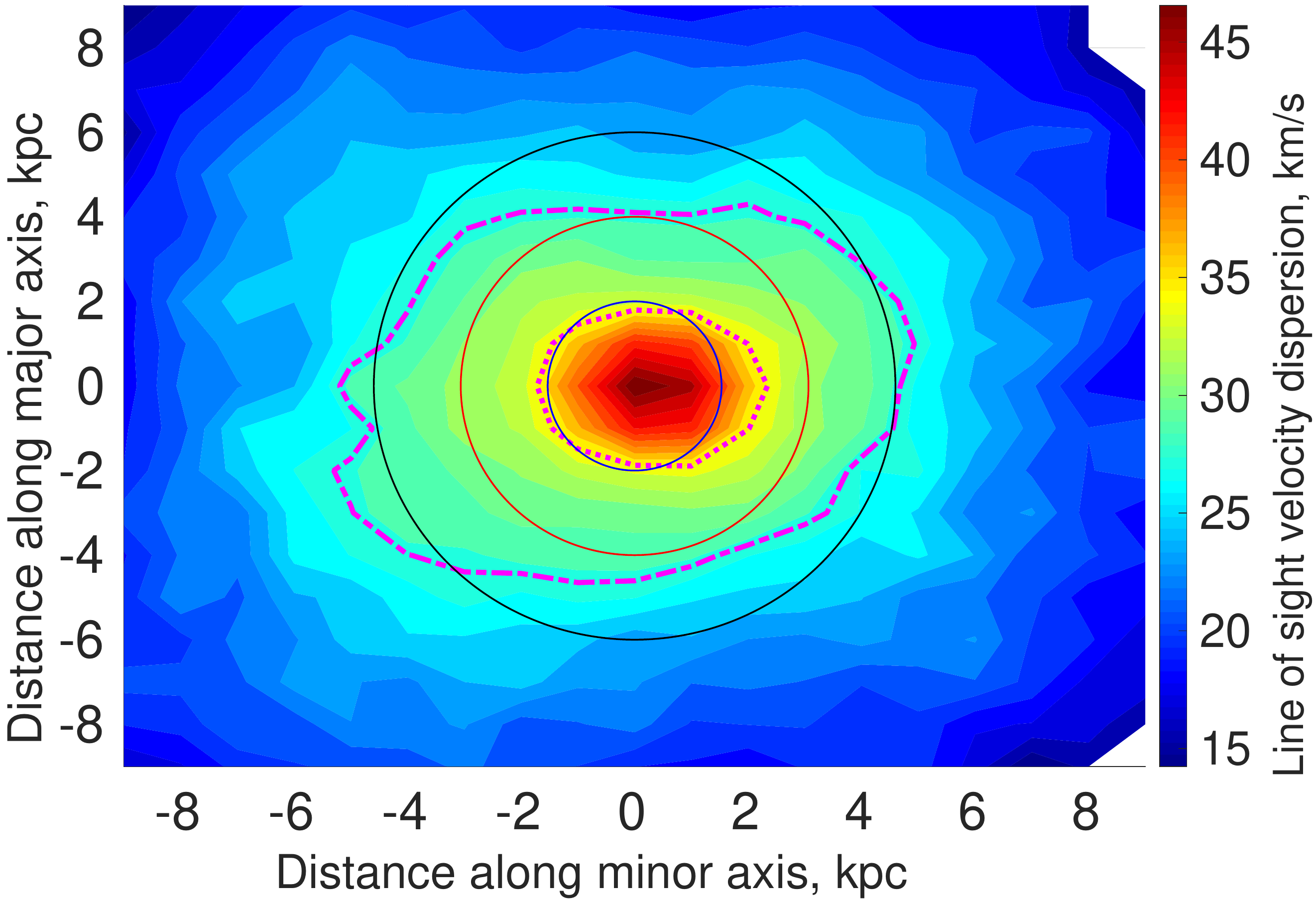}
	\caption{Similar stellar $\sigma_{_{LOS}}$ map to Figure \ref{sigma_LOS}, but for our model with the EFE. The slight asymmetry along $\pm \bm{x}$ is likely due to the EFE, which is partly towards $+\bm{x}$ (Equation \ref{g_ext_value}). Notice that $\sigma_{_{LOS}}$ is lower in this model, with the central $\sigma_{_{LOS}}$ of 48~km/s now much closer to the observational range (see text). The actual LOS velocities in this pixel closely follow a Gaussian of this width (Figure \ref{LOSVD_central_pixel_EF30}).}
	\label{sigma_LOS_EF}
\end{figure}

The stellar $\sigma_z$ is similar between our 25~kK models with and without the EFE (Figure \ref{sigma_z}), but the slightly lower $\sigma_z$ when including the EFE suggests that M33 is more stable in this case. Although $\sigma_z$ is not directly observable, $\sigma_{_{LOS}}$ is. Observationally, $\sigma_{_{LOS}} = \left( 28 - 35 \right)$~km/s \citep[section 3.2 of][]{Corbelli_2007_image}. In our isolated hydro simulation, the central pixel has $\sigma_{_{LOS}} = 57$~km/s, well above this range (Figure \ref{sigma_LOS}). When we include the EFE, this decreases to 48~km/s (Figure \ref{sigma_LOS_EF}). While still rather high, the agreement is much better. The exact choice of snapshot and pixel size would also have some effect, with the observing direction at fixed $i$ contributing an uncertainty of $\approx 1$~km/s.

\begin{figure}
	\centering
	\includegraphics[width = 8.5cm] {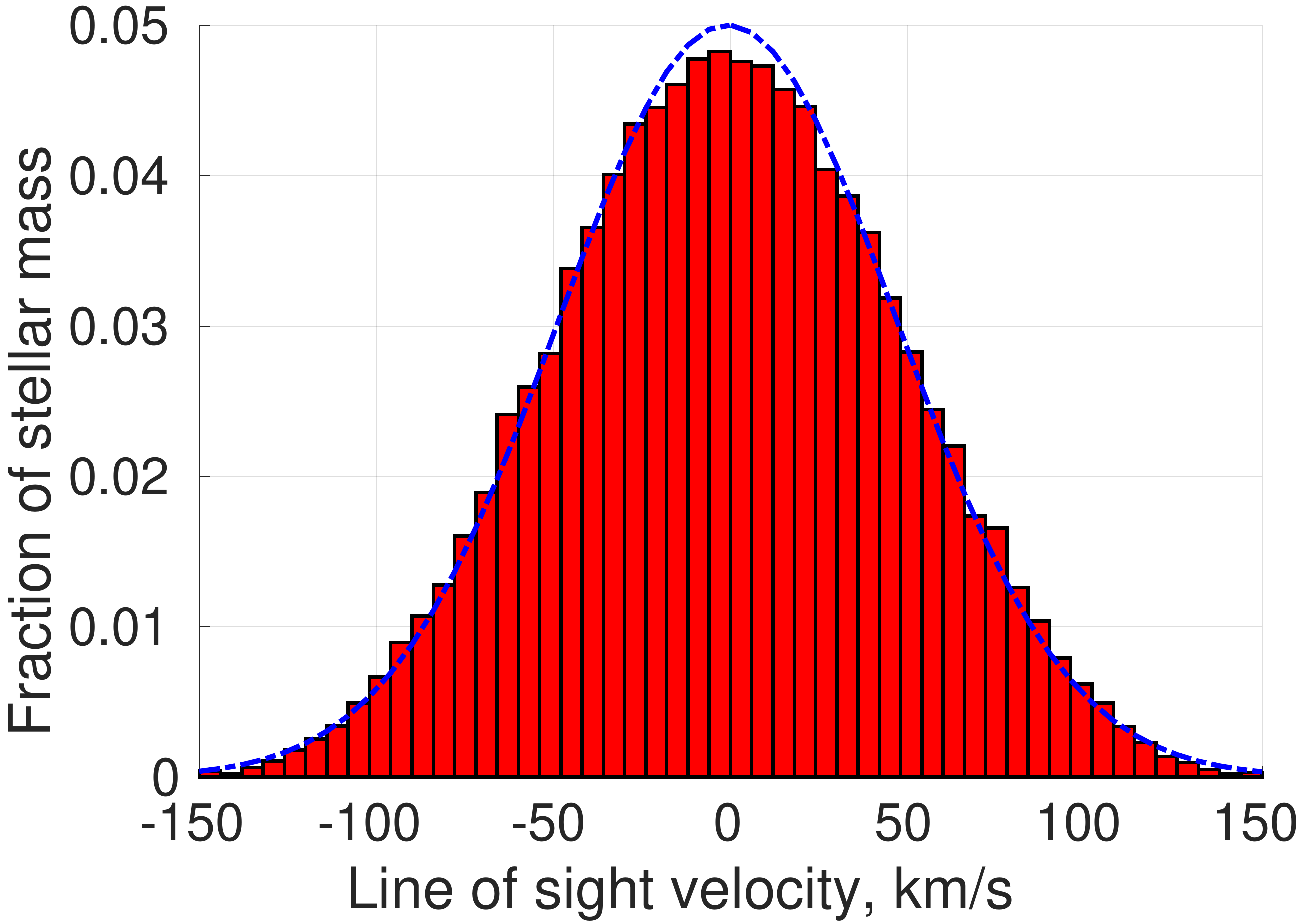}
	\caption{Distribution of the LOS velocity in the central pixel of Figure \ref{sigma_LOS_EF} (red bars). The blue dot-dashed line is a Gaussian of width $\sigma_{_{LOS}} = 47.8$~km/s, demonstrating the adequacy of a standard mass-weighted variance (Equation \ref{sigma_LOS_mass_weighted}). To ensure the robustness of our results, we use $5\sigma$ outlier rejection, but this has almost no effect on $\sigma_{_{LOS}}$.}
	\label{LOSVD_central_pixel_EF30}
\end{figure}

In addition to $\sigma_{_{LOS}}$, observations can also tell us the distribution of $v_{_{LOS}}$, which we show in Figure \ref{LOSVD_central_pixel_EF30} for the central pixel. A Gaussian of width $\sigma_{_{LOS}}$ provides a very good description of the results. This also demonstrates that a mass-weighted variance calculated using standard techniques is sufficient for the central pixel.

The discrepancy in $\sigma_{_{LOS}}$ could again probably be addressed by a small adjustment of the initial conditions or a time-varying EFE, though this still needs to be demonstrated. In addition, including stellar feedback processes would help to limit the central accumulation of material, which is probably also responsible for the mild tension with the RC (Figure \ref{M33_rotation_curve_EF}). As in the RC case, the comparison with observations could be problematic. One important reason is that our calculated $\sigma_{_{LOS}}$ is a mass-weighted velocity dispersion (Equation \ref{sigma_LOS_mass_weighted}), whereas the observational estimate is based on the broadening of spectral lines. Since stars do not have an exactly linear mass-luminosity relation, this can create a slight mismatch between mass- and luminosity-weighted velocity dispersions \citep{Aniyan_2016}. This issue was also discussed in section 4.3 of \citetalias{Sellwood_2019}.\footnote{They concluded that the uncertainties were probably not enough to accommodate $\sigma_{_{LOS}} = 43$~km/s.} Since the luminosity of a star generally rises faster than its mass, the bias is likely to be in the right direction, especially in a galaxy that is still forming stars \citep{Verley_2009}. This is different to our simulations where star formation is disabled, so they are not directly comparable to observations even if these give a true mass-weighted $\sigma_{_{LOS}}$ for the stars.


Another interesting feature of our $\sigma_{_{LOS}}$ map is the slight left-right asymmetry in our model with the EFE (Figure \ref{sigma_LOS_EF}), something not evident in our analogous isolated model (Figure \ref{sigma_LOS}). This is because the underlying gravitational physics is symmetric with respect to $\pm \bm{x}$ without any EFE, but this symmetry is broken by an external field with non-zero component towards $+\bm{x}$ (Equation \ref{g_ext_value}). The potential of a point mass is symmetric with respect to $\bm{g}_{ext}$ when $\bm{g}_{ext}$ dominates (Equation \ref{Boundary_Phi_EFE}), but there is an asymmetry between $\pm \bm{g}_{ext}$ if the external field is weak. This effect can induce asymmetry in tidal streams of satellite galaxies \citep{Thomas_2018}, and in galaxies orbiting within a galaxy cluster \citep{Wu_2010, Wu_2017}. At large distances, $\bm{g}_{ext}$ becomes dominant and this asymmetry vanishes. The asymmetry is discussed further in Section \ref{Different_g_ext_directions}.

\subsubsection{Face-on view and the bar}
\label{Bar_EF}

\begin{figure*}
	\centering
	\includegraphics[width = 17.7cm] {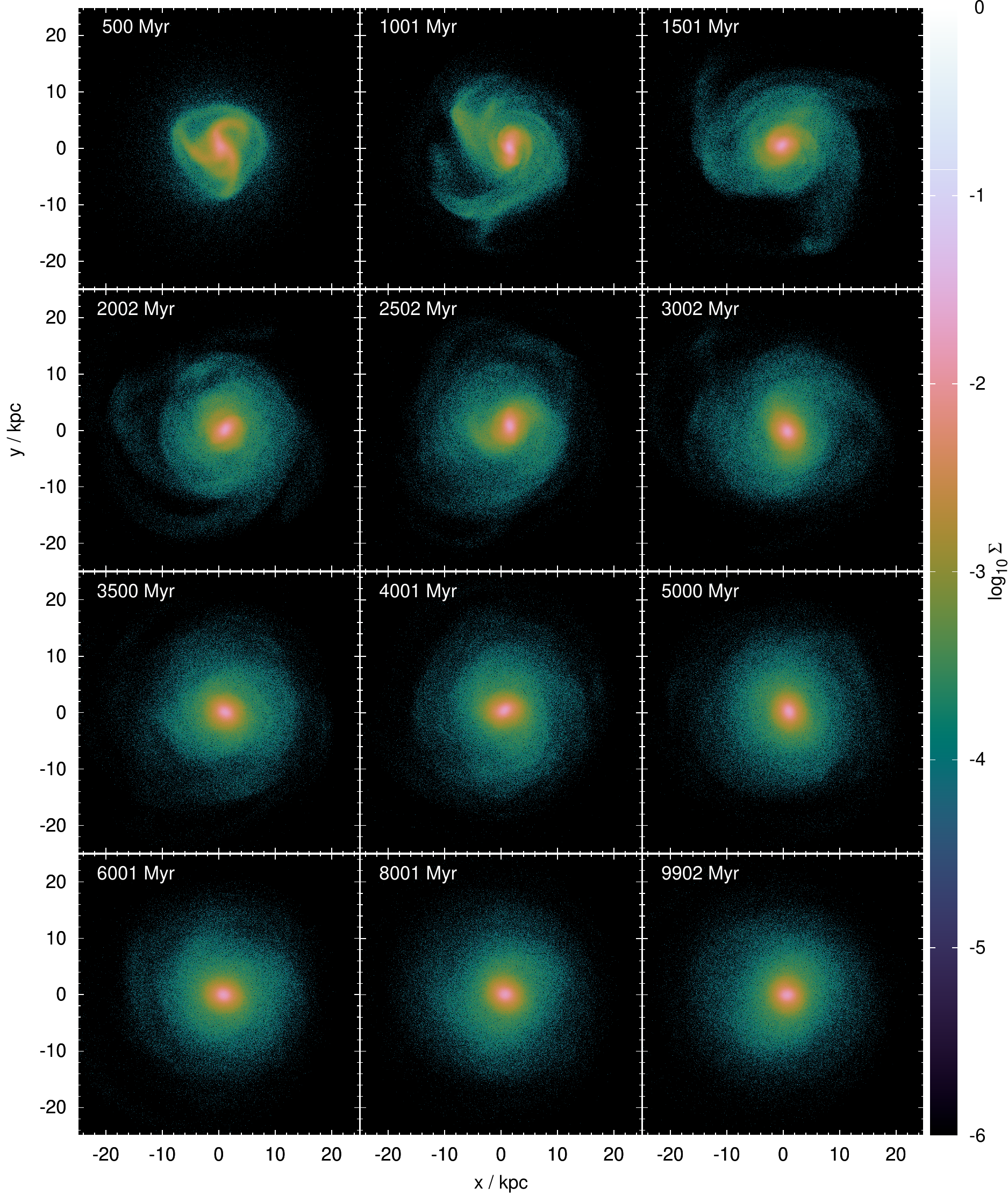}
	\caption{Similar stellar $xy$ view to Figure \ref{Particles_face_on}, but for our hydro model with the EFE. In both cases, the outer regions look nearly circular after 10~Gyr, indicating a genuinely weak bar and rather little non-circular motion (Section \ref{Non_circular_motions}).}
	\label{Particles_face_on_EF}
\end{figure*}

The face-on appearance of M33 remains rather circular once the EFE is included (Figure \ref{Particles_face_on_EF}). A weak bar is evident, whose length is related to $\Omega_p$. This is not much affected by the EFE (Figure \ref{M33_pattern_speed}). Regardless of whether we use the simulated or observed RC, $\Omega_p = 30$~km/s/kpc corresponds to a corotation radius of $\approx 3$~kpc (Figure \ref{M33_rotation_curve_EF}). As argued in Section \ref{Bar_pattern_speed}, this is quite reasonable.

\begin{figure}
	\centering
	\includegraphics[width = 8.5cm] {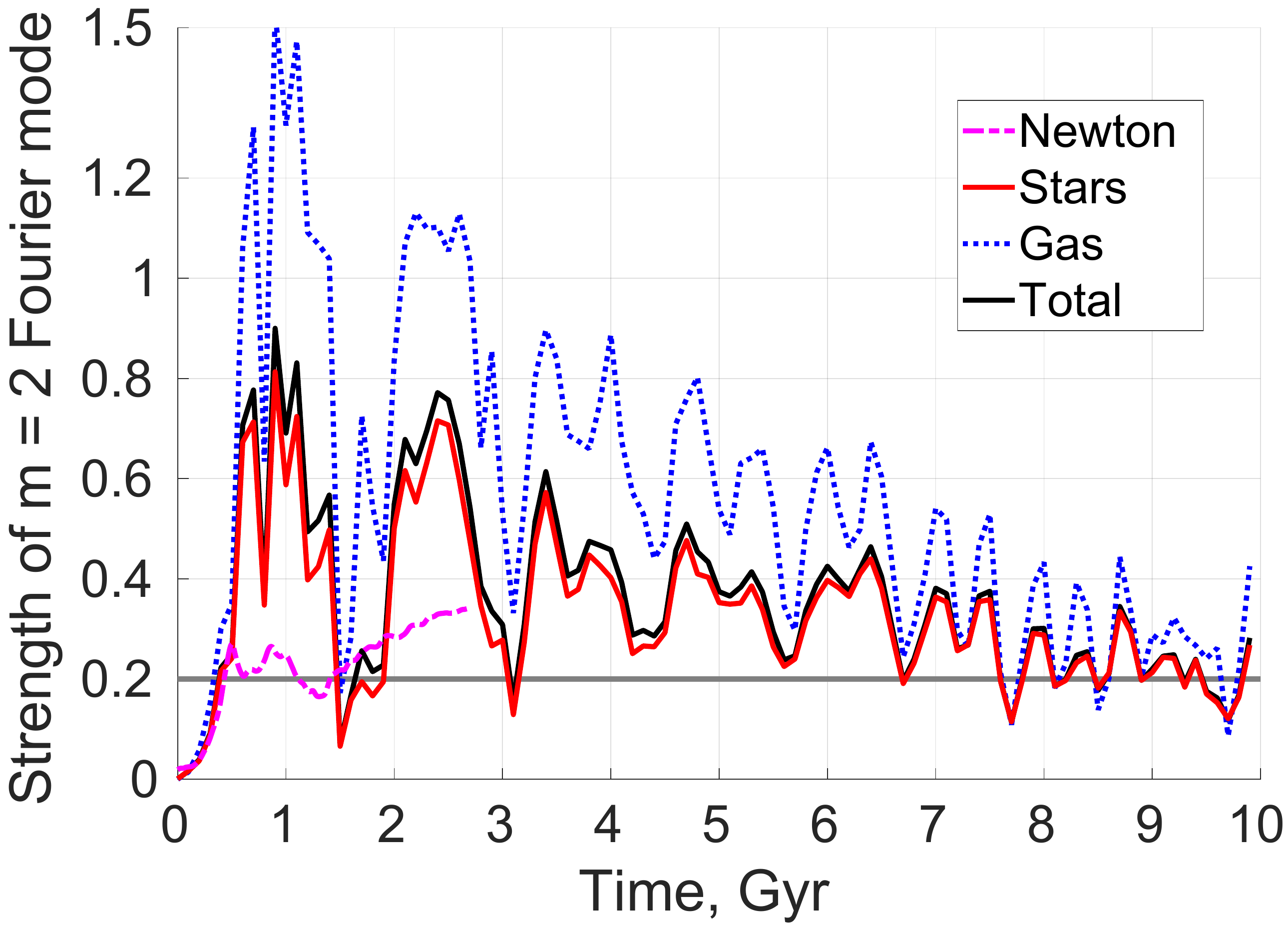}
	\caption{Similar to Figure \ref{Mode_2_strength}, but for our model with the EFE. The larger oscillations mean the bar is typically stronger than the observed 0.2 (grey line), so the weak bar of M33 might be atypical.}
	\label{Mode_2_strength_EF}
\end{figure}

In both our isolated and non-isolated models, the bar strength is similar to the observed value of 0.2 by the end of the simulation. The main difference is that the typical strength is larger when the EFE is included (Figure \ref{Mode_2_strength_EF}). This is more in line with the observed population of spiral galaxies, which often have strong bars \citep{Laurikainen_2002, Garcia_Gomez_2017}. However, a meaningful comparison would need to restrict the observational sample to non-interacting galaxies with a similar surface density to M33, since the stability properties of Milgromian disks depend on their central surface density relative to a particular threshold (Equation \ref{Sigma_MOND}).

\subsubsection{Non-circular motion}
\label{Non_circular_motions}

\begin{figure}
	\centering
	\includegraphics[width = 8.5cm] {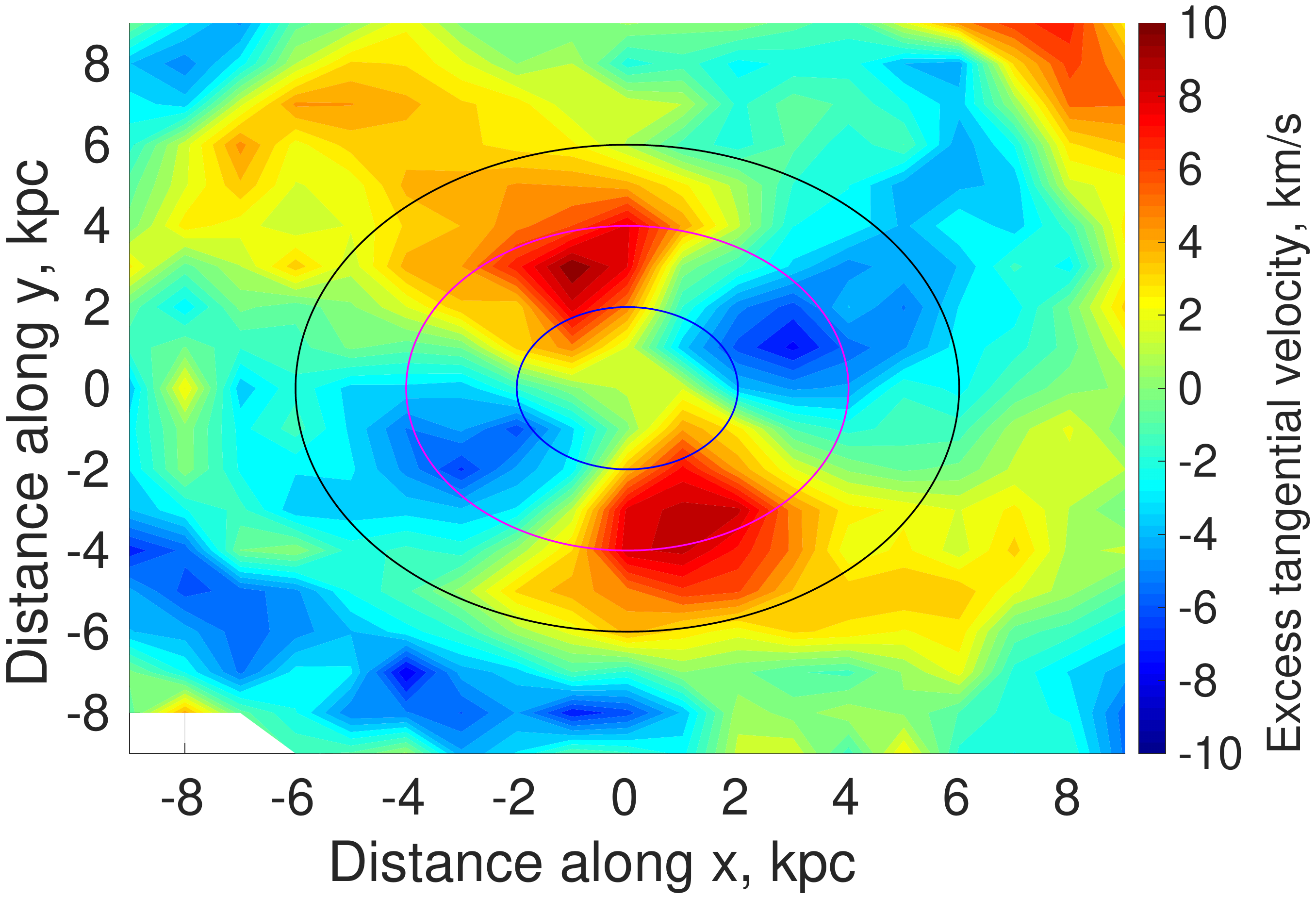}
	\includegraphics[width = 8.5cm] {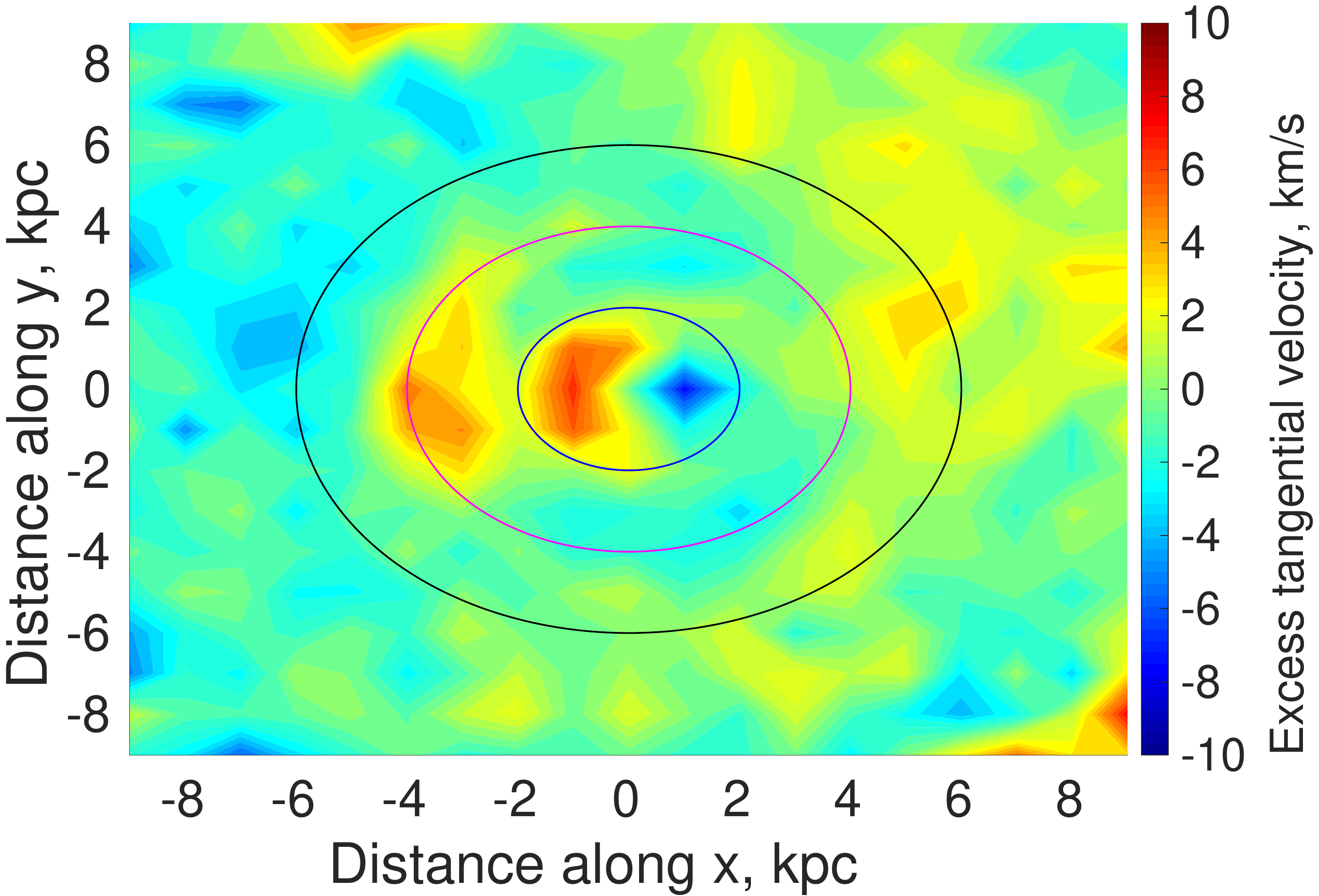}
	\caption{Face-on view of M33 showing the mean tangential velocity within its disk plane less the average for particles at similar $r$ (Equation \ref{Delta_v_t}) at the end of our 25~kK hydro simulation in isolation (top) and with the EFE set by Equation \ref{g_ext_value} (bottom). The curves show circles of radius 2, 4, and 6~kpc. Our binning and outlier rejection procedures are described in Section \ref{Vertical_behaviour}. The disks have been rotated into the plane defined by their overall angular momentum, which is important for our model with the EFE (Section \ref{Warp}). Both panels use the same colour scale, highlighting the much smaller non-circular motions in our model with the EFE. If we use snapshots at different times (not shown), the pattern in the isolated model appears to rotate, while results with the EFE change very little and do not circulate.}
	\label{vt}
\end{figure}

Our isolated model of M33 appears mildly non-circular when viewed face-on, especially if we consider just the stars (Figure \ref{Particles_face_on}). Non-circular motions are necessary to sustain a non-circular shape. We quantify this using the in-plane tangential velocity
\begin{eqnarray}
	v_t ~\equiv~ \frac{x v_y - y v_x}{r} \, .
\end{eqnarray}
Most of this is just circular motion, so we first determine the mean $v_t$ for particles in different radial bins. This yields a list of mass-weighted mean radii $\overline{r}$ and tangential velocity $\overline{v}_t$. We interpolate this to obtain the expected $\overline{v}_t$ at any ${r \la 25}$ kpc. We then subtract $\overline{v}_t$ to obtain the excess tangential velocity:
\begin{eqnarray}
	\Delta v_t ~\equiv~ v_t - \overline{v}_t \, .
	\label{Delta_v_t}
\end{eqnarray}
We find $\Delta v_t$ for different parts of M33 using the same binning procedure as described in Section \ref{Vertical_behaviour}. To better illustrate the dynamics of M33, we now use a face-on view. The resulting map of $\Delta v_t$ is shown in Figure \ref{vt}. We previously estimated that the corotation radius is 3~kpc in our 25~kK models regardless of the EFE (Section \ref{Bar_EF}). Thus, non-circular motions should mostly be restricted to the central 3~kpc, which is approximately the case. The non-circular motions as quantified by $\Delta v_t$ are much smaller when the EFE is included (bottom panel). The EFE is therefore able to suppress the bar instability somewhat. Another difference is that the isolated model is close to symmetric with respect to $\phi \to \phi + \mathrm{\pi}$, indicating a dominant ${m = 2}$ mode. The EFE breaks the symmetry between $\pm \bm{x}$ but preserves that between $\pm \bm{y}$, at least for $r \la 6$~kpc. This is no doubt a consequence of the particular direction we adopted for $\bm{g}_{ext}$ (Equation \ref{g_ext_value}). Indeed, the pattern of $\Delta v_t$ appears to rotate continuously with time in the isolated model, but undergoes rather little variation once the EFE is included.




\subsubsection{Disk precession and warping}
\label{Warp}

\begin{figure}
	\centering
	\includegraphics[width = 8.5cm] {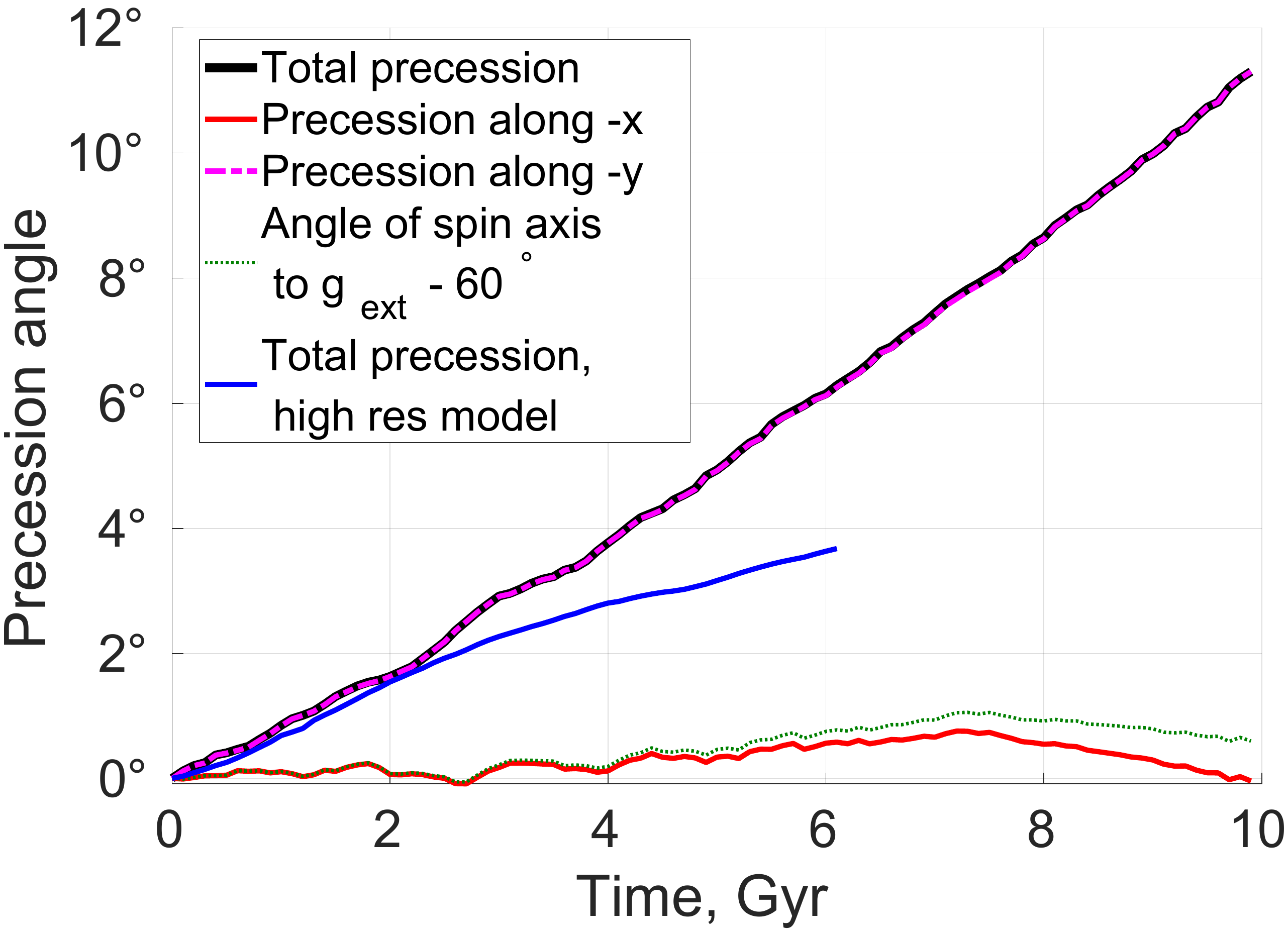}
	\caption{Time evolution of the disk spin axis $\widehat{\bm{h}}$ in our model with the fiducial EFE (changes are negligible in other models, not shown). The solid red and dot-dashed magenta curves show the individual components of $\widehat{\bm{h}}$. The solid black line shows the total change in $\widehat{\bm{h}}$ from its initial orientation (along $+\bm{z}$), while the dotted green line shows how much of this is a nutation rather than precession around $\bm{g}_{ext}$. The behaviour of $\widehat{\bm{h}}$ is very similar in our higher resolution simulation with the EFE (Section \ref{Numerical_convergence}), so we show only the total precession angle in this case (solid blue). The precession is almost entirely towards $-\bm{y}$ in both cases (solid black and dot-dashed magenta curves coincide). The direction and approximate magnitude of the precession are explained in the text using analytic arguments.}
	\label{Disk_precession}
\end{figure}

The EFE breaks the axisymmetry in our isolated models of M33. One consequence is that the whole disk precesses by a small amount over the course of the simulation. We quantify this based on the total angular momentum of all material at $\left| z \right| < 20$~kpc. The direction of this vector is the disk spin axis, which we denote $\widehat{\bm{h}}$. The time evolution of $\widehat{\bm{h}}$ is shown in Figure \ref{Disk_precession}, which reveals a very slow precession by just over $1^\circ$/Gyr.

Since the external field picks out only one preferred direction (Equation \ref{g_ext_value}), we expect $\widehat{\bm{h}} \cdot \bm{g}_{ext}$ to remain constant. Indeed, the angle between these vectors (which is initially $60^\circ$) changes by $<1^\circ$ over 10~Gyr, much less than the change in $\widehat{\bm{h}}$. Thus, we can consider $\widehat{\bm{h}}$ as precessing around $\bm{g}_{ext}$. Since $\bm{g}_{ext}$ is within the $\bm{x}\bm{z}$ plane (Equation \ref{g_ext_value}), this implies $\widehat{\bm{h}}$ mostly precesses within the $\bm{y}\bm{z}$ plane. In our simulation, the precession is towards $-\bm{y}$. Our isolated model of M33 exhibits only a very small amount of precession ($<0.01^\circ$), which we attribute to numerical noise. Thus, the more significant precession evident in Figure \ref{Disk_precession} is caused by the EFE, and remains if the simulation is run at higher resolution (Section \ref{Numerical_convergence}).

We can understand this by considering Equation \ref{Boundary_Phi_EFE} and approximating the disk as a point mass. Since $K_0 < 0$, the potential is deepest along the $\bm{g}_{ext}$ axis. As a result, stars at $x > 0$ would feel an extra force towards $+\bm{z}$, while stars at $x < 0$ would feel a force towards $-\bm{z}$. Stars at $\pm \bm{y}$ in general also feel an extra force out of the disk plane (Section \ref{Different_g_ext_directions}), but this would be in the same direction due to the EFE preserving reflection symmetry with respect to the $\bm{x}\bm{z}$ plane containing $\bm{g}_{ext}$. Thus, the net torque would come only from the asymmetry between $\pm \bm{x}$ (there is none between $\pm \bm{y}$). Our preceding argument suggests that the torque would cause $\widehat{\bm{h}}$ to precess towards $-\bm{y}$, which is indeed what happens in our simulation. This is a manifestation of the `external field torque', which arises because the force between two masses is in general misaligned with their separation once the EFE is considered (e.g. Equation \ref{Boundary_Phi_EFE} is not spherically symmetric).

We can obtain a rough estimate of the precession rate using appendix A of \citet{Banik_Ryan_2018}, which provides a general equation for the tangential force on a test particle due to a point mass $-$ this is non-zero only if we consider the EFE. Their result is based on direct numerical integration of Equation \ref{QUMOND_governing_equation_EFE} for a point mass in an external field, with all accelerations in the DML.\footnote{\citet{Banik_2019_spacecraft} provides a visualization of the point mass gravitational field when $g_{ext} = 1.78 \, a_{_0}$, the value appropriate to the Solar neighbourhood of the Milky Way \citep{McMillan_2017}.} The results are reliably known only for the case of a point mass $M$ orbited by a test particle, but this will be sufficient for our purposes. In this case, the angular precession rate for a ring of material at some radius $r$ is approximately:
\begin{eqnarray}
	\dot{\Omega} ~=~ \frac{f_{geom} r \sqrt{GMa_{_0}} \sin \theta \cos \theta}{2 v_c \, {R_{_{EFE}}}^2} \, ,
	\label{Precession_rate_estimate}
\end{eqnarray}
where $M = 6.5 \times 10^9 \, M_\odot$ is the mass of M33, its $R_{_{EFE}} = 41$~kpc (Equation \ref{r_EFE}), $\theta$ is the minimum angle between $\bm{g}_{ext}$ and a vector within the disk plane, and $f_{geom} = 1/2$ is a geometric factor that accounts for azimuthal averaging, i.e. the fact that the torque on a particle at $\left(x = r, \, y = z = 0 \right)$ is not representative over all $\phi$. We consider only precession towards $-\bm{y}$ because other components cancel out by symmetry. Equation \ref{Precession_rate_estimate} suggests a precession rate of $1.66^\circ$/Gyr for a ring at $r = 4$~kpc with $v_c = 100$~km/s. This is a rather rough estimate because we have neglected disk self-gravity and the choice of $r$ is arbitrary (though it should be similar to the disk scale length). Nonetheless, the precession rate of $1.1^\circ$/Gyr in our \textsc{por} simulation is not too far off, and is in the predicted direction (Figure \ref{Disk_precession}).

The precession of the M33 disk would have a small impact on our previous results, since ideally we should rotate M33 into a reference frame aligned with $\widehat{\bm{h}}$ before doing further analyses. Fortunately, the effect is rather small because $\cos 10^\circ = 0.985$, so even after 10~Gyr, derived quantities like the RC would be only slightly affected by the precession. An interesting situation where disk precession does matter is the bifurcation apparent at large $r$ in our cylindrical $rz$ projection (Figure \ref{Particles_radial_EF}). This is because the mean height of the disk is $\overline{z} \propto \sin \phi$, where $\phi$ is the azimuthal angle measured from the line of nodes with respect to the initial disk plane (in this case, the $\bm{x}$-axis). The intensity of the $rz$ projection is highest when $\overline{z}$ is stationary with respect to $\phi$, which occurs when $\phi = \pm \mathrm{\pi}/2$ and $\left| z \right|$ is maximal. As a result, sharp ridges appear in the $rz$ projection, with both ridges having the same angle to the $r$-axis (Figure \ref{Particles_radial_EF}). However, as the main purpose of this work is to consider the stability of M33's central regions, this does not affect our results very much.

Since different parts of the disk are subject to different amounts of internal gravity, we expect that not all parts precess at exactly the same rate, which could cause the disk to warp. This possibility was explored previously by \citet{Brada_2000} using integration of test particle orbits in a potential generated by an exponential disk.\footnote{They put $\bm{g}_{ext}$ at $45^\circ$ to the disk rather than our adopted $30^\circ$.} Their work neglected self-gravity of the disk, so was unable to explore how its stability might be influenced by the EFE. Since then, no other works have used self-consistent simulations to explore how a MONDian thin disk galaxy would be influenced by a weak external field. The effect on elliptical galaxies was explored in \citet{Wu_2017}, while \citet{Candlish_2018} investigated the impact of using $g_{ext} \approx a_{_0}$, thereby addressing how disk galaxies might evolve in a cluster environment. Such a strong EFE is seriously damaging to the disk, as might be expected $-$ MOND effects are greatly suppressed in these circumstances, essentially reducing the problem to a purely Newtonian disk, which is very unstable \citep{Hohl_1971}. However, our results suggest that a weak external field can actually make the disk \emph{more} stable, reducing the concentration of material at low $r$ (Figure \ref{Sigma_star_profile}).

\begin{figure}
	\centering
	\includegraphics[width = 8.5cm] {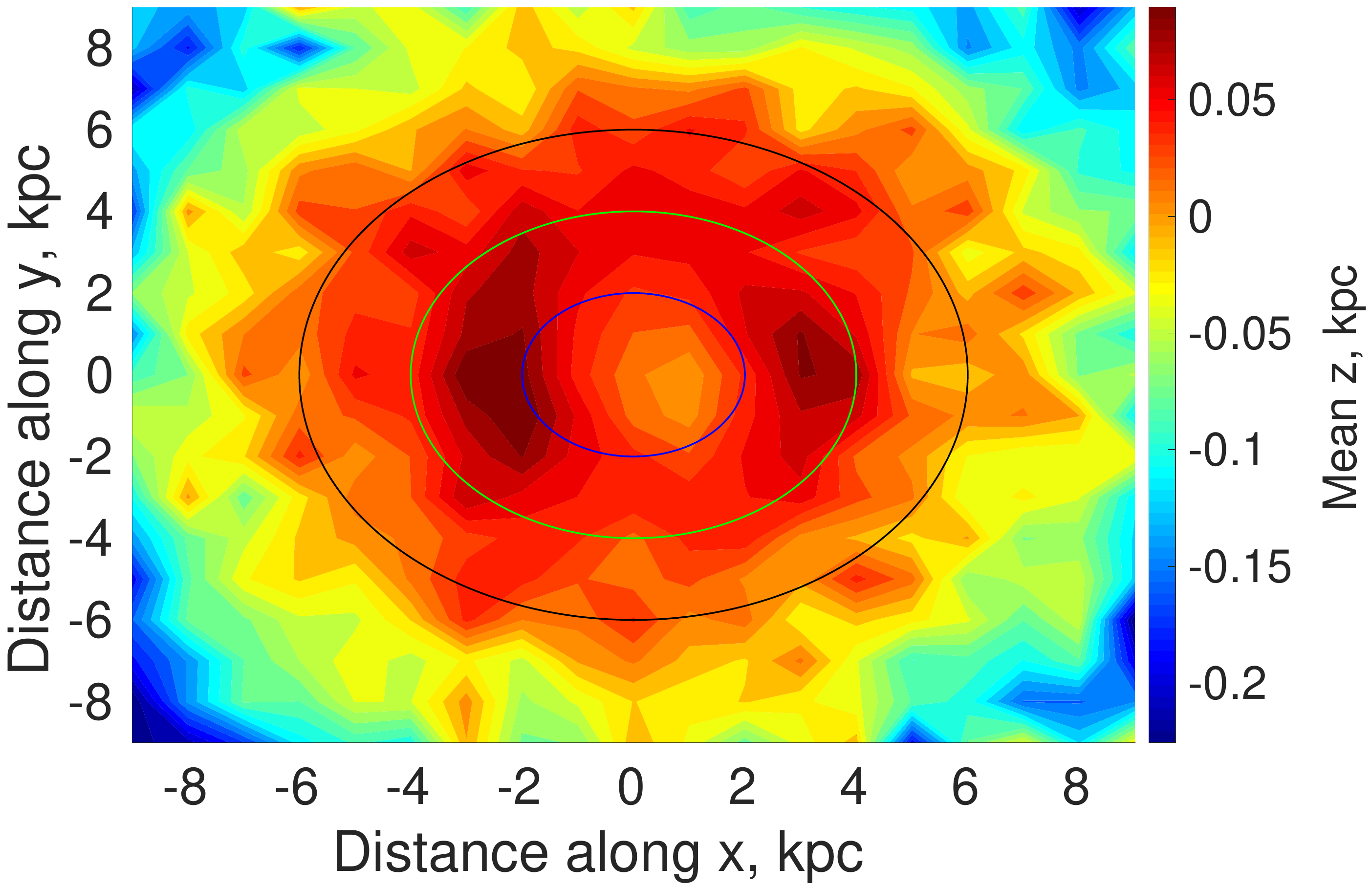}
	\caption{The mean height $\overline{z}$ of stars in the M33 disk after 9.9~Gyr, found after applying a rotation as shown in Figure \ref{Disk_precession}. The circles have radii of 2, 4, and 6~kpc. The external field is in the initial $\bm{x}\bm{z}$ plane (Equation \ref{g_ext_value}).}
	\label{Mean_z}
\end{figure}

To explore how the EFE might achieve this, we use the same binning and outlier rejection procedure described in Section \ref{Vertical_behaviour} to show $\overline{z}$, the mass-weighted mean $z$ of the stellar particles in each pixel. The effects of disk precession are first removed by an appropriate rotation to align the co-ordinate system with $\widehat{\bm{h}}$, allowing us to see what M33 would look like face-on if we had reliable depth perception. Its disk is indeed mildly warped (Figure \ref{Mean_z}). Since the only two vectors in the problem are $\widehat{\bm{h}}$ and $\bm{g}_{ext}$, we expect the $\bm{x}\bm{z}$ plane to be a plane of symmetry (Equation \ref{g_ext_value}). This is why the results are nearly symmetric with respect to $\pm \bm{y}$. However, they are not symmetric with respect to $\pm \bm{z}$ because this symmetry is broken by the $z$-component of $\bm{g}_{ext}$. The resulting asymmetry in the potential is discussed further in Section \ref{Different_g_ext_directions}.

The mild warping evident in Figure \ref{Mean_z} is not sufficient to explain the $\approx 5^\circ$ warp inferred by \citet{Corbelli_2007} $-$ a change in $\overline{z}$ of 0.1~kpc between $r = 0$ and 3~kpc implies a slope of only $2^\circ$. The more significant warp identified by \citet{Corbelli_2007} could be due to recent gas accretion or because of a stronger historical EFE on M33 arising from a smaller separation with M31. Tidal effects might also have been important $-$ tides naturally pull down on one side of M33 while pulling the other side up, which is difficult to do with a constant EFE. However, if tides were able to significantly affect the M33 disk at $r = 3$~kpc, then they would have much more significant effects further out, contradicting the regular appearance of the M33 disk. Thus, gas accretion or a stronger EFE in the past appear to be more promising explanations for its warping.\footnote{It is of course not possible to disentangle tides from the EFE when following the orbit of a single satellite, but this can be done using numerical experiments, or through comparison with other systems.}

This is related to the M31-M33 orbit, which could be constrained with better proper motion data $-$ especially for M31. In addition, it would also be important to consider the Milky Way, which in a MOND context must have previously interacted with M31 \citep{Zhao_2013}. Their satellite galaxy planes \citep{Ibata_2013, Kroupa_2013, Pawlowski_2020, Sohn_2020} might have condensed out of tidal debris expelled during this interaction \citep{Banik_Ryan_2018, Bilek_2018}.

\subsubsection{Numerical convergence}
\label{Numerical_convergence}

Our simulation with the EFE uses a lower resolution compared to our isolated simulations, which may affect our results (Section \ref{Reduced_resolution_EFE}). To check if they are numerically converged, we run a higher resolution simulation with a box size of only 512~kpc and allow up to $levelmax = 13$ levels of refinement. The box size and resolution settings are thus the same as in our isolated simulations (Section \ref{Simulation_setup}). We only advance our high-resolution hydro simulation with the EFE for 6.1~Gyr due to significant numerical drift of M33, which is much less pronounced in our lower resolution model with a larger box size (Section \ref{Reduced_resolution_EFE}).

\begin{figure}
	\centering
	\includegraphics[width = 8.5cm] {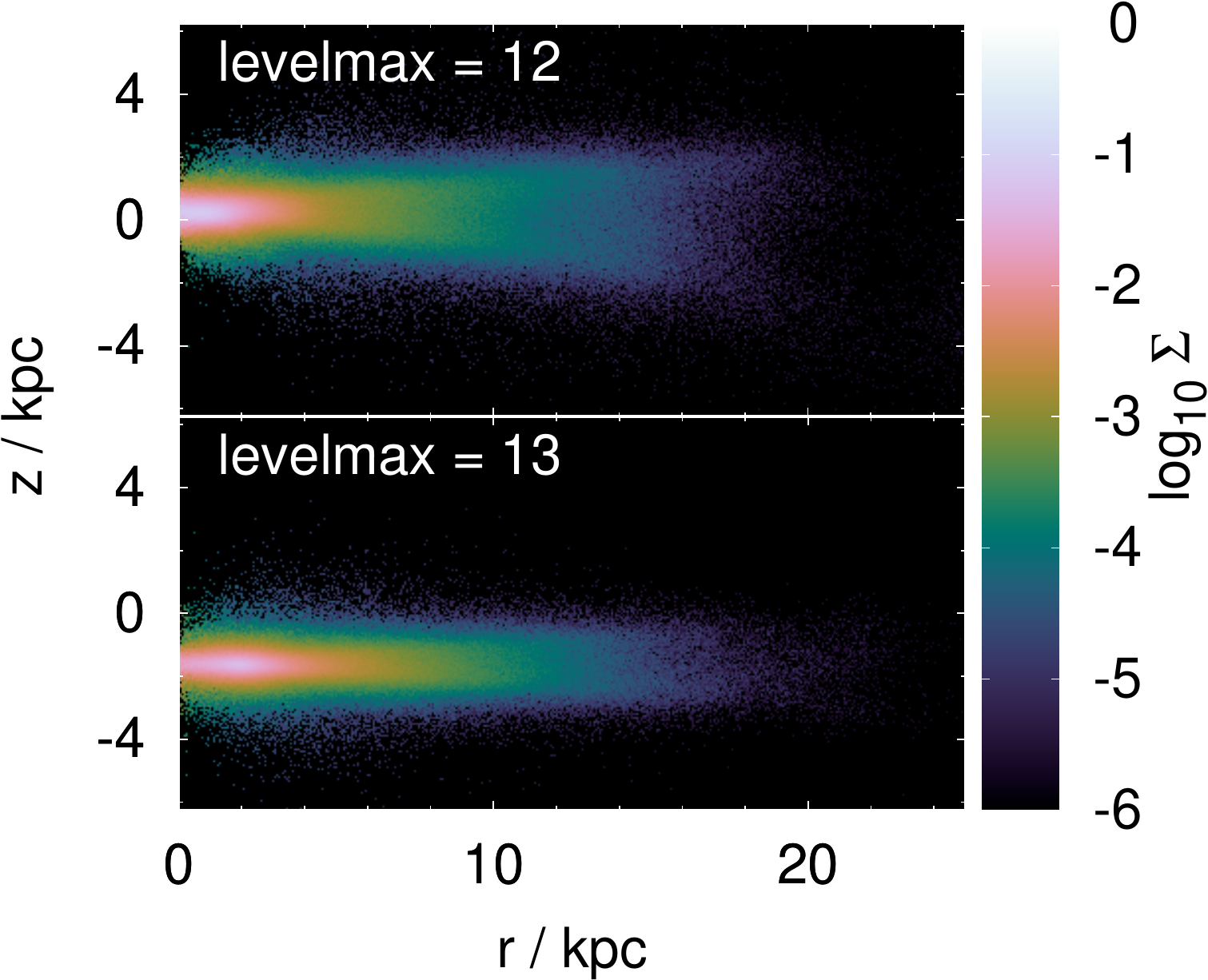}
	\caption{Similar to Figure \ref{Particles_face_on}, but now showing the stellar $rz$ projection at 6.1~Gyr for our lower resolution (\emph{top}) and higher resolution (\emph{bottom}) simulation with the EFE from Equation \ref{g_ext_value}. The latter has the same resolution settings and box size as our isolated simulations, but was not advanced further due to significant numerical drift of M33. The slightly smaller disk precession in our higher resolution simulation (Figure \ref{Disk_precession}) causes the disk to appear thinner at large $r$.}
	\label{Resolution_test_rz}
\end{figure}

We checked that the increased resolution has only a small effect on the RC at 6.1~Gyr. At $r \la 3$~kpc, both models including the EFE behave rather similarly, and have a lower $v_c$ than our isolated model. Another important consideration is whether a central bulge develops. We test this by using Figure \ref{Resolution_test_rz} to show $rz$ projections of our M33 simulations with the EFE. The results look rather similar, with no sign of a central bulge in either case. This is also true in our isolated model at the same $T = 25$~kK (Figure \ref{Particles_radial}).

We therefore conclude that the lack of development of a central bulge in our hydro models at 25~kK is a numerically converged result. If anything, the disk appears slightly thinner in our higher resolution model with the EFE (bottom panel of Figure \ref{Resolution_test_rz}). Thus, neither the resolution nor the EFE affect our ability to stabilize a thin bulge-free Milgromian disk initialized to the M33 surface density profile. We therefore use the lower resolution settings when varying the direction of $\bm{g}_{ext}$ (Section \ref{Different_g_ext_directions}).

\subsection{Reducing the Toomre parameter}
\label{Different_Q_lim}

Our simulations use a disk template where the Toomre parameter $Q$ \citep[generalized to QUMOND in][]{Banik_2018_Toomre} has a lower limit of $Q\_lim = 1.25$ (Section \ref{DICE}). Using $Q\_lim > 1$ ensures some margin of safety, but the exact choice is arbitrary. To explore the impact of $Q\_lim$, we run two more versions of our hydro simulation with the EFE at $30^\circ$ to the disk (Equation \ref{g_ext_value}). In these simulations, $Q\_lim$ is reduced to 1.1 and then 1.

\begin{figure}
	\centering
	\includegraphics[width = 8.5cm] {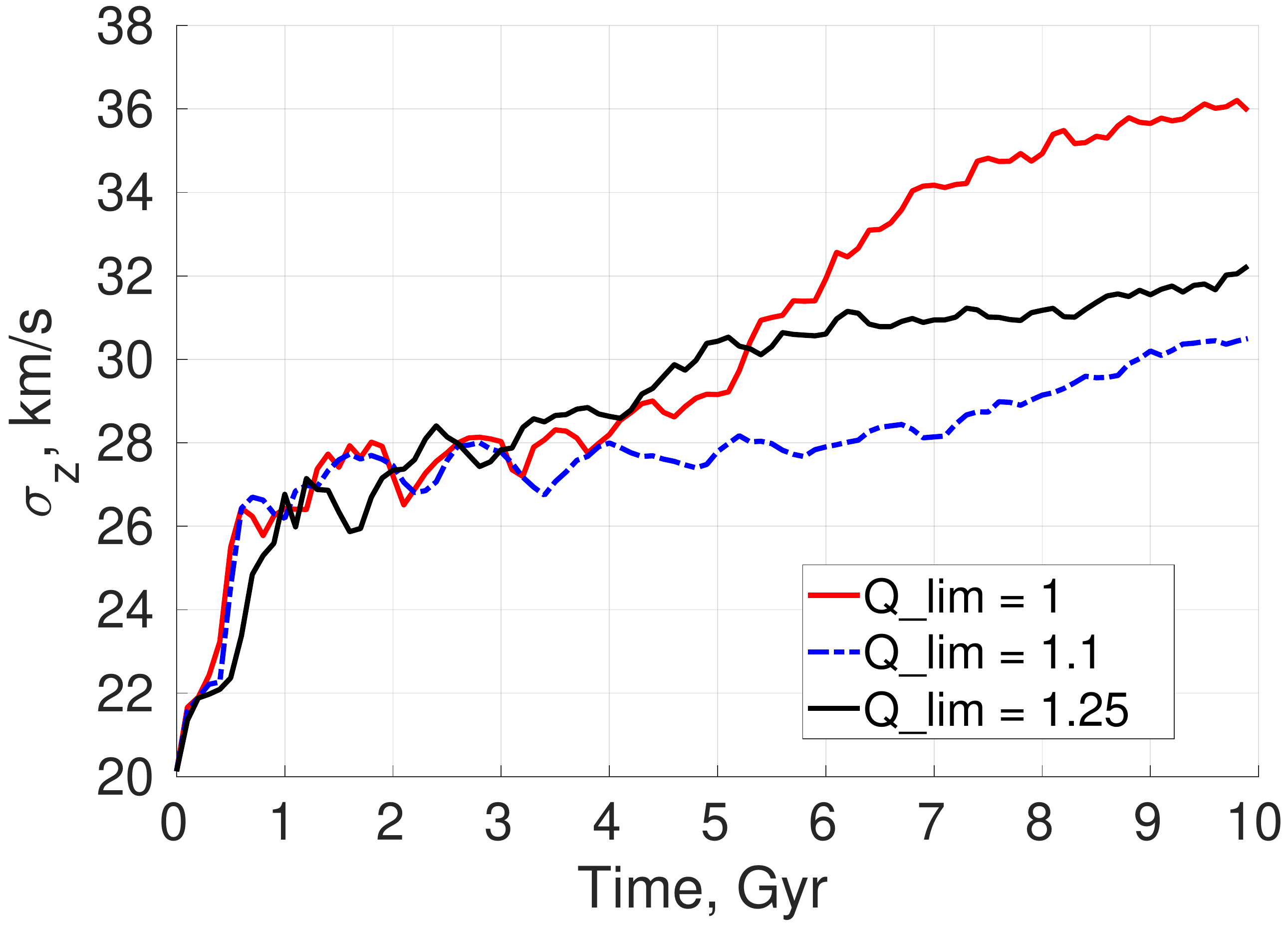}
	\caption{Similar to Figure \ref{sigma_z}, but now showing $\sigma_z$ for the stars at $r = \left( 0.5 - 3 \right)$~kpc in our hydro simulation with the EFE at $30^\circ$ to the disk (Equation \ref{g_ext_value}) in which the floor on the Toomre $Q$ parameter is varied to $Q\_lim = 1$ (solid red), $Q\_lim = 1.1$ (dot-dashed blue), or has the fiducial value of $Q\_lim = 1.25$ (black) used in all other simulations. The simulation at $Q\_lim = 1$ appears to become unstable after $\approx 5$~Gyr, while the other cases remain stable. Results are very similar if the full mass distribution is used instead of just the stars (not shown).}
	\label{sigma_z_Q}
\end{figure}

The effect of varying $Q\_lim$ is apparent in Figure \ref{sigma_z_Q}, which shows $\sigma_z$ for stars with $r = \left( 0.5 - 3 \right)$~kpc. Reducing $Q\_lim$ from 1.25 to 1.1 reduces $\sigma_z$ somewhat, as might be expected. In the model with $Q\_lim = 1$, $\sigma_z$ is never constant for an extended period, though it rises more slowly when $t \approx \left( 2 - 4 \right)$~Gyr. The model appears to be marginally stable, and starts rapidly heating when $t \approx 5$~Gyr. This suggests that our analytic Toomre criterion (Equation \ref{K_0_definition}) is actually a rather good estimate of what initial conditions would be stable.

These results on $\sigma_z$ are mirrored in $\sigma_{_{LOS}}$, which for the central kpc square is 48~km/s for our fiducial model with $Q\_lim = 1.25$ (Figure \ref{sigma_LOS_EF}). This drops to 47~km/s when $Q\_lim = 1.1$, but rises to 53~km/s when $Q\_lim$ is reduced to 1. The $\mathrm{m} = 2$ mode strength also drops somewhat when $Q\_lim$ is reduced from 1.25 to 1.1, but then becomes higher if $Q\_lim$ is reduced further to 1. Therefore, changes to $Q\_lim$ do not have a significant impact on our results for a reasonable choice of $Q\_lim$, i.e. for a value slightly above 1 to ensure local stability.

\subsection{Different external field orientations}
\label{Different_g_ext_directions}

To gain a deeper insight into how the EFE affects our M33 model, we run two more simulations where we put $\bm{g}_{ext}$ in two special orientations. We use the same simulation duration, numerical settings, and external field strength of $0.07 \, a_{_0}$, but set it to point either within the disk plane (along $+\bm{x}$) or along the disk spin axis (along $+\bm{z}$). We call these the disk-aligned and axially aligned models, respectively. The barycentre has a numerical acceleration of $2.3 \times 10^{-3} \, a_{_0}$ in the axially aligned case and $3.9 \times 10^{-4} \, a_{_0}$ in the disk-aligned case, with the drift being very nearly in the opposite direction to $\bm{g}_{ext}$. The disk-aligned model has a numerical drift similar in magnitude to the model with intermediate $\bm{g}_{ext}$ discussed in the preceding section, but the numerical drift is greater for the axially aligned case. 

The overall appearance of the disk and its RC are quite similar in all our models with the EFE. They all have a stellar $\sigma_z$ that is $\approx \left( 3 - 5 \right)$~km/s lower than in the isolated case, signifying somewhat more stability. As a result, the central $\sigma_{_{LOS}}$ differs little between our simulations with the EFE $-$ the axially aligned case yields 49~km/s while the disk-aligned case gives 47~km/s, bracketing the 48~km/s result for the intermediate orientation.

\begin{figure}
	\centering
	\includegraphics[width = 8.5cm] {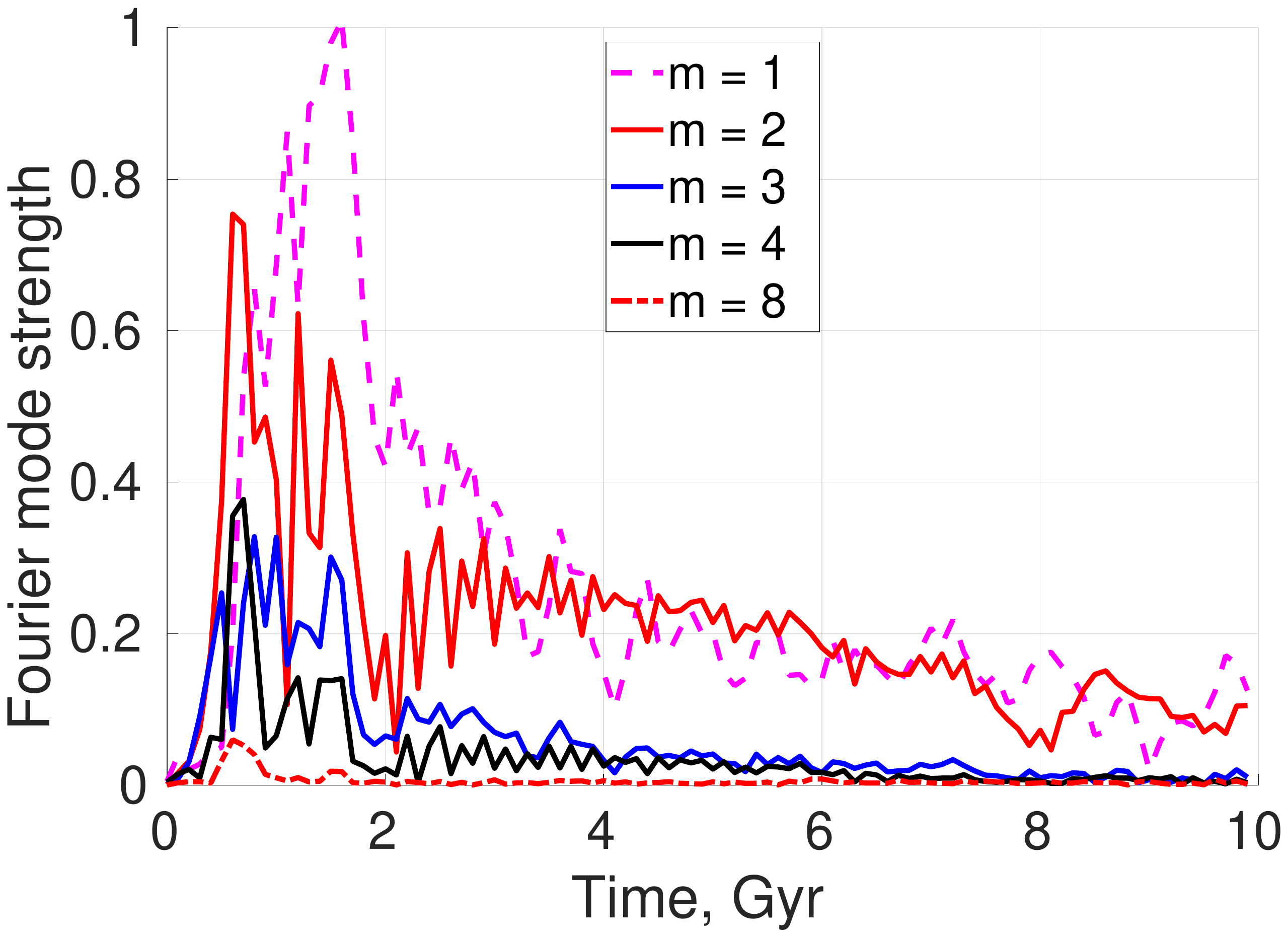}
	\caption{Similar to Figure \ref{Mode_strengths_stellar_only}, but now showing Fourier analysis of the stars at the end of our hydro simulation with axially aligned $\bm{g}_{ext}$. This is the only model to avoid a dominant $\mathrm{m} = 1$ mode at all times. Its strength in the gas is also rather weak (not shown).}
	\label{Mode_strengths_EF_axial}
\end{figure}

Some differences between these simulations are revealed upon Fourier analysis. The relatively strong $\mathrm{m} = 1$ mode evident in Figure \ref{Mode_strengths_stellar_only} also arises in our models with the EFE $-$ but not for the axially aligned case. Figure \ref{Mode_strengths_EF_axial} shows the corresponding results for this model. Depending on how well $A_1/A_0$ is constrained observationally, this could mean the axially aligned $\bm{g}_{ext}$ case is more realistic.

\begin{figure}
	\centering
	\includegraphics[width = 8.5cm] {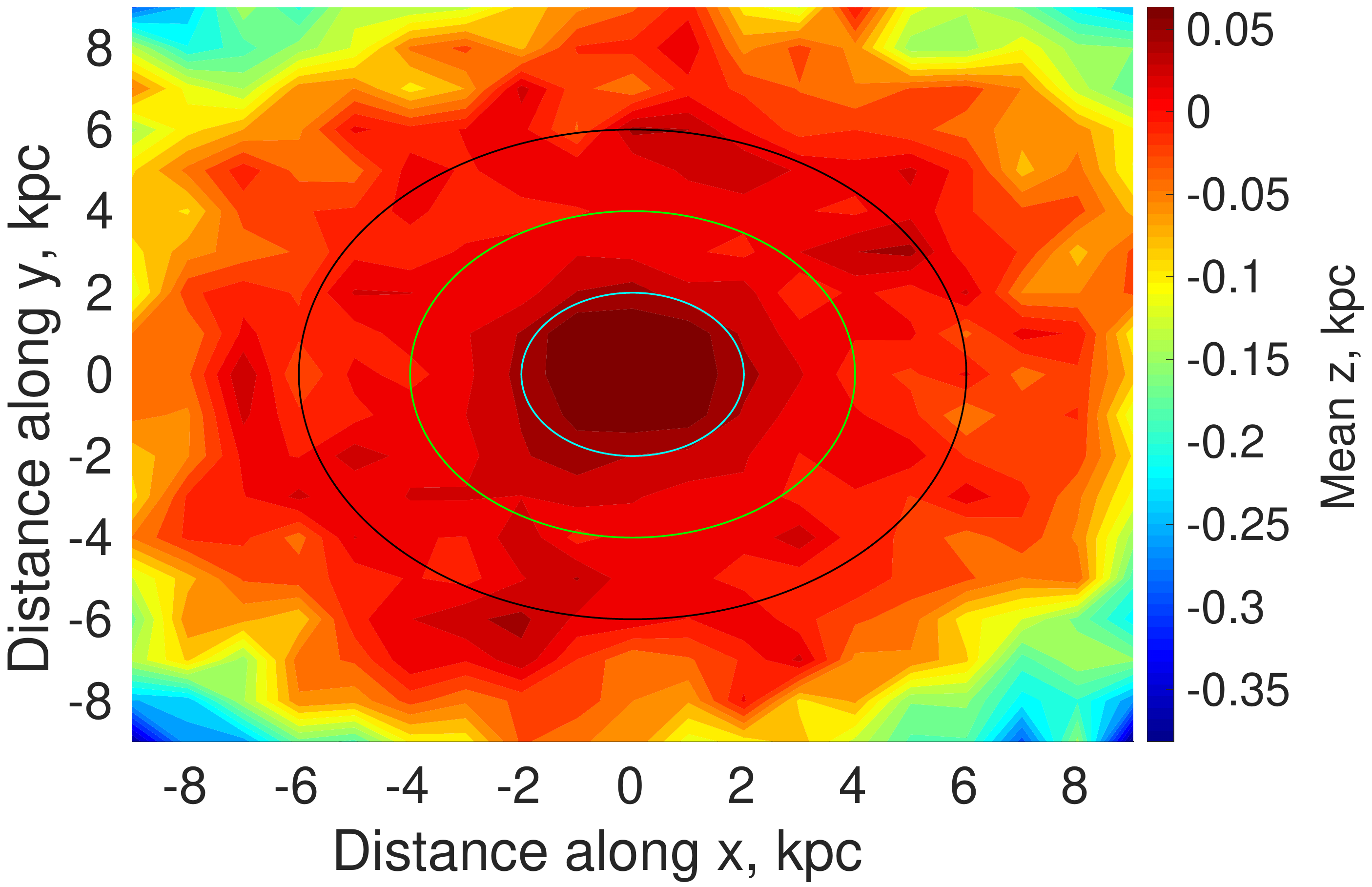}
	\caption{Similar to Figure \ref{Mean_z}, but now showing $\overline{z}$ for our simulation with axially aligned $\bm{g}_{ext} \propto +\bm{z}$. Notice that the warping is now axisymmetric, and occurs in the direction opposite to $\bm{g}_{ext}$. Results remain fairly similar despite the orientation of $\bm{g}_{ext}$ differing by $60^\circ$.}
	\label{Mean_z_g_ext_axial}
\end{figure}

Due to the different symmetry properties of our models with differently aligned $\bm{g}_{ext}$, the behaviour of the warp differs significantly, as discussed next. Warping of the disk is negligible ($\la 0.03$~kpc) in the disk-aligned case, which is expected as the problem is symmetric with respect to $\pm \bm{z}$. Figure \ref{Mean_z_g_ext_axial} shows the disk warp for the axially aligned case. Since the problem is axisymmetric, it is not surprising that the warp also retains approximate axisymmetry. Relative to the centre, the outer parts of the disk become warped in the direction opposite to $\bm{g}_{ext}$. The potential is symmetric with respect to $\pm \bm{g}_{ext}$ in both the analytically tractable isolated and $\bm{g}_{ext}$-dominated cases (Equations \ref{Boundary_Phi_iso} and \ref{Boundary_Phi_EFE}, respectively). This is not true for the intermediate case, as pointed out in earlier works \citep[e.g.][]{Thomas_2018}.

\begin{figure}
	\centering
	\includegraphics[width = 8.5cm] {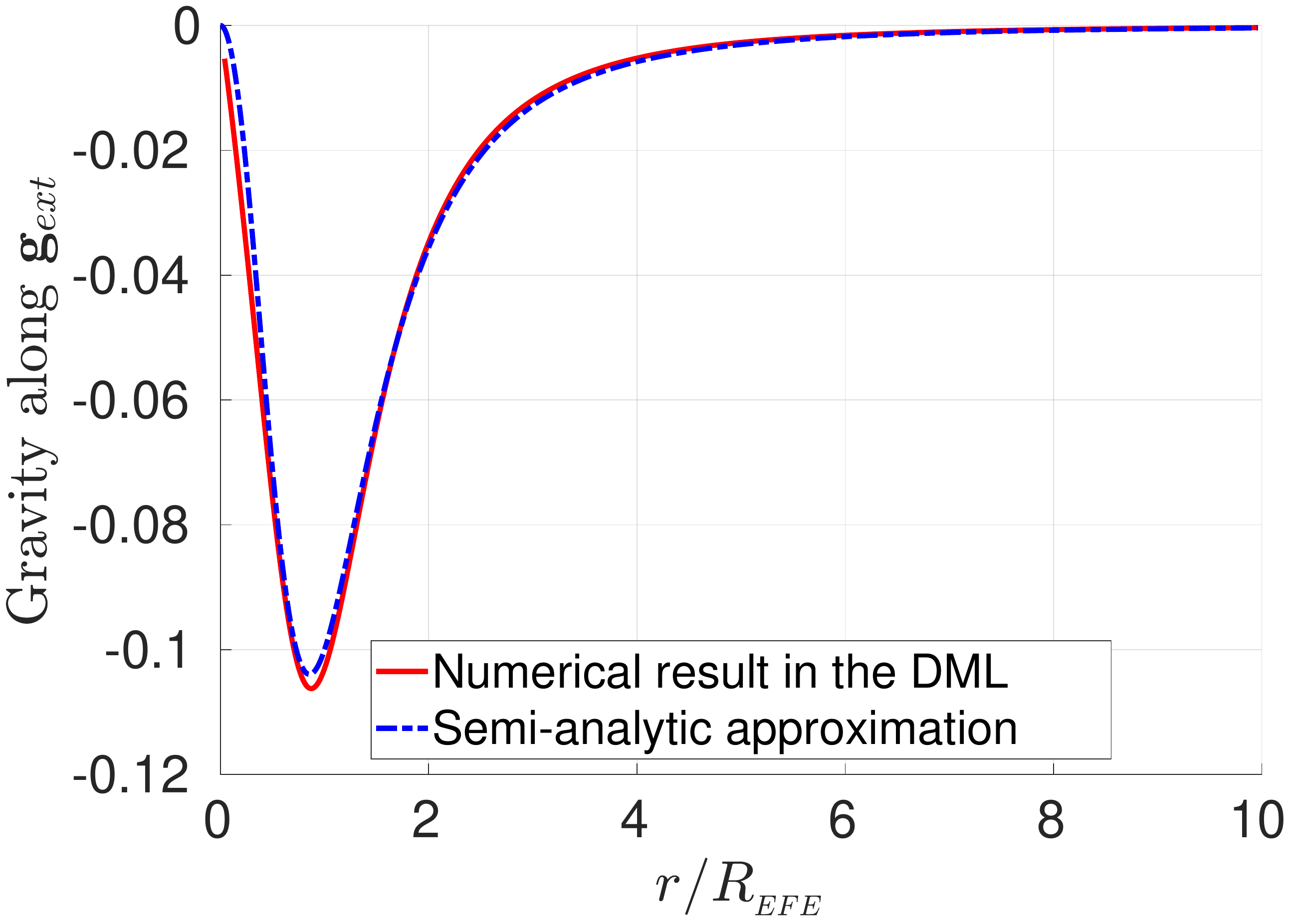}
	\caption{The gravity in the direction parallel to $\bm{g}_{ext}$ experienced by a test particle due to a point mass located in the orthogonal direction. This is a measure of asymmetry in the gravitational field between $\pm \bm{g}_{ext}$, since results would be zero in the symmetric case $-$ which arises in the EFE-dominated limit (Equation \ref{Boundary_Phi_EFE}). The results shown in solid red are from numerical force calculations in the DML of a point mass embedded in a uniform external field \citep[section 2 of][]{Banik_2018_escape}. The units are such that $G = M = a_{_0} = g_{ext} = 1$, so $R_{_{EFE}} = 1$ (Equation \ref{r_EFE}). Results shown here can be scaled to other $g_{ext}$ if the problem remains in the DML \citep{Milgrom_2009_DML}. The dot-dashed blue line shows the semi-analytic fit given in appendix A of \citet{Banik_Ryan_2018}, with the $-$ sign changed to $+$ in their equation A2 to fix an error in the original. The gravity in the radial direction is not shown here, but is the dominant component at all radii.}
	\label{Force_z_at_theta_half_pi}
\end{figure}

To gain insight into the potential, we make use of the DML numerical force library described toward the end of section 2 in \citet{Banik_2018_escape}. This provides calculations of the gravitational field from a point mass in the DML, with the EFE rigorously accounted for by direct numerical integration of Equation \ref{QUMOND_governing_equation_EFE}. The force evaluations relevant to our discussion here are for points at $\theta = \mathrm{\pi}/2$, since all points in the $\bm{x}\bm{y}$ plane are at right angles to $\bm{g}_{ext}$ as perceived from the centre of M33. Their numerical calculations show that at these points, the force is mostly toward the centre, but there is a small additional component along $-\bm{g}_{ext}$. The strength of this symmetry-breaking component is shown in Figure \ref{Force_z_at_theta_half_pi} as a function of radius.\footnote{The fact that the tangential component of the gravity is anti-aligned with $\bm{g}_{ext}$ is also evident in figure 2 of \citet{Banik_2019_spacecraft}, though their results are for a much stronger $g_{ext} = 1.78 \, a_{_0}$ as appropriate for the Solar neighbourhood.}

The DML force library in \citet{Banik_2018_escape} was later fit using the fitting function given in appendix A of \citet{Banik_Ryan_2018}, bearing in mind the gravitational field at all locations. We noticed that the $-$ sign in their equation A2 must be replaced by $+$ to get the correct results. With this correction and restricting to the case of a test particle at $\theta = \mathrm{\pi}/2$ relative to a point mass $M$ embedded in an external field along $+\bm{z}$, we expect a vertical force of:
\begin{eqnarray}
	\frac{\bm{g}_z}{g_{ext}} ~=~ -\frac{2 \, {\widetilde{r}}^2}{5 \left( 1 + {\widetilde{r}}^2 \right) \left( 1 + \widetilde{r}^3 \right)} \, ,
	\label{g_z_EFE_induced}
\end{eqnarray}
where $\widetilde{r} \equiv r/R_{_{EFE}}$ is the radius scaled to $R_{_{EFE}}$ (Equation \ref{r_EFE}), the only physical scale in this DML problem. Using $M = 6.5 \times 10^9 \, M_\odot$ (Table \ref{Parameters}) and $g_{ext} = 0.07 \, a_{_0}$ as before, we get that $\bm{g}_z = -0.0017 \, a_{_0}$ at $r = 10$~kpc.

We can estimate the height of the induced warp by approximating that the EFE-induced vertical gravity from M33 must be balanced by a geometric term similar to that in Section \ref{Gas_thickness_profile}. Treating the disk as having an extent $\ll 10$~kpc, the gravity it exerts directly towards itself is very nearly ${v_f}^2/r$, with $v_f$ given by Equation \ref{BTFR}. We expect this to be approximately valid if $\bm{g}_{ext}$ is sub-dominant to the disk gravity ($\widetilde{r} \ll 1$), which is true for $r \ll R_{_{EFE}} = 41$~kpc (Equation \ref{r_EFE}). With this approximation, the vertical component of the disk gravity near the $z = 0$ plane is $-\left( {v_f}^2/r \right) \left( z/r \right)$ even without the EFE. We assume that in equilibrium, this approximately balances the additional $\bm{g}_z$ induced by the EFE (Equation \ref{g_z_EFE_induced}). As a result, the equilibrium height of the warp can be estimated as:
\begin{eqnarray}
	\overline{z}_{eq} ~\approx~ \frac{r^2 \bm{g}_z}{{v_f}^2} < 0 \, .
	\label{z_eq}
\end{eqnarray}
For the case of M33, we get that $\overline{z}_{eq} = -0.061$~kpc. This is fairly close to the actual warp of $\approx -0.15$~kpc evident in Figure \ref{Mean_z_g_ext_axial}, comparing $\overline{z}$ at the disk centre with its value at $r = 10$~kpc.

Our semi-analytic approach does not capture the full complexity of an extended disk. If we focus on a point $P$ within the disk mid-plane (not necessarily $z = 0$) at $r = 10$~kpc, regions close to $P$ do not exert a net gravitational force along $\bm{z}$. The regions at lower $r$ do, but all the material is not concentrated at $r = 0$. Moreover, even the inner regions of the disk become warped to some extent, making the $z$ co-ordinate of $P$ closer to the barycentric $\overline{z}$ of the whole system. This reduces the geometric $\left( z/r \right)$ factor assumed in Equation \ref{z_eq}, so the outer regions of the disk must warp more to compensate. As a result, we expect the true warping of the disk to exceed our semi-analytic estimate, which is indeed the case. Nonetheless, it is possible to determine the approximate extent of disk warping without $N$-body or hydro simulations.

\subsection{Broader implications}
\label{Broader_implications}

Although the EFE is a natural part of MOND (Section \ref{External_field_effect}) and follows inevitably from its governing equations \citep{Milgrom_1986}, its impact on our simulations is rather surprising given that it is sub-dominant at $r \la 41$~kpc. In QUMOND, any possible effect on the dynamics must come through the value of $\nu$, which depends solely on the Newtonian gravity (Equation \ref{QUMOND_governing_equation_EFE}). Since $g_{_N} \appropto 1/r^2$ over the range $r = \left( 4 - 40 \right)$~kpc, we see that $g_{_{N, ext}}$ should be sub-dominant by a factor of 100 at 4~kpc. In particular, a point mass of $M = 6.5 \times 10^9 \, M_\odot$ creates a Newtonian gravity of $g_{_N} = 0.472 \, a_{_0} = 103 \, g_{_{N, ext}}$ at this distance. A similar level of isolation is evident if we consider the $z$-component of the internal and external Newtonian gravity for our intermediately aligned $\bm{g}_{ext}$: at 4~kpc, the former dominates by a factor of 101 if we assume $g_{_{N,z}} = 2 \mathrm{\pi} G \Sigma$ (Figure \ref{M33_surface_density}). Thus, the EFE could in principle have a significant impact on the secular evolution of M33 after 100 dynamical times, i.e. after about $100 \, \left( r_d/v_f \right) \approx 2$~Gyr, well within the duration of our simulations.

The EFE alters the resonant structure of the disk by breaking axisymmetry. This likely inhibits the usual process where a whole ring of material moves inward and ends up feeling even more radial gravity, with build-up of random motions eventually halting the inward migration. Our simulations are initially stable according to the QUMOND Toomre condition (Section \ref{DICE}), but radial migration always occurs to a limited extent. In addition to breaking the axisymmetry of the disk, the EFE generally also breaks its up-down symmetry and causes the disk to warp (Figure \ref{Mean_z}). This makes it even more difficult to form resonances as different rings of material are in different planes. Restoring axisymmetry or up-down symmetry with a special alignment of $\bm{g}_{ext}$ has little effect on our results (Section \ref{Different_g_ext_directions}), suggesting that the more efficient radial migration in our isolated simulation is related to it having both these symmetries. The EFE necessarily breaks at least one of these symmetries. Therefore, the EFE has very important implications for the development of instabilities, especially the bar instability (Figure \ref{vt}). This in turn affects the efficient redistribution of disk material, which is apparent in Figure \ref{Sigma_star_profile}.


While our model may provide a viable scenario for the weak bar and bulge in M33, bars are often much stronger \citep{Delmestre_2007}. Using isophotal ellipticity as a proxy for bar strength, they identified that only $\approx 40\%$ of galaxies have a very weak bar that might be undetected in their analysis. Thus, it is necessary to get a mixture of both strong and weak bars. Our results suggest that this could partly arise from differences in the EFE.


Some evidence for the EFE has already been found in that galaxies with declining RCs in their outskirts generally have an identifiable perturber \citep{Haghi_2016, Chae_2020}. Our results show that even a weak perturber would also affect the overall level of dynamical stability, an issue which has not previously been discussed. Our work is the first to conduct a self-consistent hydro simulation of how a weak external field affects the secular evolution of a Milgromian disk. The implications for the bar strength are somewhat ambiguous and degenerate with other properties like the gas temperature (Section \ref{Results}). A much more powerful test could come from the warp induced by the EFE, as discussed next.

\subsubsection{The external field warp}
\label{External_field_warp}

Our results on disk warping in Section \ref{Different_g_ext_directions} highlight that when the internal and external gravitational fields are comparable, the gravity due to a mass distribution becomes asymmetric with respect to $\pm \bm{g}_{ext}$. Though this was known in prior work and is evident from the point mass potential (Figure \ref{Force_z_at_theta_half_pi}), ours is the first to demonstrate the impact on a self-consistent numerical simulation of a disk galaxy. The asymmetry also has implications for tidal streams, as discussed further in \citet{Thomas_2018} using the example of Palomar 5. Consider first the leading arm, which consists of material lost from the satellite in the direction towards the host galaxy. Suppose that a particle in the leading arm is almost directly ahead of the satellite in its orbit, so that as perceived from the satellite, the particle and the host are at right angles. As shown in Figure \ref{Force_z_at_theta_half_pi}, the asymmetric potential causes the gravity on the particle to receive a slight excess contribution in the direction opposite to $\bm{g}_{ext}$, which in this case is provided by the host galaxy. Thus, the particle feels an extra force in the radially outward direction as viewed from the host. Due to angular momentum conservation, the angular velocity of the particle then decreases, causing it to move back towards the satellite. This has the effect of shortening the leading arm. The above logic remains exactly the same for the trailing arm, except for the last step $-$ since the material is already trailing, a further decrease in its angular velocity relative to the host \emph{increases} the length of the trailing arm. This is critical to the results shown in figure 5 of \citet{Thomas_2018}, and the putative comparison they draw with observations. Note however that the asymmetry relies on having comparable gravity from the satellite and the host. For a very low mass satellite, the host gravity would completely dominate, causing the satellite to experience a dominant external field. This would not only weaken the self-gravity of the satellite, but would also make it much more symmetric with respect to $\pm \bm{g}_{ext}$. Therefore, tidal tails are not guaranteed to be asymmetric in MOND. Semi-analytic arguments can be used to narrow down the most promising targets to search for such asymmetry.

Tidal tails would be difficult to observe in this level of detail beyond the Local Group, limiting the statistics. Our work suggests another promising line of investigation $-$ external disk galaxies viewed nearly edge-on should curve away from $\bm{g}_{ext}$. The recent results of \citet{Chae_2020} already indicate that disk galaxies are sensitive to the magnitude of $\bm{g}_{ext}$, so a natural extension would be to test whether they are also sensitive to its direction. Comparing the $\overline{z}$ maps in Figures \ref{Mean_z} and \ref{Mean_z_g_ext_axial}, we see that even when $\bm{g}_{ext}$ is significantly misaligned with the disk spin axis, the warp is still nearly axisymmetric and mostly in the opposite hemisphere to $\bm{g}_{ext}$, with approximately the same strength.\footnote{The only difference is a slight central depression for the misaligned case, which could be incorporated into a more sophisticated analysis.} For a much more significant misalignment, $\bm{g}_{ext}$ would by definition lie very close to the disk plane, making the situation similar to our disk-aligned EFE model. As expected on symmetry grounds, there is no warp in this case, so we do not expect a warp to develop in all disks experiencing a non-negligible EFE.

In cases with a suitable $\bm{g}_{ext}$, the main expected signal would be for the disk to warp away from the nearest major galaxy. This would be quite unusual conventionally, since any linear gravity theory is not compatible with a star orbiting around a mass located outside the orbital plane of the star. Even if a galactic disk were initialized to have a bowl-shaped warp, the usual expectation would be for the warp to disappear on a dynamical timescale due to the imbalance of forces along the spin axis. Any observation of such a feature would therefore be quite remarkable, especially if the direction follows expectations based on the large scale gravitational field. This was mapped by \citet{Desmond_2018} and used in the analysis of \citet{Chae_2020}. While the \citet{Desmond_2018} analysis used conventional methods to estimate $\bm{g}_{ext}$ on individual galaxies, this was still sufficient for \citet{Chae_2020} because an important part of the analysis was comparing deviations from the radial acceleration relation between galaxies experiencing a weak or strong EFE. The latter generally involved a relatively massive nearby galaxy, in which case $\bm{g}_{ext}$ would be closely aligned with the direction towards it. Moreover, the critical quantity in QUMOND is the Newtonian external field (Equation \ref{QUMOND_governing_equation_EFE}). It should therefore be quite feasible to estimate the direction of $\bm{g}_{ext}$ on SPARC galaxies by adapting existing work. In principle, this could lead to a detection of the EFE through three main channels:
\begin{enumerate}
	\item The existence of a nearly axisymmetric bowl-shaped warp in edge-on disk galaxies.
	\item A correlation between the direction of curvature and that of the external field.
	\item The expected magnitude of the curvature could be estimated based on $\bm{g}_{ext}$ and parameters of the disk, allowing a comparison with observations.
\end{enumerate}

\begin{figure}
	\centering
	\includegraphics[width = 8.5cm] {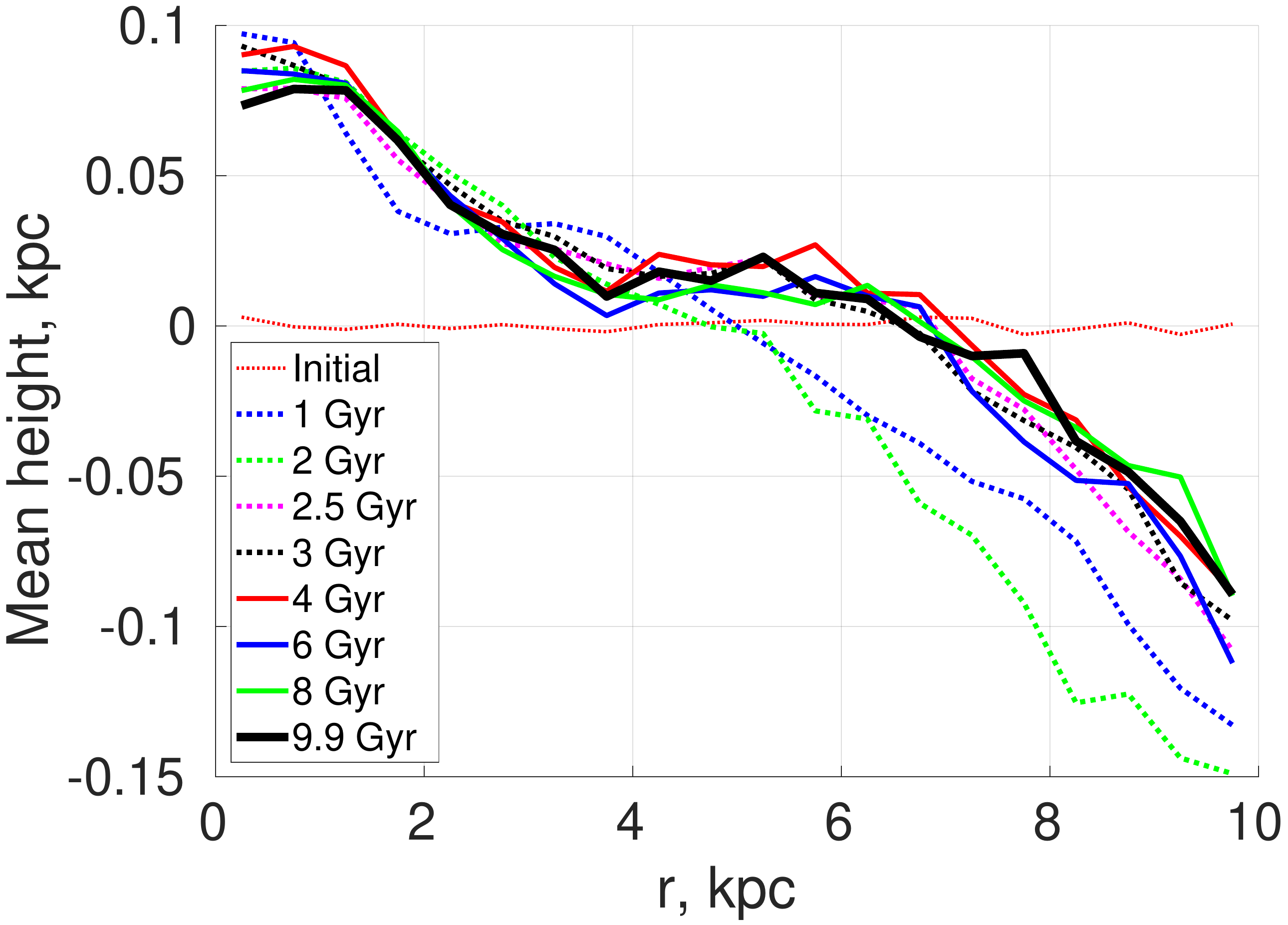}
	\caption{The mean height $\overline{z}$ of different annuli in the M33 disk in an axially aligned external field. Different colours and linestyles show different times, as indicated in the legend. The 1D projection used here loses little information because the warp is nearly axisymmetric (Figure \ref{Mean_z_g_ext_axial}). Notice that the warp changes very little after 2.5~Gyr, with the main change before this being the development of a `knee' at $r \approx \left( 4 - 6 \right)$~kpc. There is no warp initially, so results at this time give an idea of the numerical noise (thin dotted red line).}
	\label{Warp_profiles_g_ext_axial}
\end{figure}

To address the last issue, we use Figure \ref{Warp_profiles_g_ext_axial} to show the warp profile as a function of $r$, using annular bins aligned with the disk spin axis. The results could be scaled to different galaxies based on Equations \ref{g_z_EFE_induced} and \ref{z_eq}, exploiting the scale invariance of dynamics in the DML \citep[which should be valid as typically $g_{ext} \la 0.1 \, a_{_0}$,][]{Chae_2020}. Since nearby galaxies move over long periods, an important issue is the time required for the warp to develop as a result of the EFE. We address this by showing the warp profile $\overline{z} \left( r \right)$ at different times (Figure \ref{Warp_profiles_g_ext_axial}). From this, it is clear that the warp is already very apparent after just 1~Gyr, before then settling into what is presumably the equilibrium configuration after $\approx 2.5$~Gyr. At $r = 10$~kpc, the orbital period is $2 \mathrm{\pi} r/v_f = 0.61$~Gyr, so the warp is apparent after only a few revolutions. This short timescale means it should be sufficient to predict $\overline{z} \left( r \right)$ using the present $\bm{g}_{ext}$. Indeed, MOND simulations of DF2 showed that memory effects due to even a quite strongly time-varying $\bm{g}_{ext}$ play a rather small role in its internal velocity dispersion, which can be estimated quite accurately by considering it to be in equilibrium with the present $\bm{g}_{ext}$ \citep[figure 5 of][]{Haghi_2019}. Therefore, orbital motion of galaxies and the resulting changes in $\bm{g}_{ext}$ would not much affect the above-mentioned disk warp effect. Another very useful feature is that it does not require very precise kinematics of the disk $-$ a rough estimate of the RC amplitude would be helpful, but the test is based mainly on the shape of the galaxy rather than subtle features in its RC. The main observational input would thus be high-resolution photographs, with a 0.1~kpc warp in a galaxy 50~Mpc away requiring an angular resolution better than $0.4\arcsec$.

Searches for warps of this sort have previously been conducted to test fifth-force theories in which galaxies are almost Newtonian and reside in CDM halos \citep{Ferreira_2019}. These models generically violate the weak equivalence principle, creating an offset between the baryonic disk and the CDM halo \citep{Desmond_2018_offset}. This would cause the disk to warp, an effect which those authors and \citet{Desmond_2018_warp} claimed to detect. Recent results suggest that the signal was driven by only a small number of galaxies, so an appropriate choice of quality cuts would eliminate them $-$ and therewith the signal \citep{Desmond_2020}.

It would be very interesting to revisit these analyses in a MOND context, since the environmental dependence would be rather different. In addition, the same analysis should be applied to mock galaxy images from a hydro $\Lambda$CDM cosmological simulation \citep[e.g. Illustris,][]{Pillepich_2018}. This would clarify whether $\Lambda$CDM contains processes which can create a warp correlated with galaxy and environmental properties in a manner analogous to MOND. In both cases, it would be necessary to apply careful quality control measures, e.g. excluding interacting galaxies significantly affected by tidal forces. One important outcome of our simulations is that the MOND-predicted warp should be nearly axisymmetric (Figures \ref{Mean_z} and \ref{Mean_z_g_ext_axial}), but a random perturbation like a satellite would generally affect one side much more than the other. Therefore, systematic errors could be reduced by focusing only on galaxies whose image is nearly symmetric about the sky-projected minor axis. Conventional warps are not symmetric in this sense as the warp material would still orbit the galactic centre.

\section{Conclusions}
\label{Conclusions}

In MOND, disk galaxies are self-gravitating and lack DM halos, leading to different stability properties compared to $\Lambda$CDM \citep{Milgrom_1989, Banik_2018_Toomre}. Our main objective was to see if these differences could help alleviate reported difficulties reproducing the observed dynamical properties of M33 with a live DM halo, where a strong bar tends to form in disagreement with observations \citepalias{Sellwood_2019}. Those authors could stabilize M33 to have a realistic morphology over at least 3~Gyr if the Toomre parameter $Q = 2$ (Equation \ref{Q_Newton}), which would make it difficult to explain why M33 is currently forming stars \citep{Verley_2009}.


To investigate the global stability of M33 in MOND, we set up $N$-body and hydro simulations of M33 in MOND using a modified version of \textsc{dice} \citep{Perret_2014}, which we make publicly available (Section \ref{DICE}). We ran them with and without the estimated EFE from M31 using the publicly available \textsc{por} code \citep{Lughausen_2015}. \textsc{por} implements the rather computer-friendly QUMOND formulation of MOND \citep{QUMOND}.

We ran isolated stellar-only and hydro simulations for just over 6~Gyr using the same total surface density profile initially. The stellar-only and $T = 100$~kK simulations become substantially non-circular even in the outer regions, indicating that the whole galaxy is essentially a very long bar (Appendix \ref{100kK_iso_face_on}) with pattern speed $\Omega_p \approx 10$~km/s/kpc, implying a corotation radius of 10~kpc. These models also develop a substantial central bulge, unlike the observed M33 (Figure \ref{Particles_radial}). This is related to $\sigma_{_{LOS}}$ being $62$~km/s at the centre of M33, much higher than observed.

Most of these problems can be resolved by using a lower gas temperature of 25~kK, which is more in line with the 12~kK adopted in section 3.2 of \citetalias{Sellwood_2019}. In this cooler isolated model, no bulge forms. The stellar disk appears fairly circular in the outer regions (Figure \ref{Particles_face_on}), with a clear bisymmetric spiral evident in the gas when viewed face-on (Figure \ref{Gas_face_on}). A strong bar forms in the first Gyr, but it then loses strength for a variety of reasons, including buckling, radial gas inflow, and chaos \citep[see also][]{Tiret_2008_gas}. As a result, the observed weak bar of M33 is recovered by the end of this simulation. Our Fourier analysis indicates that the dominant mode of non-axisymmetry is $\mathrm{m} = 1$, with a significant $\mathrm{m} = 2$ component but very little strength in higher harmonics (Figure \ref{Mode_strengths_stellar_only}). The $A_2/A_0$ ratio for material with $r = \left( 0.5 - 3 \right)$~kpc is sometimes similar to the observed 0.2 \citep[section 4.3 of][]{Corbelli_2007_image}, though it is generally smaller. Since the outskirts appear nearly circular, the bar is genuinely weak. Its pattern speed of 30~km/s/kpc implies a corotation radius of 3~kpc, so the bar is fairly short. Despite these promising results, there is still significant radial redistribution of material, causing the inner RC of M33 to rise more steeply than observed (Figure \ref{M33_rotation_curves_iso}). The central $\sigma_{_{LOS}}$ of 57~km/s is also above the observational range of ${\left( 28 - 35 \right)}$~km/s \citep[section 3.2 of][]{Corbelli_2007_image}. We conclude that allowing a more dissipative gas component greatly helps to stabilize the thin observed disk of M33 against the formation of a very strong bulge and bar, but further improvements to the model are still necessary.


In MOND, the EFE from M31 can have a percent-level effect on the dynamics of M33. We included the EFE with $\bm{g}_{ext} = 0.07 \, a_{_0}$ at $30^\circ$ to the M33 disk. We demonstrated for the first time that this affects its secular evolution after a few Gyr, with the inward radial migration of material significantly suppressed (Figure \ref{Sigma_star_profile}). This makes the RC much closer to observations (Figure \ref{M33_rotation_curve_EF}), with the minor ($\la 10$~km/s) differences possibly being due to a warp that is not included when deriving the observed RC. The central $\sigma_{_{LOS}}$ is also much lower at 48~km/s by the end of our simulation (9.9~Gyr), with the LOS velocity distribution closely following a Gaussian of this width (Figure \ref{LOSVD_central_pixel_EF30}). This is much more consistent with the observational range. The non-circular motions are also much smaller (Figure \ref{vt}).

To test the numerical convergence of these results, we ran a higher resolution simulation for 6.1~Gyr. No central bulge formed (Figure \ref{Resolution_test_rz}), consistent with our lower resolution model and our isolated 25~kK simulation at this time (Figure \ref{Particles_radial}). However, the isolated 100~kK model develops a substantial bulge. We therefore argue that the observed configuration of M33 is inherently stable in MOND if a realistic gas temperature is adopted, though the EFE helps to improve the agreement with some observables. Despite the low observed gas fraction near the centre, its dissipative nature is seemingly important to our main result that no central bulge forms, in agreement with observations \citep[e.g.][]{Kam_2015}. Since an important reason for the success of our model is a reduction in $T$ from 100~kK to 25~kK, a colder gas component may help to further stabilize the stellar disk. However, our results from a very cold 10~kK model indicate that the disk becomes unstable once $T$ is reduced this much. It is unclear whether stellar feedback would change this picture, though the ejection of material should make it even more difficult to form a substantial bulge. A reduced central surface density would cause the RC to rise more gradually in the central few kpc, which can also be achieved by starting with a larger scale length than observed. This would make the disk more stable by pushing it deeper into the MOND regime (Equation \ref{Sigma_MOND}).

In general, the EFE breaks both the axisymmetry and up-down symmetry of the underlying gravitational physics. To isolate the effect of each in turn, we ran simulations where $\bm{g}_{ext}$ is aligned with the disk spin axis or put in the disk plane (Section \ref{Different_g_ext_directions}). The suppression of radial migration is similar for all three considered external field orientations, so we conclude that this arises from breaking the combination of axisymmetry and up-down symmetry present only in the isolated case. In models with the EFE where one of these symmetries is preserved, the restoration of some symmetry prevents the disk as a whole from precessing, which it does with intermediately aligned $\bm{g}_{ext}$ (Figure \ref{Disk_precession}). The disk also becomes warped if the up-down symmetry is broken. We relate the warp to previous DML calculations of the gravity from a point mass in a weak external field, focusing on the extent to which results are asymmetric with respect to $\pm \bm{g}_{ext}$ (Figure \ref{Force_z_at_theta_half_pi}). Possible observational signatures of this asymmetry are discussed in Section \ref{Broader_implications}, focusing in particular on the induced warp in disk galaxies viewed close to edge-on (Section \ref{External_field_warp}). The warp develops in only a few dynamical times (Figure \ref{Warp_profiles_g_ext_axial}) in the sense that the outskirts curve towards $-\bm{g}_{ext}$ or away from the mass sourcing the EFE, potentially offering a clear and highly specific observational signature of the EFE. This would extend the previous result of \citet{Chae_2020} that its magnitude has a detectable impact on galaxy RCs, violating the strong equivalence principle because the reported effect is not due to tides in conventional gravity (see their section 4).

MOND has enjoyed a great deal of success predicting the RCs of spiral galaxies, including M33 \citep{Sanders_1996, Famaey_McGaugh_2012, Kam_2017}. It can also form such galaxies out of a collapsing gas cloud \citep{Wittenburg_2020}, which may well fit into a broader cosmological context that better explains the large scale structure and expansion rate of the local Universe without violating early Universe constraints \citep{Haslbauer_2020}. Using a numerical implementation of MOND, we were able to match the leading-order non-axisymmetric features of M33 reasonably well once its gas is assigned a not too high temperature. Our model produces no significant central bulge, but the simulated RC still rises more steeply than observed due to efficient inward radial migration. This process is hampered when we include the MOND-predicted weak EFE representative of the estimated gravity exerted by M31. This yields a much better representation of the observed RC. The central velocity dispersion $\sigma_{_{LOS}}$ is however still too high, but this could be due to possible systematics in the measurements and their comparison to our simulations (Section \ref{Bulge_EF}). Further improvements to our models should be possible by e.g. including stellar feedback to reduce the central surface density, or forcing this to a lower initial value by starting with a more extended mass distribution. 

Our work highlights for the first time the role of a weak external field on the stability and evolution of disk galaxies in MOND. Further simulations with a time-varying external field, modeling the full orbit of M33, will be needed to confirm, and perhaps refine, its resemblance to observations.

\section*{Acknowledgements}

IB is supported by an Alexander von Humboldt Foundation postdoctoral research fellowship. BF and RI acknowledge funding from the Agence Nationale de la Recherche (ANR project ANR-18-CE31-0006 and ANR-19-CE31-0017) and from the European Research Council (ERC) under the European Union's Horizon 2020 research and innovation programme (grant agreement number 834148). GC acknowledges support from FONDECYT Regular No. 1181708. IB thanks Stacy McGaugh for providing the M33 baryonic mass model and RC measurements. He also thanks Jerry Sellwood for providing bar strength time series in several Newtonian simulations of M33 embedded in a DM halo. The authors are very grateful to the referee for helping to significantly improve this manuscript and the simulations it presents.

\bibliographystyle{aasjournal}
\bibliography{M33_bbl}

\appendix 
\section{Effect of temperature on the gas disk}
\label{Gas_edge_on}

\begin{figure*}
	\centering
	\includegraphics[width = 17.7cm] {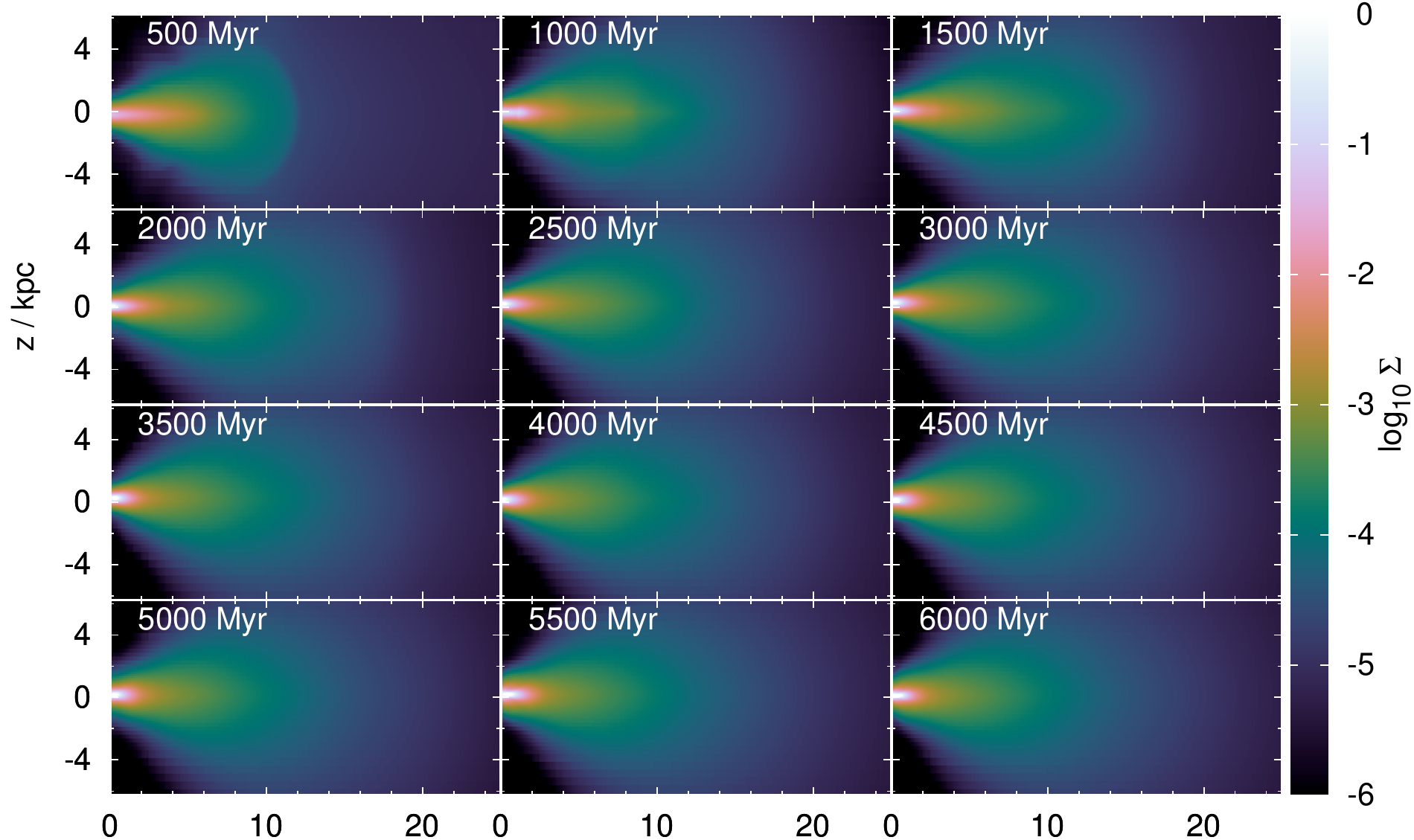}
	\includegraphics[width = 17.7cm] {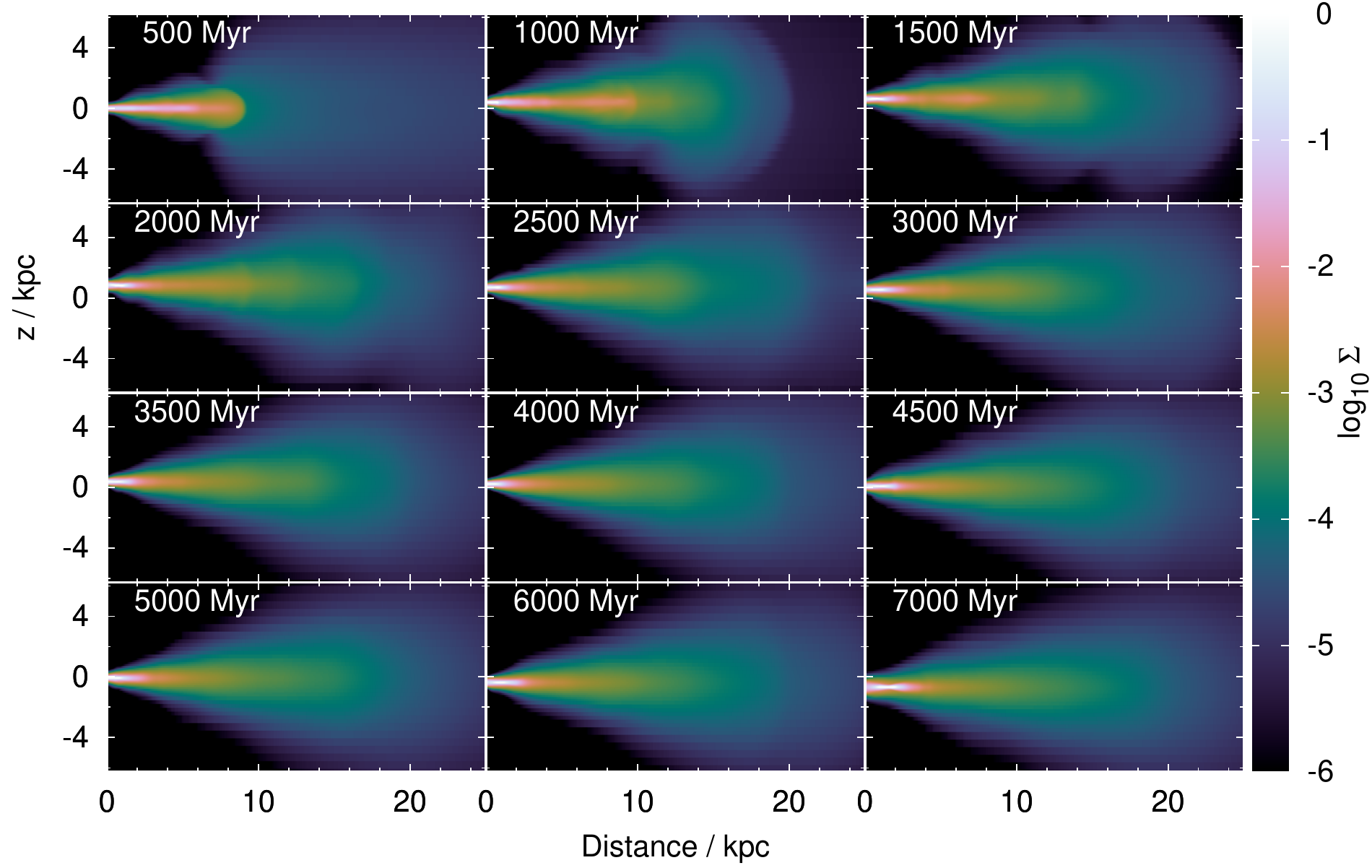}
	\caption{Similar to Figure \ref{Particles_radial}, but now showing the gas in our isolated simulation at 100~kK (top) and 25~kK (bottom). The thinner gas disk in the cooler model is expected from the initial thickness profile (Figure \ref{Gas_initial_thickness}), though some thickening occurs in both models.}
	\label{Gas_radial}
\end{figure*}

Figure \ref{Gas_radial} shows the gas in our isolated hydro simulations at different times using the cylindrical $rz$ projection (Section \ref{Cylindrical_view}). In the outskirts, the gas expands outward to a small extent in both models. Meanwhile, the density in the central pixel rises slightly, especially in the hotter (100~kK) model. Notice that the appearance changes little after the first Gyr, which is already enough to discern a substantial difference between the models. The retention of a thin gas disk in the cooler model is important to the global stability of M33, as discussed in the main article.

\section{Face-on views of the 100~kK isolated simulation}
\label{100kK_iso_face_on}

\begin{figure*}
	\centering
	\includegraphics[width = 17.7cm] {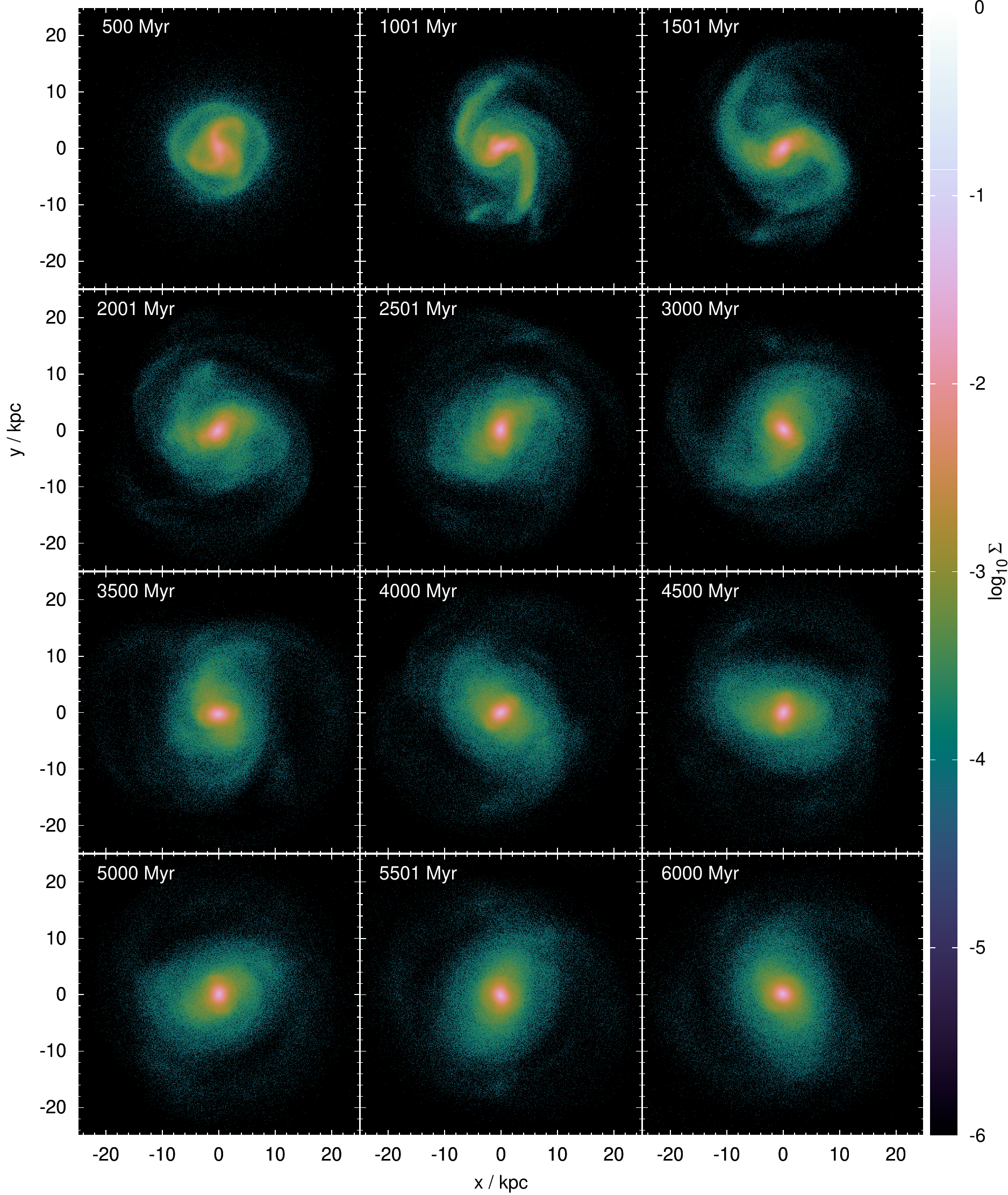}
	\caption{Similar to Figure \ref{Particles_face_on}, but now showing the particles in our isolated simulation at 100~kK. Notice the significantly more non-circular shape of M33 in its outskirts, signifying a very strong and extended bar quite inconsistent with observations.}
	\label{Part_face_on_100kK}
\end{figure*}

Figure \ref{Part_face_on_100kK} shows the stars in our isolated 100~kK simulation as viewed face-on at different times. As discussed in Section \ref{Results}, the outer parts appear significantly non-circular, to the extent that the whole galaxy is essentially one giant bar. This is consistent with the corotation radius of 10~kpc implied by the bar pattern speed (Figure \ref{M33_pattern_speed}).

\begin{figure*}
	\centering
	\includegraphics[width = 17.7cm] {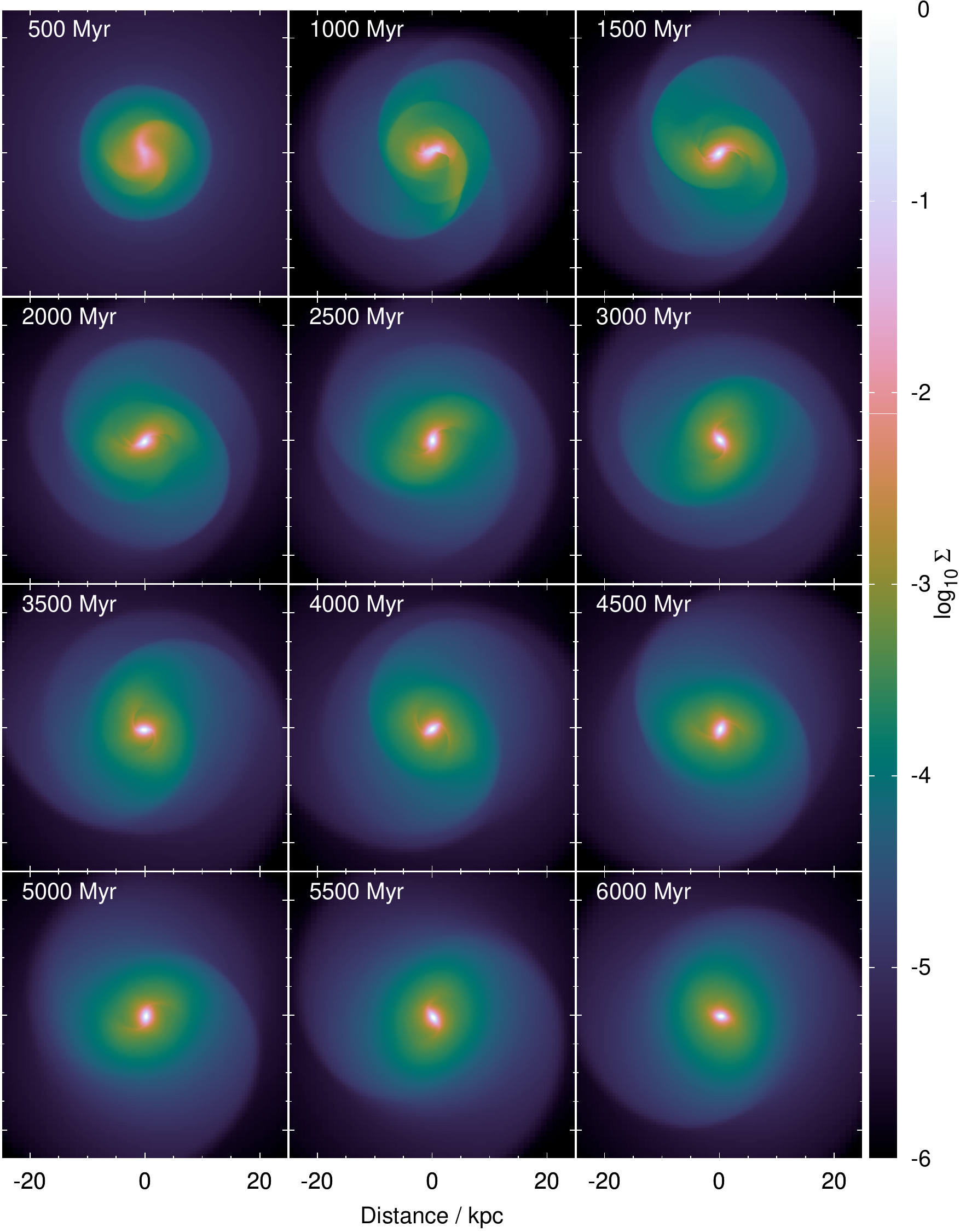}
	\caption{Similar to Figure \ref{Gas_face_on}, but now showing the gas in our isolated 100~kK simulation. Notice the lack of small-scale features, an expected consequence of a much hotter and thicker disk.}
	\label{Gas_face_on_100kK}
\end{figure*}

The gas in this simulation is shown in Figure \ref{Gas_face_on_100kK} using the same face-on view. The high temperature causes a distinct lack of small-scale features, making the appearance quite different to our cooler isolated model at 25~kK (Figure \ref{Gas_face_on}).

\end{document}